\documentclass[a4paper,11pt]{article}
\pdfoutput=1 
\usepackage{jcappub} 
\usepackage[T1]{fontenc}

\usepackage{amsmath}
\usepackage{bm}
\usepackage{graphicx,color}
\usepackage{tikz}
\usepackage{multirow}
\usepackage{makecell}
\usepackage{mathtools}

\usetikzlibrary{arrows.meta} 
\usepackage{layouts}

\renewcommand{\vec}[1]{\boldsymbol{\bm{#1}}}
\newcommand{\vecs}[1]{\tilde{\vec{#1}}}
\newcommand{\dg}{\delta_\mathrm{g}}
\newcommand{\dgd}{\delta_\mathrm{g,det}}
\newcommand{\dgds}{\tilde{\delta}_\mathrm{g,det}}
\newcommand{\tracer}{\mathrm{g}}
\newcommand{\pp}{{||}}
\newcommand{\op}{\mathcal{O}}
\newcommand{\uop}{\mathcal{U}}
\newcommand{\los}{\hat{\vec{n}}}
\newcommand{\ha}{\mathcal{H}}
\newcommand{\ee}{\mathrm{e}}
\newcommand{\dirac}{\mathrm{D}}
\newcommand{\lagrangian}{^\mathrm{L}}
\newcommand{\lin}{^{(1)}}
\newcommand{\Rstar}{R_*}
\newcommand{\Rvstar}{R_\mathrm{v*}}
\newcommand{\grad}{\vec{\nabla}}
\newcommand{\lpt}{\mathrm{LPT}}
\newcommand{\fwd}{\mathrm{G,fwd}}
\newcommand{\eul}{\mathrm{G,Eul}}
\newcommand{\mpc}{\mathrm{Mpc}}
\newcommand{\LCDM}{$\Lambda$CDM}
\newcommand{\tr}{\mathrm{tr}}
\newcommand{\oporderPT}{n_\mathrm{pt}}
\newcommand{\oporderDeriv}{n_\mathrm{deriv}}
\newcommand{\oporderV}{n_\mathrm{v}}
\newcommand{\udgpp}{u_{\tracer,\mathrm{det}\pp}}

\title{Cosmology inference at the field level from biased tracers in redshift-space}

\author[a,b]{Julia Stadler,}
\author[a,b]{Fabian Schmidt,}
\author[a]{Martin Reinecke}
\affiliation[a]{Max-Planck-Institut für Astrophysik, Karl-Schwarzschild-Str. 1, 85748 Garching, Germany}
\affiliation[b]{Excellence Cluster ORIGINS, Boltzmannstr. 2, 85748 Garching, Germany}
\emailAdd{jstadler@mpa-garching.mpg.de}
\emailAdd{fabians@mpa-garching.mpg.de}
\emailAdd{martin@mpa-garching.mpg.de}

\abstract{Cosmology inference of galaxy clustering at the field level with the EFT likelihood in principle allows for extracting all non-Gaussian information from quasi-linear scales, while robustly marginalizing over any astrophysical uncertainties. A pipeline in this spirit is implemented in the \texttt{LEFTfield} code, which we extend in this work to describe the clustering of galaxies in redshift space. Our main additions are: the computation of the velocity field in the LPT gravity model, the fully nonlinear displacement of the evolved, biased density field to redshift space, and a systematic expansion of velocity bias. We test the resulting analysis pipeline by applying it to synthetic data sets with a known ground truth at increasing complexity: mock data generated from the perturbative forward model itself, sub-sampled matter particles, and dark matter halos in N-body simulations. By fixing the initial-time density contrast to the ground truth, while varying the growth rate $f$, bias coefficients and noise amplitudes, we perform a stringent set of checks. These show that indeed a systematic higher-order expansion of the velocity bias is required to infer a growth rate consistent with the ground truth within errors. Applied to dark matter halos, our analysis yields unbiased constraints on $f$ at the level of a few percent for a variety of halo masses at redshifts $z=0,\,0.5,\,1$ and for a broad range of cutoff scales $0.08\,h/\mpc \leq \Lambda \leq 0.20\,h/\mpc$. Importantly, deviations between true and inferred growth rate exhibit the scaling with halo mass, redshift and cutoff that one expects based on the EFT of Large Scale Structure. Further, we obtain a robust detection of velocity bias through its effect on the redshift-space density field and are able to disentangle it from higher-derivative bias contributions.
}

\begin{document}
\maketitle
\flushbottom

\section{Introduction}
\label{sec:intro}

The universe's Large Scale Structure (LSS) carries abundant cosmological information and can, among other tracers, be probed by the clustering of galaxies. After the recent success of SDSS BOSS/eBOSS \cite{BOSS:2016wmc,eBOSS:2020yzd}, a new generation of (spectroscopic) surveys either has started data taking recently or will soon commence operations, such as DESI \cite{DESI:2016fyo}, Euclid \cite{Amendola:2016saw}, PFS \cite{2014PASJ...66R...1T} and SphereX \cite{Dore:2014cca}. The new data sets provide a considerable improvement in survey volume and depth. Fully exploiting the data, however, imposes considerable challenges for the theoretical modeling. Firstly, galaxies are biased tracers of the underlying density field \cite{Desjaques_2018}; connecting the two becomes more complicated as one proceeds to smaller scales. Any uncertainties in this relation need to be marginalized over reliably in a cosmology analysis. Secondly, the density field becomes increasingly nonlinear towards smaller scales, and the resulting non-Gaussianities contain information which is not captured by an analysis in terms of the power spectrum alone. Both issues can be addressed by a field-level analysis within the framework of Effective Field Theory (EFT) of LSS \cite{Baumann:2010tm, Carrasco:2012cv}.

The EFT analysis operates on quasi-linear scales, introducing a cutoff wavenumber $\Lambda$ up to which the density contrast of galaxies, $\dgd$, is expanded perturbatively. Wavenumbers larger than $\Lambda$ are ignored in the analysis. The baryonic and gravitational back-reaction of small scales onto the large modes and the bias of galaxies is captured by an expansion of the form (see  \cite{Desjaques_2018} for a review)
\begin{equation}
\dgd(\vec{x}) = \sum_\op b_\op\, \op(\vec{x})\,.
\label{eq:intro-bias-expansion}
\end{equation}
Here, the operators $\op$ are constructed from local gravitational observables, that is from the density and tidal field as well as products and spatial derivatives thereof. They can be ranked by their orders in perturbations and derivatives. Each operator comes with an a-priori unknown coefficient, whose value needs to be fitted to the data. Since galaxy formation is a local process and higher-order operators are successively suppressed in the perturbative regime, this expansion is expected to converge, provided the cutoff is chosen smaller than the nonlinear scale (i.e. $\Lambda\lesssim0.25\,h/\mpc$ at $z=0$).

The field-level approach then compares the observed galaxy density to theory predictions on a pixel-by-pixel basis over the full survey volume. As such, it extracts all possible information from the observations, up to the cutoff. The challenge of the method is that it requires a statistical description of the galaxy density in the late-time universe, conditional on the cosmological parameters. No such closed-form expression currently exists. However, CMB observations have shown that primordial density perturbations are very close to a Gaussian random field \cite{Planck:2018nkj}. The analysis therefore starts from a Gaussian prior on the initial density contrast and follows the gravitational evolution of dark matter into present-day structures. Following the principles of the EFT, eq.~(\ref{eq:intro-bias-expansion}) can be used to relate the evolved dark matter distribution to the density contrast of biased tracers, such as galaxies or halos (throughout this work we use all three terms interchangeably, depending on the context). The stochasticity of tracers around the predicted mean is captured by the EFT likelihood \cite{Schmidt:2018bkr, Elsner:2019rql, Cabass:2019lqx}. In the forward model, cosmological parameters enter through their impact on the spectrum of primordial fluctuations and on their gravitational evolution. The exploration of the posterior, i.e. the joint distribution of prior and likelihood, usually is done by sampling algorithms such as Hamiltonian Monte Carlo (HMC) \cite{2011hmcm.book..113N}. In summary, the method yields a fully Bayesian analysis of galaxy clustering and constrains the initial conditions in the survey volume alongside with the cosmological parameters and bias coefficients.

The recovery of initial conditions from simulated mock galaxies at fixed cosmology has been explored with various forward models, such as Lagrangian Perturbation Theory (LPT) \cite{2013MNRAS.432..894J, 2013MNRAS.429L..84K, 2013ApJ...772...63W}, full particle-mesh simulations \cite{Wang:2014hia} or neural networks \cite{Modi:2018cfi, Shallue:2022mhf, Modi:2022pzm, Dai:2022dso, Qin:2023dew, Jindal:2023qew, Charnock:2019rbk}. Field-level analyses have also been applied to infer initial conditions in the nearby universe \cite{Jasche:2018oym} and from the BOSS SDSS-III catalog \cite{Lavaux:2019fjr} at fixed cosmology. These studies used a parametric bias expansion obtained from N-body simulations \cite{Neyrinck:2013ezr} and a bias expansion in powers of the (locally averaged) density contrast, respectively. The ability of a perturbative bias expansion together with the LPT gravity model to describe a biased tracer field has been studied for halos in the rest frame \cite{Schmidt:2018bkr, Schmittfull:2018yuk} as well as in redshift space \cite{Schmittfull:2020trd}. These works found the residuals of the best-fit bias expansion consistent with the theoretical noise expectation. Recently, the joint inference of cosmology parameters and initial conditions was explored for mock data sets of biased tracers, generated with a LPT gravity model \cite{Kostic:2022vok}.

In this work, we continue the development of a field-level analysis for cosmological parameters \cite{Schmidt:2018bkr, Elsner:2019rql, Cabass:2019lqx, Schmidt:2020viy, Schmidt:2020tao, Schmidt_2021, Babic:2022dws, Kostic:2022vok} in the \texttt{LEFTfield} code. This pipeline uses the EFT bias expansion and likelihood  as outlined above to allow for rigorous control of bias and the back-reaction from small scales. 
The sampling over initial conditions increases the uncertainty since it incorporates cosmic variance (apart from imposing substantial computational demands). Thus, a very stringent test is the analysis of simulations with known ground truth at fixed initial conditions.
Indeed, analyzing dark matter halos from N-body simulations in their rest frame demonstrated unbiased constraints on the amplitude of density perturbation $\sigma_8$, at a level of 4-8\% \cite{Schmidt:2020viy,Schmidt:2020tao}. Since $\sigma_8$ and the linear bias parameter are perfectly degenerate in the linear regime, this test explicitly demonstrates the extraction of information from quasi-linear scales. 

As the next step to bring the technique closer to its application to survey data, we here extend the forward model to account for redshift-space distortions (RSD). The apparent displacement of galaxies in spectroscopic surveys due to their peculiar motions leads to enhanced clustering along the line of sight (LOS) \cite{1987MNRAS.227....1K,1992ApJ...385L...5H}. The displacement depends on the growth rate of structure $f$, and as such carries cosmological information itself. Since the forward-modeling, field-level analysis follows the evolution of density perturbations in the survey volume, it provides all information required to predict the large-scale velocity field. With this, the field of biased tracers can be transformed to redshift space self-consistently, where one performs the comparison with observations. The statistics of biased tracers in redshift space have been studied previously and are captured by the EFT likelihood in redshift space \cite{Cabass:2020jqo}. 

The main objective of this work thus is twofold. First, we develop the numerical model to compute the density of biased tracers in redshift space (section \ref{sec:model}). This also includes a systematic expansion of velocity bias. Second, we test the model at fixed initial conditions by applying it to a set of tracer fields with increasing complexity: mock data generated from our forward model (section \ref{sec:mocktests}), and matter particles (section \ref{sec:matter-inference}) and halos from N-body simulations (section \ref{sec:results}). We conclude in section \ref{sec:conclusions}. 

\section{Modeling biased tracers in redshift space}
\label{sec:model}

The forward model needs to predict the late-time density of biased tracers for any given realization of the initial conditions. Here, we extend the gravity and bias model introduced in \cite{Schmidt_2021} to provide the tracer field in redshift space. This includes the computation of the large-scale matter velocity field (section \ref{sec:model-lpt}), the introduction of an expansion for the velocity bias (section \ref{sec:model-bias}) and the transformation of the biased density field from the tracer rest frame to redshift space (\ref{sec:model-ztransform}). The transformation to redshift space also affects the tracers' stochasticity \cite{Cabass:2020jqo} and thus the EFT-likelihood (section \ref{sec:model-likelihood}). The implementation is summarized in section \ref{sec:model-summary}, where we also comment on numerical details, and we compare it to previous approaches in the literature in section \ref{sec: model-comparison}.

\subsection{Matter density and velocity fields in Lagrangian Perturbation Theory}
\label{sec:model-lpt}

Our forward model for the gravitational evolution of the matter field is based on n-th order Lagrangian Perturbation Theory (nLPT) \cite{Baldauf:2015tla}. In contrast to the Eulerian description \cite{Taruya:2018jtk}, LPT shows better correlation with the full nonlinear density contrast \cite{2012JCAP...04..013T, Tassev:2013rta} and straightforwardly allows to incorporate a bias expansion of the form of eq.~(\ref{eq:intro-bias-expansion}). A further advantage is the computation of the velocity field at little additional cost, which allows for the effective modeling of RSDs. Here, we recall the main steps required to predict the density evolution \cite{Schmidt_2021}, and explain how the velocity field is obtained.

LPT describes the gravitational evolution of the cosmic density field in terms of particle trajectories,
\begin{equation}
\vec{x}(\tau) = \vec{q} + \vec{s}(\vec{q},\tau)\,,
\label{eq:model-lagrangian-shift}
\end{equation}
where $\tau$ denotes conformal time, and $\vec{q}$ is the initial or Lagrangian position of the particle, defined at time $\tau_0$. The displacement vector $\vec{s}(\vec{q},\tau)$ vanishes as $\tau$ approaches $\tau_0$, and its evolution is determined by the continuity, Euler and Poisson equations. To find a perturbative solution, $\vec{s}$ is expanded as \cite{Buchert:1992ya}
\begin{equation}
\vec{s}(\vec{q},\tau) = \sum_{n=1}^\infty \vec{s}^{(n)}(\vec{q},\tau)\,.
\label{eq:model-desplacement-expansion}
\end{equation}
Further, it is useful to decompose the displacement into a curl-free and a divergence-free contribution,
\begin{equation}
\vec{s}(\vec{q},\tau) = \frac{\vec{\nabla}}{\nabla^2} \sigma(\vec{q},\tau) - \frac{1}{\nabla^2} \vec{\nabla}\times \vec{t}(\vec{q},\tau)\,,
\end{equation} 
which both can be expanded analogously to eq.~(\ref{eq:model-desplacement-expansion}). This allows to derive recursion relations \cite{Rampf:2012up, Zheligovsky:2013eca, Matsubara:2015ipa} and to solve for $\sigma$ and $\vec{t}$ to any desired order. The solution is possible for any cosmic expansion history, given it starts from an initial epoch of matter-domination during which $\tau_0$ is chosen \cite{Schmidt_2021, Ehlers:1996wg}. Here, however, we assume a purely matter-dominated Einstein-de Sitter (EdS) universe and further neglect the curl contributions. These approximations greatly simplify the computation at a small loss in accuracy: the transverse component only begins at third order and deviations from an EdS expansion history affect the final density contrast at the sub-percent level \cite{Schmidt_2021}. Switching the time coordinate from $\tau$ to $\lambda=\ln D$, where $D$ is the growth factor, the time evolution of the displacement is
\begin{equation}
\sigma^{(n)}(\vec{q},\lambda) = \ee^{n\lambda}\, \sigma^{(n)}(\vec{q},\lambda_0)\,.
\label{eq:model-sigma-timedependence}
\end{equation}
At lowest order, $\sigma\lin=-\delta\lin$ gives the Zel’dovich approximation \cite{1970A&A.....5...84Z, White:2014gfa}; the higher-order terms can be found in appendix A of \cite{Schmidt_2021}.

Now, it follows directly from the geodesic equation of non-relativistic particles that the velocity is $\vec{v}\left(\vec{q},\tau\right) = \dot{\vec{s}}\left(\vec{q},\tau\right)$. From eq.~(\ref{eq:model-sigma-timedependence}) we then obtain
\begin{equation}
\frac{\dot{\vec{s}}(\vec{q},\lambda)}{\ha} = f\,\vec{s}'(\vec{q},\lambda) 
= f\,\sum_{n=1}^\infty n\, \vec{s}^{(n)}
\left(\vec{q}, \lambda\right)\,,
\label{eq:model-s-prime}
\end{equation} 
where derivatives with respect to $\lambda$ are denoted by a prime. Further, $\ha = aH$ is the conformal Hubble rate, $f = \partial \ln D/\partial \ln a$ the growth rate, and $a$ the scale factor. Hence, once the displacement field has been constructed, summing up all contributions weighted by their perturbative order readily yields the desired velocity field in Lagrangian space.

We compute the displacement field and the velocity on a uniform grid in Lagrangian space from the initial conditions smoothed at the cutoff scale $\Lambda$ with a sharp-$k$ filter. However, we need the evolved matter density contrast $\delta$ and the velocity field in terms of their Eulerian-frame coordinate $\vec{x}$. The coordinate transformation between Lagrangian and Eulerian frame defined in eq.~(\ref{eq:model-lagrangian-shift}) is mass conserving, and hence its Jacobian is given by
\begin{equation}
1 + \delta(\vec{x},\tau) =  \left|\frac{\partial\vec{x}\left(\vec{q},\tau\right)}{\partial\vec{q}}\right|^{-1}_{\vec{x}=\vec{x}(\vec{q},\tau)}\,.
\end{equation}
More generally, we can write the transformation of any Lagrangian-space field $\op\lagrangian\left(\vec{q}\right)$ as
\begin{equation}
\left[1 + \delta\left(\vec{x}, \tau \right) \right]\,\op\left(\vec{x}\right) 
= \int d^3\vec{q}~ \op\lagrangian\left(\vec{q}\right)\delta^\dirac\left[\vec{x} - \vec{q} - \vec{s}\left(\vec{q},\tau\right)\right]\,.
\label{eq:model-shift-to-eul}
\end{equation}
We implement this transformation by generating an ensemble of $N_\eul^3$ uniformly-spaced pseudo-particles, whose mass is $\op\lagrangian\left(\vec{q}\right)$ at the particles' location. The particles are then displaced by $\vec{s}(\vec{q},\tau)$ and their density is re-assigned to a grid of size $N_\eul$ with a cloud-in-cell (CIC) kernel \cite{Hockney_1988}. The use of pseudo-particles in the transformation from Lagrangian to Eulerian coordinates ensures mass conservation at machine precision. Importantly, it prevents the creation of spurious noise on large scales which could not be guaranteed if the displacement instead was expanded perturbatively \cite{Schmidt_2021}. Now, from eq.~(\ref{eq:model-shift-to-eul}) the displacement of a unit field obviously yields the evolved density contrast, and the Eulerian-frame momentum density is obtained by weighting the particles with $\vec{s}'\left(\vec{q},\tau\right)$. Since the displacement to redshift space depends on the LOS-projected velocity, we evaluate the displacement for this component of $\vec{s}'$ only and subsequently divide out the Jacobian.

\subsection{Bias Expansion}
\label{sec:model-bias}
Having computed the nonlinear matter density field, we next need to specify the bias operators in eq.~(\ref{eq:intro-bias-expansion}). The rest-frame density of biased tracers can be expanded in terms of invariants of the Lagrangian distortion tensor \cite{Schmidt_2021}, as we briefly review here. Further, velocity bias terms become important when the tracer density is transformed to redshift space (section \ref{sec:model-ztransform}), and consequently, we introduce a systematic expansion in the subsection \ref{sec:model-bias-vbias}.

\subsubsection{Expansion in perturbations and higher derivatives}
\label{sec:model-bias-general}

At leading order in spatial derivatives, the Lagrangian distortion tensor $\vec{H}(\vec{q},\tau)$ captures all gravitational observables for an observer comoving with the matter trajectory \cite{Mirbabayi:2014zca}. It is closely related to the Jacobian of the transformation between Lagrangian and Eulerian space and defined as,
\begin{equation}
H_{ij}(\vec{q},\tau) = \frac{\partial s_j(\vec{q},\tau)}{\partial q_i}\,.
\end{equation}
The galaxy density in Lagrangian coordinates can be formally expressed as a functional in time of the symmetric part of $\vec{H}$, denoted by $\vec{M}$,
\begin{equation}
\dgd = \int_{\tau_0}^{\tau} d\tau'~F_\tracer\left[\vec{M}\left(\vec{q},\tau'\right); \tau', \tau \right]\,.
\label{eq:model-bias-general-time-functional}
\end{equation}
Note that $\vec{H}$ is symmetric anyway under the assumptions made in section \ref{sec:model-lpt}, and hence the distinction between $\vec{H}$ and $\vec{M}$ is superficial in this context. We still maintain it here to keep the notation consistent with \cite{Schmidt_2021}. The functional in eq.~(\ref{eq:model-bias-general-time-functional}) can further be expanded in perturbative powers of $\vec{M}^{(n)}$, that follow from the expansion of the shift vector $\vec{s}$ in eq.~(\ref{eq:model-desplacement-expansion}). To obtain a bias expansion of the form in eq.~(\ref{eq:intro-bias-expansion}), the time integral in eq.~(\ref{eq:model-bias-general-time-functional}) is performed formally. The latter is possible if the spatial and temporal dependency of $\vec{H}$ factorize, such as in eq.~(\ref{eq:model-sigma-timedependence}). The bias expansion in Lagrangian space then is \cite{Schmidt_2021}
\begin{equation}
\dgd(\vec{q},\tau) = \sum_\op b_\op\left(\tau\right)\, \op^\mathrm{L}(\vec{q},\tau)\,,
\end{equation}
where the Lagrangian bias operators, $\op^\mathrm{L}$, are given by all rotational invariants that one can construct from the symmetric part of the distortion tensor $\vec{M}^{(n)}$ and of products thereof. Since each order in perturbation theory obeys a different time-dependence, they also enter the expansion with independent bias coefficients $b_\op$. 

At lowest order, there is only one rotational invariant, namely $\mathrm{tr}[\vec{M}^{(1)}] \propto \delta^{(1)}$. We replace it by $\delta$ to keep the familiar linear bias relation explicit. This change can be viewed as a rotation of the basis of bias operators and is absorbed in a redefinition of the bias coefficients. To obtain the higher-order operators, we first construct all relevant scalar invariants from $\vec{M}^{(n)}$ and then form all independent products, up to the desired order $\oporderPT$. At second order, this yields in total three bias operators,
\begin{equation}
\delta\,,\quad\sigma^2\,,\quad \tr[M^{(1)} M^{(1)}]\,.
\label{eq:model-list-bias-2order}
\end{equation}
In appendix \ref{sec:appendix-bias-list}, we list all operators up to third order, which are the operators relevant for this work.

After constructing the set of Lagrangian bias operators, each of them is transformed to the Eulerian frame according to eq.~(\ref{eq:model-shift-to-eul}), again by computing the displacement of a set of weighted pseudo-particles. The transformed operators,
\begin{equation}
\op\left(\vec{x},\tau\right) = \left[1 + \delta\left(\vec{x},\tau\right)\right] \op^\mathrm{L}\left(\vec{q}\left[\vec{x}\right],\tau\right)\,,
\label{eq:model-lagrangian-bias-expansion}
\end{equation}
correspond to the desired result $ \op^\mathrm{L}\left(\vec{q}\left[\vec{x}\right]\right)$ at leading order. At higher order, the Jacobian prefactor can be absorbed by a redefinition of the bias parameters. 

In addition to the leading-order operators, there are higher-order-in-derivatives contributions, which capture the non-locality of galaxy formation. Their relevance scales as $(\Lambda \Rstar)^2$, where $\Rstar$ is a tracer-specific length scale. For halos, $\Rstar$ is expected to roughly coincide with the Lagrangian radius $R_\mathrm{L}$, i.e. the comoving radius of a sphere containing the mass of the halo \cite{Desjaques_2018},
\begin{equation}
R_\mathrm{L}(M) = \left(\frac{3\,M}{4\pi\, \Omega_\mathrm{m}\bar{\rho}}\right)^{1/3}\,.
\label{eq:model-lagrangian-radius}
\end{equation}
Here, $\Omega_\mathrm{m}\, \bar{\rho}$ 
is the mean comoving matter density. The corresponding operators in the bias expansion are constructed from $\grad ... \grad \vec{M}^{(n)}$, whereby the total number of derivatives needs to be even so that all indices can be contracted. This corresponds to the absence of a preferred direction in the rest frame. Still, there can be an odd number of derivatives acting on any single instance of $\vec{M}^{(n)}$.

The memory required for constructing the tensor $\grad ... \grad \vec{M}^{(n)}$ scales exponentially with the number of derivatives and there are degeneracies between these higher-order operators, which are difficult to remove. For this reason, we only consider a subset of derivative operators in this work, namely those that can be constructed by applying the Laplace operator, possibly multiple times, to the leading-order in derivative rotational invariants.\footnote{Note that this is a subset of the higher-derivative operators considered in \cite{Schmidt_2021}, who also included operators of the form $\partial_i \op \partial^i \op'$ for invariants $\op,\op'$.} As in \cite{Schmidt_2021}, we construct the higher-derivative operators in the Eulerian frame, i.e. we compute $\grad^2_{\vec{x}} \op(\vec{x})$ rather than $\grad_{\vec{q}}^2 \op^\mathrm{L}(\vec{q})$ followed by a displacement. For a complete set of operators, this change would be equivalent to a rotation of the operator basis, and is absorbed by the coefficients.

To decide which higher-order-in-derivative operators need to be included, we compare their relevance to that of the highest-order leading operator, characterized by $\oporderPT^\mathrm{max}$. That is, we include a higher-derivative operator in our forward model if it obeys
\begin{equation}
\left(\frac{\Lambda}{k_\mathrm{NL}}\right)^{\oporderPT(\op)\times\left(3 + n_\mathrm{L}\right)/2}
\times
\left(\Lambda \Rstar\right)^{2\, \oporderDeriv(\op)}
\ge
\left(\frac{\Lambda}{k_\mathrm{NL}}\right)^{\oporderPT^\mathrm{max}\times\left(3 + n_\mathrm{L}\right)/2}\,.
\end{equation}
Here, $\left. n_\mathrm{L}=\partial \ln P(k)/\partial \ln k\right|_{k=\Lambda}$ and $\oporderDeriv(\op)$ is the number of Laplace operators applied to an operator of order $\oporderPT(\op) \leq \oporderPT^\mathrm{max}$. 

In practice, this would mean that for a fixed expansion order the set of higher-derivative operators changes depending on the cutoff scale and mass bin. Instead, we would like to be able to compare directly between all results at identical expansion order. Therefore, we determine the set of higher-order in derivative operators at fixed parameter values which are
\begin{equation}
z=0\,,
\quad
\Lambda = 0.10\,h/\mpc\,,
\quad
k_\mathrm{NL} = 0.25\,h/\mpc\,,
\quad
R_* = 5\,\mpc/h\,.
\end{equation}
At $n^\mathrm{max}_\mathrm{pt}=3$, which we adopt as default for this study, and for all halo mass bins except the highest, this corresponds to one higher-order-in-derivative operator, $\nabla^2\delta$. Higher-order operators that our convention neglects arise for $\Lambda \geq 0.18\,h/\mpc$ and for the highest of the four halo mass bins which we consider in section \ref{sec:results}. Thus, the distinction between the full- and the sub-set of higher-derivative operators does not make a difference for the bulk of the results presented below.

\subsubsection{Velocity bias}
\label{sec:model-bias-vbias}

On large scales, the equivalence principle ensures that galaxies and matter move along the same trajectories, and hence velocity bias arises only as a higher-derivative effect. In the case of galaxies, velocity bias is produced by baryonic effects such as stellar winds and supernova feedback. More importantly, also halos, which form in special, biased, regions of the density and velocity field over some finite-size volume, are subject to velocity bias \cite{1986ApJ...304...15B,peacock/lumsden/heavens:1987,percival/schaefer:2008,Desjacques:2008jj}. In the following sections, we show that velocity bias significantly impacts the inference of the growth rate. Here, we therefore introduce a systematic velocity bias expansion. Since the transformation to redshift space depends on $\udgpp = \vec{v}_{\tracer,\mathrm{det}} \cdot \los/\ha$, where $\los$ denotes the LOS-direction and projections onto the LOS are indicated by a subscript parallel symbol, we formulate the expansion for this quantity directly.

The relative velocity between galaxies and matter is in principle a local observable; as such its large-scale expansion can only depend on gravitational observables. Further, the velocity bias operators need to be vectorial quantities. A general expansion of the LOS velocity,
\begin{equation}
\udgpp\left(\vec{x},\tau\right) = u_\pp\left(\vec{x},\tau\right) + \sum_\uop \beta_\uop\left(\tau\right)\, \uop\left(\vec{x},\tau\right)\,,
\label{eq:model-vbias-expansion}
\end{equation}
thus considers the LOS component $\uop$ of all 3-vectors that can be constructed from contracting the distortion tensors $M^{(n)}_{ij}$ with themselves and spatial derivatives. 

We generate the velocity bias operators in a similar way as the higher-derivative expansion. A non-complete set can be obtained by acting with an odd number of derivatives along the LOS on the operators of the isotropic bias expansion in the Eulerian frame. The relevance of these new operators scales as
\begin{equation}
\left(\frac{\Lambda}{k_\mathrm{NL}}\right)^{\oporderPT(\op)\times\left(3 + n_\mathrm{L}\right)/2}
\times
\left(\Lambda \Rstar\right)^{2\, \oporderDeriv(\op)}
\times
\left(\Lambda \Rvstar\right)^{ \oporderV(\op)}
\,,
\label{eq:model-vbias-relevance}
\end{equation}
where we have introduced a new length scale $\Rvstar$ and $\oporderV(\op)$ denotes the number of LOS derivatives. For dark matter halos, we expect $\Rvstar$ to coincide with the Lagrangian radius, eq.~(\ref{eq:model-lagrangian-radius}), and thus also with the scale $\Rstar$ that controls the relevance of higher-derivative operators.

At leading order, there is a single operator for which several (at leading order) equivalent expressions (cf. eq.~\ref{eq:model-linear-velocity} and below eq.~\ref{eq:model-sigma-timedependence}) are possible, namely
\begin{equation}
\partial_\pp\delta\,,\quad
\nabla^2 u_\pp\,,\quad 
 \partial_\pp\sigma\,.
\label{eq:model-vbias-linear}
\end{equation} 
Since the rest-frame density bias expansion (cf. eq.~\ref{eq:model-list-bias-2order}) contains $\delta$, our construction automatically picks the first of these three options. At $\oporderPT^\mathrm{max}=3$, the order we usually work in, the LOS derivatives of density bias operators do not yield a complete set of operators anymore. Indeed, given the list of second order bias operators in eq.~(\ref{eq:model-list-bias-2order}), one can construct four relevant LOS contractions,
\begin{equation}
\partial_\pp\delta\,,\quad
\partial_\pp\left(\sigma^2\right)\,,\quad
\partial_\pp \mathrm{tr}\left[M^{(1)} M^{(1)}\right]\,,\quad
M^{(1)}_{ji} \partial_i M^{(1)}_{\pp j} \,.
\label{eq:model-vbias-list}
\end{equation}
Only the first three operators are generated automatically, but we have also implemented the fourth in our forward model to be able to study its impact.

\subsection{Transformation to redshift space}
\label{sec:model-ztransform}

The mapping between the galaxy rest-frame or Eulerian frame coordinate $\vec{x}$ and redshift space position $\vecs{x}$ is a coordinate transformation given by 
\begin{equation}
\vecs{x}\left(\tau\right) = \vec{x} + u_\pp(\vec{x}, \tau)\, \los(\vec{x})\,.
\label{eq:model-rsd-displacement}
\end{equation}
On scales large enough that no shell crossing has occurred, this transformation is a one-to-one mapping, and its Jacobian is
\begin{equation}
\left|\frac{\partial\vecs{x}\left(\tau\right)}{\partial\vec{x}\left(\tau\right)}\right| = 1 + \partial_{\vec{x}_\pp} u_\pp \left(\vec{x},\tau\right) \equiv 1 + \eta\left(\vec{x},\tau\right)\,.
\end{equation}
The transformation between rest frame and redshift space conserves mass, or equivalently the tracer number density, and hence the density contrast of any biased tracer in redshift space, $\dgds$, is
\begin{equation}
\dgds\left(\vecs{x},\tau\right) +1 = \left. \frac{1 + \dgd\left(\vec{x},\tau\right)}{1 + \eta\left(\vec{x},\tau\right)} \right|_{\vec{x}=\vec{x}(\vecs{x},\tau)}\,.
\label{eq:model-redshift-space-density-contrast}
\end{equation}

So far, these arguments apply at any order in perturbation theory. The leading-order impact of RSDs follows from the linear density-velocity relation,
\begin{equation}
\vec{u}\lin\left(\vec{k},\tau\right) = -f\left(\tau\right)\, \frac{i\vec{k}}{k^2}\, \delta\lin\left(\vec{k}\right)\,,
\label{eq:model-linear-velocity}
\end{equation}
in combination with the linear bias relation $\dgd = b_\delta \delta$. If one further assumes a constant LOS, this yields the Kaiser formula \cite{1987MNRAS.227....1K} for the density contrast in redshift space
\begin{equation}
\dgds\lin\left(\vecs{x},\tau\right) = b_\delta\, \delta\lin\left(\vecs{x}, \tau\right) - \eta\lin\left(\vecs{x},\tau\right)\,. 
\label{eq:model-kaiser-formular-for-density}
\end{equation}
Since the difference between $\vec{x}\left[\vecs{x}\right]$ and $\vecs{x}$ in the spatial argument on the right hand side becomes only relevant at higher orders, we have written both sides of this expression as a function of $\vecs{x}$. Going to higher order, the expansion of $\vec{x}\left[\vecs{x}\right]$ leads to additional contributions, and eq.~(\ref{eq:model-redshift-space-density-contrast}) can be written as \cite{Desjaques_2018}
\begin{equation}
\dgds\left(\vecs{x},\tau\right) = \dgd\left(\vecs{x},\tau\right) + \sum_{n=1}^\infty \frac{(-1)^n}{n!}\, \partial_\pp^n\left[\udgpp^n\left(\vecs{x},\tau\right)\left(1+\dgd\left(\vecs{x},\tau\right)\right)\right]\,.
\label{eq:model-rsd-expansion}
\end{equation}

Rather than using this perturbative expression, we compute the transformation to redshift space fully non-linearly. As we did for the density contrast, we generate an ensemble of equally-spaced, weighted pseudo-particles in the Eulerian rest frame, shift each by the vector $\udgpp\left(\vec{x}\right)\,\los\left(\vec{x}\right)$ at its position and re-assign the density to a grid using a CIC kernel. This procedure transforms any field defined in the rest frame, $\op\left(\vec{x},\tau\right)$, to
\begin{equation}
\frac{\tilde{\op}\texttt{}\left(\vecs{x}, \tau\right)}{1 + \eta\left(\vecs{x},\tau\right)} 
=
\int d^3\vec{x}\, \op\left(\vec{x},\tau\right)\, \delta^\dirac\left[\vecs{x}
- \vec{x} - u_\pp(\vec{x},\tau)\,\los\left(\vec{x}\right)\right]\,.
\label{eq:model-rsd-displacement-genericop}
\end{equation}
The straightforward way to construct $\dgds$ would now be to perform the bias expansion in Eulerian space as described in section~\ref{sec:model-bias-general}, followed by a displacement of $\dgd+1$ to redshift space. This, however, is not ideal for computational efficiency. In the inference, we also need to vary the bias coefficients, and one would need to evaluate the redshift-space displacement anew for each proposed bias parameter value $b_\op$. Since the displacement is the most costly part of our forward model, it is more efficient to decompose the redshift-space density contrast as
\begin{equation}
\dgds\left(\vecs{x}, \tau\right) + 1= 
\frac{1 + b_\delta\,\delta\left(\vecs{x}, \tau\right)}{1 + \eta\left(\vecs{x}, \tau\right)} + \sum_{\op\neq\delta} b_\op\, \frac{\op\left(\vecs{x}, \tau\right)}{1 + \eta\left(\vecs{x}, \tau\right)}\,.
\label{eq:model-deltaDet-zspace}
\end{equation}
By displacing each term on the right side of eq.~(\ref{eq:model-deltaDet-zspace}) individually, the coefficients of the higher-order bias operators can be sampled more efficiently, without the need to re-evaluate the displacement to redshift space. This simplification, by construction, does not work for the linear bias parameter. Similarly, the displacement vector in eq.~(\ref{eq:model-rsd-displacement-genericop}) is given by the biased tracer velocity, and hence the displacement has to be re-evaluated whenever the velocity bias coefficients $\beta_\uop$ change.

In the numerical implementation we chose the LOS direction $\los\left(\vec{x}\right)$ to be constant and aligned with one of the coordinate axes, corresponding to the flat-sky approximation. However, this choice is only motivated by simplicity and convenience. No aspect of the computations outlined above explicitly demand a constant $\los$, and the forward model would generalize readily to a position-dependent LOS.

To discern the effect of the velocity bias operators onto the final redshift-space density contrast it is useful to consider the perturbative expansion in eq.~(\ref{eq:model-rsd-expansion}). At leading order, all velocity bias terms that we list in eq.~(\ref{eq:model-vbias-list}) pick up a second derivative along the line of sight. From the lowest-order velocity operator $\partial_\pp \delta$ then arises a contribution $\propto \partial_\pp^2\delta$ and three more terms at next to leading order, $(\partial_\pp^2\delta)\delta$, $(\partial_\pp\delta)^2$ and $\partial_\pp^2 \left[ u_\pp \left( \partial_\pp \delta \right) \right]$. They all are multiplied by the same coefficient, $\beta_{\partial_\pp\delta}$. In general, all velocity-induced contributions to the density contrast have an explicit dependence on the LOS, thus they are not included in the isotropic rest-frame expansion of section \ref{sec:model-bias-general}.

Bias terms with an explicit LOS dependence can also arise from \emph{selection effects}, which arise if the probability of observing a given galaxy depends on the line of sight to it \cite{Desjacques:2018pfv}. If this is not the case, then the transformation from the galaxy rest frame to redshift space is number-conserving, which imposes constraints on coefficients of different line-of-sight dependent contributions to the density field in redshift space, as we have seen in the previous paragraph. As an example, the leading-order selection contribution is $b_\eta \eta$, which, in the absence of selection effects, has a fixed coefficient $b_\eta = -1$. Higher-order selection operators can be constructed from the distortion tensors in the same fashion as the isotropic density operators in section \ref{sec:model-bias-general}, now also contracting indices with the LOS as preferred direction. The next-to-leading selection operator, $\partial_\pp^2\delta$ would then be degenerate with the leading-order impact of velocity bias. Since we expect that all tracer samples considered in the following -- biased mocks, matter particles and halos in N-body simulations -- are free from selection effects, we only account for velocity bias operators in our forward model.

\subsection{Likelihood}
\label{sec:model-likelihood}
Up to this point, the forward model that we have discussed describes the deterministic part of the tracer density, that is it predicts the mean of the tracer density field given fixed large-scale modes. The stochasticity around that mean is captured by the EFT likelihood $\mathcal{P}\left(\left.\dg\right|\dgd\right)$. Formally, the EFT likelihood can be derived from the generating functional and its Fourier space expression for the density field in Eulerian frame is \cite{Elsner:2019rql,Cabass:2019lqx}
\begin{equation}
\ln \mathcal{P}\left(\left.\dg\right|\dgd\right) = -\frac{1}{2} \int_{|\vec{k}|<k_{\max}}\, \frac{d^3\vec{k}}{(2\pi)^3} \left\{ \frac{\left|\dg(\vec{k})-\dgd(\vec{k})\right|^2}{P_\epsilon(k)} 
+ \ln\left[2\pi P_\epsilon(k)\right]\right\}\,.
\label{eq:model-likelihood-fourier}
\end{equation}
Here, only modes up to the cutoff scale $k_{\max}$ contribute, and from now on we suppress time arguments for clarity. Generally, $k_{\max}$ should be chosen less than or equal to $\Lambda$, the cutoff in the initial conditions, since modes beyond that limit have no linear support and are exclusively excited by the nonlinear evolution. Further, the rest-frame Fourier space noise, $P_\epsilon(k) = P_{\epsilon_\tracer}^{(0)} + P_{\epsilon_\tracer}^{(2)} k^2 + ...$, is given by an expansion in higher-derivative orders. If the scale-dependent noise contributions can be neglected, the likelihood can be transformed to real-space. Density-dependent noise, which arises from the modulation of the noise amplitude by long-wavelength perturbations, can then readily be included in the real-space version of the likelihood \cite{Schmidt:2020tao}.

For the real-space EFT-likelihood, the coordinate transformation from real- to redshift space can be done explicitly at all orders in perturbations, allowing to derive the EFT likelihood in redshift space \cite{Cabass:2020jqo},
\begin{equation}
\ln \mathcal{P}\left(\left.\tilde{\delta}_\tracer\right|\dgds\right) = -\frac{1}{2} \int d^3\vecs{x}~ \left\{ \frac{\left|\tilde{\delta}_{\tracer}(\vecs{x}) - \tilde{\delta}_{\tracer,\det}(\vecs{x})\right|^2}{P^{(0)}_\epsilon/\left[1+\eta(\vecs{x})\right]} 
- \ln\left[\frac{1+\eta(\vecs{x})}{2\pi P_\epsilon}\right]\right\}\,.
\label{eq:model-likelihood-rsd}
\end{equation}
However, despite not relying on a perturbative expansion of the transformation, the RSD likelihood still assumes a one-to-one mapping from Eulerian frame to redshift space or the absence of shell crossing. The latter is marked by $\eta=-1$, and indeed for this value the RSD likelihood is singular. 

In the form given here, the EFT likelihood in redshift space assumes white, isotropic noise $\epsilon_\tracer\left(\vec{x}\right)$ in the Eulerian frame and explicitly accounts for the noise transformation to redshift space,
\begin{equation}
\tilde{\epsilon}_\tracer(\vecs{x}) = \left.\frac{\epsilon_\tracer(\vec{x})}{1 + \eta(\vec{x})}\right|_{\vec{x}=\vec{x}(\vecs{x})}\,.
\label{eq:redshift-space-noise}
\end{equation}
Apart from the transformation of the rest-frame tracer stochasticity, the halo density in redshift space is also affected by noise in the velocities $\vec{\epsilon}_{\vec{u}}$. Since the transformation to redshift space only affects the LOS coordinate of galaxies, velocity noise becomes manifest in the tracer density contrast as an anisotropic contribution. Further, from the perturbative expansion of the redshift-space transformation in eq.~(\ref{eq:model-rsd-expansion}) it is clear that velocity noise enters the density contrast with at least one spatial derivative, acting along the LOS. Any white anisotropic noise, in contrast, would be non-local and ill defined in the $k\to 0$ limit. The noise power spectrum up to next-to-leading order then is parametrized by \cite{Perko:2016puo, Desjacques:2009kt, Desjacques:2018pfv,Cabass:2019lqx}
\begin{equation}
P_\epsilon(k) = P_\epsilon^{(0)} + P_\epsilon^{(2)} k^2 + P_{\epsilon_\mu}^{(2)} (\mu k)^2  \,,
\label{eq:model-noise-power-pectrum}
\end{equation}
where $\mu=k_\pp/k$ is the cosine of the angle between $\vec{k}$ and the LOS direction $\los$.

Apparently, the real- and Fourier-space likelihoods capture different physical effects while neglecting others -- the transformation of noise to redshift space is captured by the former while the latter allows to incorporate velocity noise. However, it is important to stress that this is not a fundamental limitation of the formalism. The differences rather arise because either version of the likelihood attempts to analytically marginalize over all noise contributions. In principle, it would also be possible to integrate out one part of the noise and sample the other explicitly. An additional problem with the RSD likelihood is how one deals with cells where the forward model predicts $\eta \leq -1$. We encountered first instances of such a behavior already for relatively mild cutoffs of $\Lambda = 0.14\,h/\mpc$. In this work, we therefore focus on the Fourier-space likelihood, and we verify this choice by testing its impact explicitly in section \ref{sec:mocktests-likelihood}. For the future, it would be interesting to explore more involved noise models that combine position-dependent and scale-dependent effects.

In the numerical implementation, it is important to ensure that the noise power spectrum cannot take on negative values. Therefore, we rewrite eq.~(\ref{eq:model-noise-power-pectrum}) as
\begin{equation}
P_\epsilon(k) = P_\epsilon^{(0)} \left[1 + \sigma_{\epsilon,\,2}\,k^2 + \sigma_{\epsilon\mu,\,2}\,\left(\mu k\right)^2 \right]^2\,.
\label{eq:model-noise-parametrization}
\end{equation}
Further, the free parameter considered in the inference is $\sqrt{P_\epsilon^{(0)}}$, such that its square is granted to be positive (see also appendix~\ref{sec:appendix-bias-list}). At leading order in derivatives, there is a one-to-one mapping between the respective parameters of eq.~(\ref{eq:model-noise-power-pectrum}) and eq.~(\ref{eq:model-noise-parametrization}).

\subsection{Summary and numerical implementation}
\label{sec:model-summary}

\begin{figure}
\begin{tikzpicture}
\newlength{\widthS}
\setlength{\widthS}{1cm}

\newlength{\widthL}
\setlength{\widthL}{2.3cm}

\newlength{\siftD}
\setlength{\siftD}{3cm}

\newlength{\shiftS}
\setlength{\shiftS}{.35cm}

\newlength{\dispArrow}
\setlength{\dispArrow}{(\textwidth -3\shiftS -2\widthS-4\widthL)/2}

\newlength{\oA}
\setlength{\oA}{.0cm}

\node[anchor=west, rounded corners, text width=\widthS, draw, align=center] at (current page.west) (n1)
{$\delta_{\mathrm{ini},\Lambda}$};

\node[anchor=west, rounded corners, text width=\widthL, draw, align=center, xshift=\shiftS] at (n1.east) (n2)
{1\\ 
$s'_\pp/f$\\
$\left\{\op^\mathrm{L}\right\}_\mathrm{LD}$};

\node[anchor=north, rounded corners, yshift=-.5\baselineskip, align=center, draw, text width=\widthL, minimum height=1.25\baselineskip] at (n2.south) (s1)
{$\vec{s}$};

\node[anchor=east, rounded corners, text width=\widthL, draw, align=center, xshift=\textwidth/2-\shiftS/2] at (n1.west) (n3)
{$1+\delta$\\ $\left(1+\delta\right)u_\pp /f$\\ 
$\left(1{+}\delta\right)\{\op^\mathrm{L}\}_\mathrm{LD}$};

\node[anchor=west, rounded corners,text width=\widthL, draw, align=center, xshift=\textwidth/2+\shiftS/2] at (n1.west) (n4)
{$1+ b_\delta\delta$\\
$\left\{\op\right\}_{\op\neq\delta}$};

\node[anchor=north, rounded corners, draw, align=center, text width=\widthL, minimum height=1.25\baselineskip] at (n4.south|-s1.north) (s2)
{$\udgpp\,\los$};

\node[anchor=east, rounded corners, text width=\widthS, draw, align=center, xshift=\textwidth] at (n1.west) (n6)
{$\dgds$};
\node[anchor=east, rounded corners, text width=\widthL, draw, align=center, xshift=-\shiftS] at (n6.west) (n5)
{$\displaystyle \frac{1+ b_\delta\delta}{1+\eta}$\\[2pt]
$\left\{\tilde{\op}\right\}_{\op\neq\delta}$};

\node[anchor=south west, yshift=2\baselineskip, text width=\widthS+\widthL+.2cm, align=center] at (n1.north west) (l1)
{Lagrangian frame};
\node[anchor=west, text width=2\widthL+.2cm, align=center] at (n3.north west|-l1) 
{Eulerian frame};
\node[anchor=west, text width=\widthS+\widthL+.2cm, align=center] at (n5.north west|-l1) 
{Redshift space};

\draw[-{Latex[open]}, double] ([xshift=.0cm]n2.east) -- ([xshift=-.0cm]n3.west);
\draw[-{Latex[open]}, double] ([xshift=.0cm]n4.east) -- ([xshift=-.0cm]n5.west);
\draw[->, dashed] ([xshift=.1cm]s1.east) to[out=0, in=-90] ([xshift=\dispArrow/4, yshift=-.1cm] n2.east);
\draw[->, dashed] ([xshift=.1cm]s2.east) to[out=0, in=-90] ([xshift=\dispArrow/4, yshift=-.1cm] n4.east);

\draw[-{Latex[length=2mm]}] ([xshift=-\oA]n1.east) -- ([xshift=\oA] n2.west);
\draw[-{Latex[length=2mm]}] ([xshift=-\oA]n3.east) -- ([xshift=\oA] n4.west);
\draw[-{Latex[length=2mm]}] ([xshift=-\oA]n5.east) -- ([xshift=\oA] n6.west);
\draw[-{Latex[length=2mm]}] ([yshift=\oA]n1.south) to[out=-90, in=180] ([xshift=\oA] s1.west);
\draw[-{Latex[length=2mm]}] ([yshift=\oA]n3.south) to[out=-90, in=180] ([xshift=\oA] s2.west);

\draw[|-|] ([yshift=-\siftD] n1.east) -- ([yshift=-\siftD]n1.west)
node[midway, fill=white,inner sep=0.1pt]{$N_\mathrm{G,ini}$};
\draw[|-|] ([yshift=-\siftD, xshift=0.1cm] n1.east) -- ([yshift=-\siftD]n2.east)
node[midway, fill=white,inner sep=0.1pt]{$N_\fwd$};
\draw[|-|] ([yshift=-\siftD, xshift=.1cm] n2.east) -- ([yshift=-\siftD, xshift=-.1cm]n5.west)
node[midway, fill=white,inner sep=0.1pt]{$N_\eul$};
\draw[|-|] ([yshift=-\siftD] n5.west) -- ([yshift=-\siftD]n6.east)
node[midway, fill=white,inner sep=0.1pt]{$N_\mathrm{G,like}$};
\end{tikzpicture}
\caption{The perturbative forward model for the density contrast of biased tracers in redshift space. Double arrows represent particle displacement operations, dashed arrows denote the corresponding displacement vectors and solid arrows indicate field-level computations.  ``LD'' stands for operators at leading order in derivatives. The likelihood of the data $\tilde{\delta}_\tracer$ given the forward-modeled field $\dgds$ is evaluated via the EFT-likelihood, see section \ref{sec:model-likelihood}. Cosmological parameters impact the computation of both shift vectors, $\vec{s}$ and $\udgpp$; their posterior is explored by a slice sampler.}
\label{fig: model-flowchart}
\end{figure}
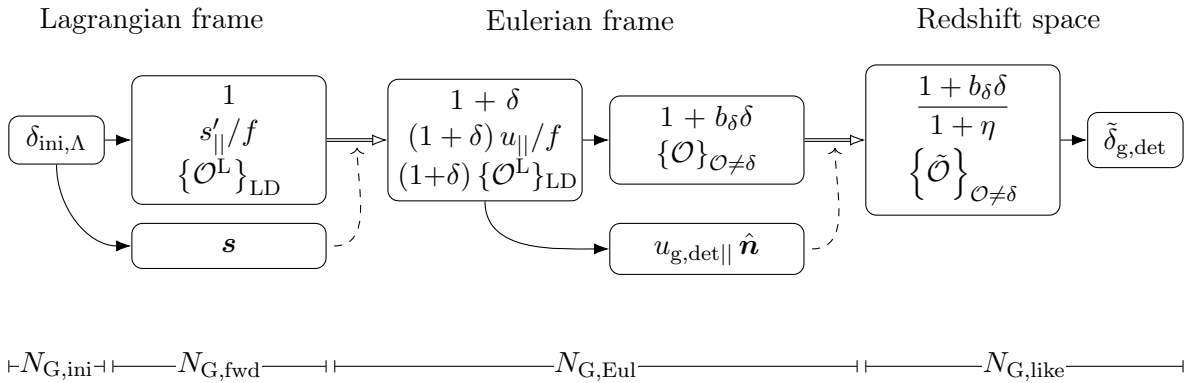

In the previous subsections, we laid out all components of the forward model. They are implemented in the C++17 code \texttt{LEFTfield} \cite{Schmidt_2021}, which, since the initial publication, has been upgraded to a more efficient, extended version. Here, we summarize the main steps (see also figure~\ref{fig: model-flowchart}) and comment on their numerical implementation.
\begin{enumerate}
\item We start from the initial conditions smoothed with a sharp-k filter at scale $\Lambda$, denoted $\delta_{\mathrm{ini},\Lambda}$. From this, we construct the curl-free part of the displacement $\vec{s}$ up to $n_\lpt$ orders in perturbations. Throughout this work we use $n_\lpt=3$. The initial grid size $N_\mathrm{G,ini}$ is chosen large enough to represent all modes up to the cutoff. For the LPT computations we copy the initial conditions to a larger grid of size $N_\fwd$. The Nyquist frequency of this larger grid corresponds to at least $n_\lpt\, \Lambda$ to ensure that all modes can be represented (see the ``no mode left behind'' option in \cite{Schmidt_2021}).
\item Next, we construct the time-derivative of the displacement $s'_\pp/f$ and the rotational invariants of the distortion tensor up to the desired order in the bias expansion $\oporderPT^\mathrm{max}$. Since $u_\pp$ is directly proportional to $f$ (see eq.~\ref{eq:model-s-prime}), we can multiply the growth rate as well in the Eulerian frame as in Lagrangian space. Choosing the former option allows to save computing time, as we do not need to re-evaluate the Lagrangian to Eulerian displacement when $f$ is varied. The set of operators is labeled $\left\{\op^\mathrm{L}\right\}_\mathrm{LD}$ in figure~\ref{fig: model-flowchart}, to indicate that the operators are all leading order in spatial derivatives. Finally, we add a constant field, which will provide the matter density field after it is transformed to the Eulerian frame.
\item We copy all operators obtained in the previous step to a grid with size $N_\eul$. For each operator, we create an ensemble of pseudo-particles whose positions coincide with the grid's nodes and whose masses are given by the operator value at their location. We then shift these particles by the displacement vector $\vec{s}$ and re-assign their density to a grid with size $N_\eul$ using a CIC kernel. Note that $N_\eul$ is a user-defined parameter which essentially controls the resolution at which the displacement operation is computed.
\item In the Eulerian frame, we compute $1 + b_\delta \delta$. We also construct all relevant higher-order derivative operators, as specified in section \ref{sec:model-bias-general}. Together with the leading-order in derivative contributions, they comprise the full set of higher-order operators $\left\{\op\right\}_{\op\neq\delta}$ in eq.~(\ref{eq:model-deltaDet-zspace}).
\item To construct the deterministic halo velocity, we divide out the Jacobian from the displaced velocity field, i.e. from eq.~(\ref{eq:model-shift-to-eul}) with $\op=u_\pp/f$, and multiply by $f$. As detailed in section~\ref{sec:model-bias-vbias}, we construct all velocity bias operators up to the same relevance as the density-bias ones. Then, $\udgpp$ follows by summing up the individual velocity operators according to eq.~(\ref{eq:model-vbias-expansion}).
\item We repeat the displacement step (3) for this new set of operators which contains the linearly biased density field $1 + b_\delta \delta$ and the higher-order bias operators in perturbations and derivatives $\left\{\op\right\}_{\op\neq\delta}$. The displacement now is in the direction of the LOS and given by $\udgpp\, \los$. The resolution at which the displacement is computed is identical to that in step (3) and again controlled by $N_\eul$.
\item In redshift space, we resize the grid to $N_\mathrm{G,like}$, such that the Nyquist frequency matches the cutoff in the likelihood $k_\mathrm{max}$. We generally set $k_\mathrm{max}=\Lambda$. We then add up all bias operators to construct $\dgds$ as given in eq.~(\ref{eq:model-deltaDet-zspace}).
\end{enumerate}
The EFT likelihood then yields the likelihood of the data given the model prediction  computed in steps (1) to (7). To construct a density cube from the observed discrete positions of tracers we use the same CIC kernel as employed in the density assignment steps of the forward model, (3) and (6). Thus, in the comparison between model and data the effect of the kernel cancels to a large extent. We discuss this cancellation further in section \ref{sec: matter-pk}.

In the following sections, we test and verify our forward model by applying it to mock data and to dark matter and halos in N-body simulations. By fixing the initial conditions to the ground truth, we eliminate a considerable source of uncertainty and perform a very stringent set of checks. The free parameters of our inference are the growth rate $f$, all bias coefficients $b_\op$ and $\beta_\uop$ and the noise amplitudes in eq.~(\ref{eq:model-noise-parametrization}). Their posterior distribution is explored with a slice sampler \cite{2000physics...9028N}. Owing to the simple form of the likelihood, it is actually possible to marginalize over the bias parameters $\{b_\op\}_{\op\neq\delta}$ analytically rather then sampling them explicitly \cite{Elsner:2019rql} (see appendix \ref{sec: appendix-marglh}). We make use of this option for inferences from halos, where we have a larger number of bias parameters. The velocity bias coefficients $\beta_\uop$ and the linear density bias $b_\delta$, on the other hand, always have to be sampled explicitly. 

\subsection{Comparison to previous approaches}
\label{sec: model-comparison}
We complete this section by comparing our forward model to other works, focusing on perturbative field-level models. Since our code is an extension of \texttt{LEFTfield}, its working principle is very similar to previous versions (e.g. as used for $\sigma_8$ inference  \cite{Schmidt_2021, Schmidt:2020tao, Schmidt:2020viy}, BAO analysis \cite{Babic:2022dws}, or for sampling the initial conditions \cite{Kostic:2022vok}). The most important additions are the calculation of the velocity field (section \ref{sec:model-lpt}), the velocity bias expansion (section \ref{sec:model-bias-vbias}), the augmentation by a second displacement step to evolve the density and bias fields from the rest frame to redshift space (section \ref{sec:model-ztransform}), and the introduction of anisotropic noise (section \ref{sec:model-likelihood}). In contrast to the fixed-initial-condition $\sigma_8$ and BAO studies mentioned above, we do not use the profile likelihood, but perform a full posterior sampling. 

The only other LPT-based field-level model for the distribution of biased tracers in redshift space that we are aware of was introduced in \cite{Schmittfull:2020trd} (though see \cite{Matsubara:2007wj, Carlson:2012bu, Wang:2013hwa, Vlah:2016bcl, Vlah:2018ygt, Chen:2019lpf, Chen:2020zjt} for LPT calculations of the redshift-space power spectrum and \cite{Pellejero-Ibanez:2022efv} for a combination of particle displacements from N-body simulations with the perturbative bias expansion in the hybrid-EFT approach). In \cite{Schmittfull:2020trd}, the biased density field is assembled in Lagrangian space and transformed to redshift space in a single displacement step. This displacement is defined by the linear-order shift vector, $\vec{s}\lin(\vec{q}) + u_\pp\lin(\vec{q})\,\los$, while higher-order contributions to the displacement are expanded perturbatively. Further, the third-order bias expansion is decomposed into contributions parallel and orthogonal to the Zel’dovich redshift-space density $\tilde{\delta}_1$. By keeping only the parallel terms, the bias can be expressed as a $k$- and $\mu$-dependent transfer function $\beta_1$ which multiplies $\tilde{\delta}_1$. The functional form of $\beta_1(k,\,\mu)$ can in principle be predicted from perturbation theory, with the bias coefficients entering as free parameters. However, reference \cite{Schmittfull:2020trd} follows a different route and first optimizes the transfer function in bins of $k$ and $\mu$, then fits it by a smooth 7-parameter function. This approach makes it somewhat hard to disentangle which bias operators precisely are included in the model. In principle, reference \cite{Schmittfull:2020trd} notes that the transfer function accounts also for higher-derivative bias and counter terms, such as 
\begin{equation}
\beta_1(k,\mu)  \supset  R_1^2\,k^2 + R_2^2\, (k\mu)^2 + R_3^4\, (k\mu)^4\,.
\label{eq:model-derivops-schmittfull}
\end{equation}
The first of these operators can be re-written as our isotropic higher-derivative bias $\nabla^2\delta$, and the second one corresponds to $\partial^2_\pp\delta$. The final contribution, $\partial^4_\pp\delta$, formally is beyond the order considered in \cite{Schmittfull:2020trd}. Nevertheless, it was shown to improve the inference of cosmological parameters from the redshift-space power spectrum \cite{Chudaykin:2020hbf}.

In our approach, anisotropic derivative operators as those appearing in eq.~(\ref{eq:model-derivops-schmittfull}) are induced by velocity bias. As we have discussed in section \ref{sec:model-ztransform}, the $\partial^2_\pp \delta$ contribution corresponds to the leading-order impact of the first velocity bias operator, $\partial_\pp\delta$. Similarly, the $\partial^4_\pp \delta$ term could arise from a $\partial^3_\pp \delta$ velocity bias. However, an important difference of our velocity bias expansion (see eq.~\ref{eq:model-vbias-list}) is that it contains three additional operators, all of them second order in perturbations and first in derivatives. Thus, these three terms all are more relevant than the $\partial_\pp^3\delta$ velocity-operator, following the scaling relation in eq.~(\ref{eq:model-vbias-relevance}).

Another important difference is that we compute the displacement from rest frame to redshift space as a fully nonlinear transformation (in terms of the relation between the displacement vector and the displaced field), using the 3LPT prediction for the velocity field and the biased velocity as shift vector. With respect to how the velocity bias affects the density contrast in redshift space, this nonlinear treatment generates additional contributions. The three next-to-leading order terms originating from $\partial_\pp\delta$ (see section \ref{sec:model-ztransform}), for example, all contribute at second-order in derivatives and perturbations. Formally, they are thus beyond the third order bias expansion which we adopt here. On the other hand, these operators are still more relevant than the $\partial_\pp^4\delta$ contribution. Thus, they should be included in some fashion, if one argues that it is necessary to account for the impact of velocity bias to a higher order.

Finally, it is interesting to compare the velocity-induced LOS-dependent terms in the density contrast of our model to the LOS-dependent counter and bias terms in the EFT calculations of the redshift-space power spectrum. In this context, reference \cite{Perko:2016puo} introduces two new terms,
\begin{equation}
\propto k^2 \mu^2 \delta\lin
\quad \mathrm{and} \quad
\propto \mu^4 k^2 \delta\lin\,.
\label{eq:model-conterterms-perko}
\end{equation}
More recent works often extend this by a higher-order contribution. For example, the LOS-dependent higher-derivative terms in \texttt{CLASS-PT} \cite{Chudaykin:2020aoj} are given by
\begin{equation}
\propto k^2\mu^2 P_\mathrm{L}(k)\,,
\quad
\propto \mu^4 k^2 P_\mathrm{L}(k)
\quad \mathrm{and} \quad
\propto \mu^4 k^4 \left(b_\delta + f\mu^2 \right) P_\mathrm{L}(k)\,.
\label{eq:model-conterterms-CLASSPT}
\end{equation}
The first term in either eq.~(\ref{eq:model-conterterms-perko}) or eq.~(\ref{eq:model-conterterms-CLASSPT}) corresponds to the lowest-order impact of velocity bias in our model, and has been discussed above. We have already noted how the third term of eq.~(\ref{eq:model-conterterms-CLASSPT}) could arise from velocity bias below eq.~(\ref{eq:model-derivops-schmittfull}). Formally, however, it is of higher order than all the contributions we consider for this work. In contrast, the second term of eq.~(\ref{eq:model-conterterms-perko}) and eq.~(\ref{eq:model-conterterms-CLASSPT}) does not arise from our velocity bias expansion. It would be the leading-order contribution to the density contrast from a velocity bias operator of the form $\partial_\pp\eta$. The status of this term remains somewhat unclear at this point. On the one hand, the LOS orientation needs to appear as preferred direction in the halo rest frame in order to construct such an operator, which corresponds to a selection effect \cite{Desjacques:2018pfv}. On the other hand, the term $\propto \mu^4 k^2 \delta$ was introduced originally to regularize the contact operators $\partial_\pp^2 [ u_\pp^2]$ and $\partial_\pp^3 [ u_\pp^3]$  \cite{Perko:2016puo}, while in our case, such contributions are kept finite and under control by the explicit cutoff $\Lambda$ in the initial conditions. In our tests, we found no clear evidence for a nonzero contribution $\dgds \supset \mu^4 k^2 \delta$, but this issue is worth revisiting in the future.

\section{Model consistency and mock analyses}
\label{sec:mocktests}

In the previous section, we have introduced the forward model to analyze the large-scale distribution of biased tracers in redshift space. We now proceed with a set of consistency checks. In particular, we apply the analysis to mock data which were generated from the forward model itself. In doing so, we can control precisely what physical effects are present in the data. We use this to explore the impact of noise modeling and to examine the back-reaction of short-scale modes onto larger ones in an idealized setting. Further, we explore the quality of the 3LPT prediction by comparing it to simulations with an identical cutoff in the initial conditions.

\subsection{Noise properties}
\label{sec:mocktests-likelihood}

The EFT likelihood can be formulated either in real- or in Fourier-space, as explained in section \ref{sec:model-likelihood}. Each choice accounts for different physical effects at the expense of neglecting others. In this work, we focus on the Fourier-space version which can accommodate scale-dependent isotropic and anisotropic noise, but not the transformation of isotropic, Gaussian noise in the Eulerian frame to redshift space.

In order to test how this choice impacts the analysis, we create two different types of mock data sets: for each we use the same forward model to predict $\dgds$, then add a noise realization. The latter is either drawn directly in redshift space,  or drawn in the Eulerian frame and subsequently transformed to redshift space. We refer to these cases as ``redshift-space noise'' and ``Eulerian-frame noise'', respectively. The redshift-space noise is consistent with our likelihood, while the Eulerian-frame noise is what we would expect for real data in redshift space according to eq.~(\ref{eq:redshift-space-noise}). We then perform the inference on these mocks and check whether the results obtained for Eulerian-frame noise significantly differ from those with redshift-space noise.

\begin{figure}
	\centering
	\includegraphics[]{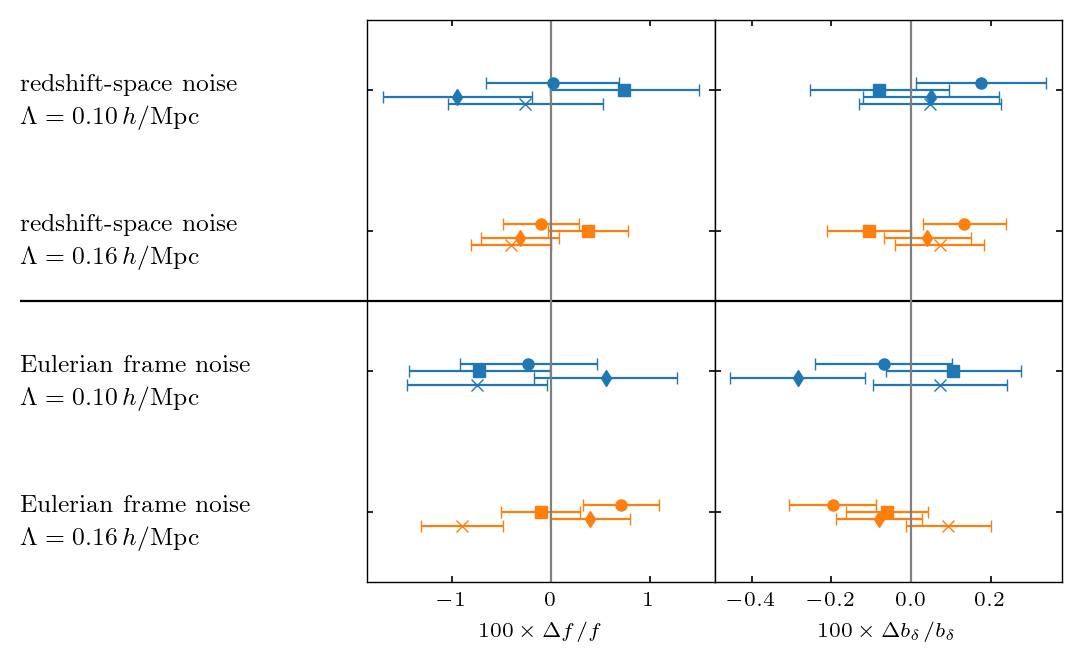}
	\caption{Effect of noise properties and the choice of the Fourier space likelihood on the inference of the growth rate and linear bias from mock data sets. We create four mock realizations (indicated by different symbols) to which we either add isotropic noise in redshift space (top) or in the Eulerian frame and subsequently transform to redshift space (bottom). For generating the mocks and in the analysis we use the same cutoff value $\Lambda$ as indicated above. The free parameters in the inference are the noise amplitudes $P_\epsilon^{(0)}$, $\sigma_{\epsilon,\,2}$, the growth rate $f$ and the linear bias $b_\delta$.}
	\label{fig:mocktest-likelihood}
\end{figure}

More concretely, for each type of mock data we generate four realizations at $z=0$ that only differ in their initial conditions and in the noise realization. The box size and fiducial cosmology used here is the same as for the N-body simulations, and the corresponding parameters are summarized in table \ref{tap: appendix-nbody-parameters}.  Our mocks are unbiased, i.e. we set $b_\delta=1$ and all other bias coefficients to zero. We consider two cutoff scales for generating the mock data, $\Lambda_0 = 0.10\,,~0.16\,h/\mpc$. The noise power spectrum (eq.~\ref{eq:model-noise-parametrization}) is given by $P_{\epsilon}^{(0)}=1.93\times10^3\,\left(\mpc/h\right)^3$, $\sigma_{\epsilon,\,2}=10$ and $\sigma_{\epsilon\mu,\,2}=0$. This choice is roughly motivated by the values we infer for halos in the lowest mass bin at $\Lambda=0.10\,h/\mpc$ and $z=0$ in section \ref{sec:results}.

In the analysis, we use the same cutoff from which the mocks where generated, i.e. $\Lambda=\Lambda_0$. Thus, the only source of inconsistency we expect between our forward model and the data are indeed the noise properties in the mocks with Eulerian-frame noise. As explained in the introduction \ref{sec:intro}, we fix the initial density perturbations to their ground truth, and we infer the growth rate, the linear bias parameter and the noise amplitudes $P_\epsilon^{(0)}$, $\sigma_{\epsilon,2}$. Since we fixed the velocity noise to zero when generating the mocks, we do not expect an anisotropic noise contribution at leading order. 
Our main criterion to judge the quality of the results is an unbiased inference of $f$ and $b_\delta$, meaning that the true values are recovered within errors. Note that the error bars returned by the EFT likelihood in the fixed-phase inference include any residual cosmic variance due to the different realizations of the small-scale modes above the cutoff, and as such should incorporate the variance between different mock realizations {\cite{Schmidt:2020viy}}. Though we only consider a relatively low number of mock realizations, we do not find indications that the error bars returned by the EFT likelihood are significnatly over- or underestimated. We can thus detect any systematic biases in the inferred parameters that are significantly larger than this statistical error bar. Through the high precision of the field-level analysis at fixed phase, this corresponds to a sensitivity to percent-level systematics.

The results of our analysis are summarized in figure \ref{fig:mocktest-likelihood}, where we show the mean and standard deviation over the posterior samples after removing the burn-in (see appendix \ref{sec:appendix-convergence} for more details on the convergence criteria for the sampling). As expected, the analysis of mocks with redshift-space noise yields an unbiased inference of $f$ and $b_\delta$; both parameters are recovered with percent-level precision and accuracy. As we go to higher cutoff values $\Lambda$, we see the expected shrinking of the error contours. Moving on to the mocks with Eulerian frame noise, we neither note a systematic shift of the inferred mean values, nor a significant difference in the width of the error bars. However, we encounter a larger scatter of the inferred mean values, relative to the inferred error bars, in particular in the high-$\Lambda$ case. This might indicate a slight under-estimation of the error bars, but for a quantitative statement a larger number of mock realizations would be required. Overall, we cannot discern a significant impact that the use of the Fourier-space likelihood would have on the analysis.

As this section demonstrates, the inference of cosmological parameters at fixed initial conditions is rather robust with respect to noise misspecifications in the likelihood. We still want to stress that this conclusion does not necessarily apply to the inference with free initial conditions, where the parameter space is larger and there is more freedom to absorb model misspecifications in physically meaningful quantities. Thus, mock tests as presented here should be repeated once one moves to the sampling of initial conditions.

\subsection{Velocity bias}
\label{sec:mocktests-vbias}
We also use mock data sets to test how well our bias expansion can absorb the back reaction of small scales onto large-wavelength modes. To this end, we create mock data sets with a large cutoff, $\Lambda_0 = 0.40\,h/\mpc$, that is beyond the nonlinear scale at $z=0$. We then analyze these data sets using more moderate values for $\Lambda$ that are in the perturbative regime. This ``$\Lambda$-mismatch'' test clearly isolates the effect of small-scale modes while not being affected by bias or potential shortcomings of the perturbative gravity model. Still, any effect that is present in the $\Lambda$-mismatch mocks, should also be important for the analysis of tracers from N-body simulations.

The four mock realizations of unbiased tracers considered here use the same fiducial cosmology and box size as in section \ref{sec:mocktests-likelihood} (see table \ref{tap: appendix-nbody-parameters}). The noise is added in Eulerian frame to mimic a realistic inference scenario and only contains a white contribution with $P_{\epsilon}^{(0)}=1.93\times10^3\,\left(\mpc/h\right)^3$. We nevertheless expect nonzero inferred values for the scale-dependent isotropic and anisotropic noise, which arise from the mismatch between $\Lambda_0$ and $\Lambda$, i.e. the cutoff used for generating the data and in the analysis.

Since the $\Lambda$-mismatch between analysis and mock data can only generate higher-derivative operators, we set the coefficients of all leading-derivative operators apart from $b_\delta$ to zero. We then investigate the accuracy at which our analysis recovers the linear bias and growth rate while simultaneously inferring the noise amplitudes $P_\epsilon^{(0)}$, $\sigma_{\epsilon,\,2}$, $\sigma_{\epsilon\mu,\,2}$ and the coefficients of higher-derivative bias contributions. Throughout, we include the leading higher-derivative term $\nabla^2\delta$, but we test different formulations to account for velocity bias as listed below.
\begin{enumerate}
\item The ``\textbf{no velocity bias}'' scenario completely neglects any contributions from velocity bias.
\item The ``\textbf{velocity-induced bias at leading order}'' scenario introduces one LOS-dependent operator in the density contrast, $\nabla^2\eta$. This operator arises from the leading-order velocity bias, when the redshift-space transformation in eq.~(\ref{eq:model-rsd-expansion}) is expanded to leading order. Note that, at leading order, $\nabla^2\eta$ and $\partial_\pp^2\delta$ are directly proportional.
\item The ``\textbf{velocity-induced bias with higher-order contributions}'' adds one additional operator, $\partial_\pp^4\delta$ to the ``velocity-induced bias at leading order'' scenario. We discuss this operator in more detail below eq.~(\ref{eq:model-derivops-schmittfull}). Formally, it is of higher order than all terms considered in the explicit scenarios below, but its effect was explored in previous studies \cite{Chudaykin:2020hbf, Schmittfull:2020trd}.
\item The ``\textbf{velocity bias at leading order}'' uses the biased velocity directly to perform the transformation to redshift space, instead of introducing velocity-induced bias operators for the redshift-space density contrast. In the leading-order scenario, we only consider a single velocity bias operator, which is $\partial_\pp\delta$.
\item The ``\textbf{baseline}'' scenario extends the velocity bias expansion to include the first three terms of eq.~(\ref{eq:model-vbias-list}). These are all terms that can be constructed by applying a single derivative along the LOS to invariants constructed from the Lagrangian distortion tensor up to second order. 
\item The ``\textbf{velocity bias at next-to-leading order}'' scenario, finally, allows for all operators present in eq.~(\ref{eq:model-vbias-list}). That is, it augments the ``baseline'' scenario by one additional operator, $\partial_i M^{(1)}_{\pp j} M^{(1)}_{ji}$. From an EFT perspective, there is no reason to exclude this operator from the ``baseline'' scenario in the first place. The reason we do so nevertheless is that marginalization over velocity bias coefficients is computationally costly, as each proposed value requires an evaluation of the displacement to redshift space in the forward model. Below, we check the impact of this operator for several test cases, and usually find it to be secondary. Therefore, we produce the majority of our results with the more economical baseline scenario.
\end{enumerate}

\begin{figure}
	\centering
	\includegraphics[]{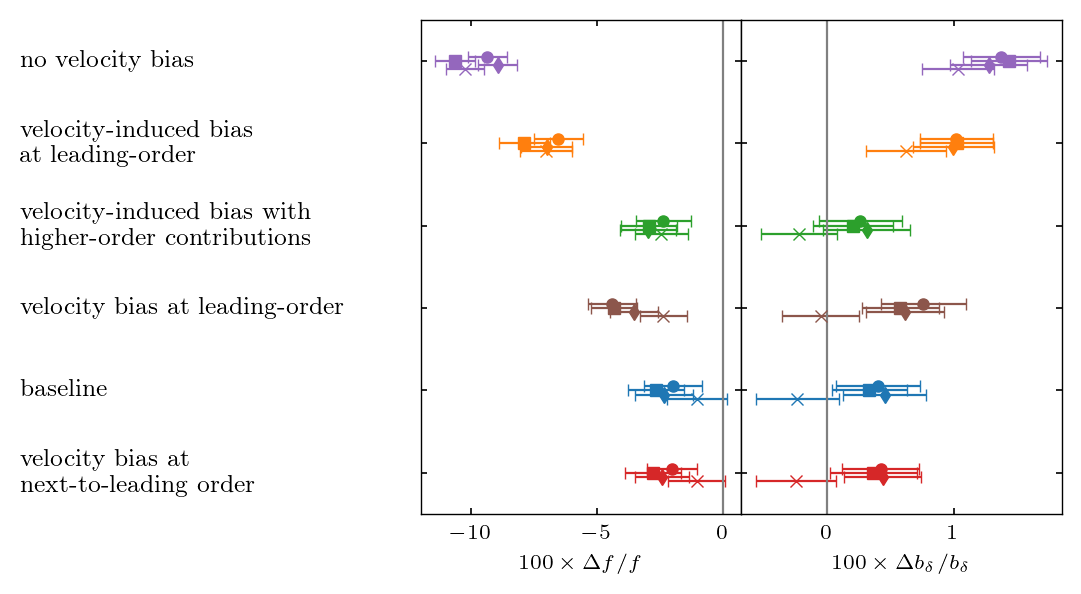}
	\caption{The effect of the velocity bias model on the inference of cosmological parameters from mock data. The data was generated with a higher cutoff, $\Lambda_0=0.40\,h/\mpc$, than used in the analysis ($\Lambda=0.10\,h/\mpc$). Four realizations of the mock data are indicated by different symbols. We test different sets of higher-derivative bias parameters as indicated above and jointly infer the cosmology parameters $f$ and $b_\delta$, the bias coefficients and the three noise amplitudes in eq.~(\ref{eq:model-noise-parametrization}).}
	\label{fig: mock-operators}
\end{figure}

\begin{figure}
	\centering
	\includegraphics[]{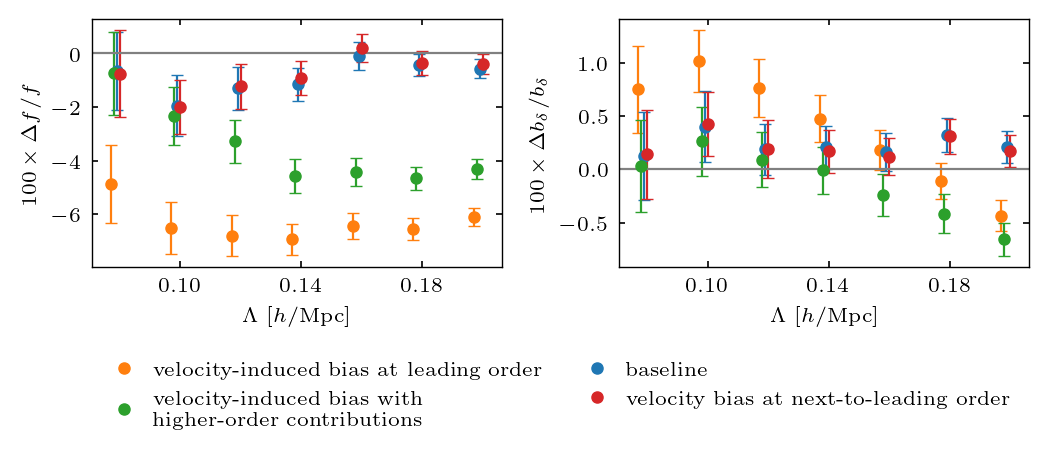}
	\caption{Cutoff dependence of the parameters inferred from mock data for different descriptions of the velocity bias. The mock data, analysis choices and color coding are identical to figure \ref{fig: mock-operators}, and we focus on the first mock data set (marked by circles in either figure).}
	\label{fig: mock-operator-scaling}
\end{figure}

The results of these analyses at $\Lambda=0.10\,h/\mpc$, each applied to four independent mock data sets, are summarized in figure~\ref{fig: mock-operators}. Evidently, the lowest order effective descriptions of velocity bias, option (1) and (2), yield strongly biased results for both $f$ and $b_\delta$ even in the case of a rather moderate cutoff. Including the higher-order velocity-induced operator $\partial^4_\pp\delta$, i.e. option (3), can considerably reduce this bias. It actually is consistent with the ``velocity bias at leading order'' scenario (4) and fares only slightly worse than the higher-order velocity descriptions (5) and (6). In the latter two cases, we still note some residual bias towards smaller values of $f$ and higher values of $b_\delta$, but now its magnitude is reduced to the level of $1-2\,\%$. Including the additional operator in option (6), on the other hand, only has a marginal impact.

It is illuminating to study how these results evolve with cutoff $\Lambda$; we do this for a subset of configurations -- the ``velocity-induced bias at leading order'' scenario (2), the ``velocity-induced bias with higher-order contributions'' (3) and the velocity expansions (4) and (5). These results are summarized in figure \ref{fig: mock-operator-scaling}. While the velocity-induced bias formulation (3) is comparable to the explicit velocity expansion for small cutoffs, its performance quickly deteriorates as one goes to higher $\Lambda$. At $\Lambda=0.12\,h/\mpc$ we already see a deviation of $\sim 4\%$, that further increases to higher cut-offs. Given the high statistical significance of this shift ($\sim4\sigma$ at $\Lambda=0.12\,h/\mpc$), we expect it to be representative of a general trend. The higher-order velocity-induced bias term $\partial_\pp^4\delta$ was introduced in \cite{Chudaykin:2020hbf}, and that analysis obtained consistent results up to a slightly larger scale, $k_{\max} = 0.14\,h/\mpc$ at $z=0$. There are several differences between these two studies which can explain the slight disagreement; among them the field level analysis versus the power spectrum monopole and quadrupole, use of mock data versus halo catalogs as tracers, and fixed initial conditions versus comparison of ensemble means. The fact that we find a deviation already for rather small values of $\Lambda$ once more illustrates the stringency of our tests. The two higher-order velocity bias expansions (5) and (6), finally, perform well up to the highest cutoff tested here, and we do not see a notable benefit of including the operator $\partial_i M^{(1)}_{\pp j} M^{(1)}_{ji}$.

Even in case of the higher-order velocity expansions, there remains a small, percent-level shift $\Delta f<0$ around $\Lambda=0.10\,h/\mpc$. Three of the four mock realizations in figure~\ref{fig: mock-operators} experience this shift, which makes a statistical fluctuation less likely. At increasing cutoffs, the inferred values again move closer to the ground truth, which essentially rules out higher-order perturbative contributions as cause. Interestingly, Ref.~\cite{Kostic:2022vok} found a similar deviation on intermediate scales $0.1 \leq \Lambda \leq 0.13$ in one of their test cases, which decreased when considering datasets with higher signal-to-noise. In any case, the discrepancy is not very significant and always remains below the level of $2\sigma$. 

\subsection{Validation of the 3LPT model against fully non-linear predictions}
\label{sec: matter-pk}

\begin{figure}
	\centering
	\includegraphics[]{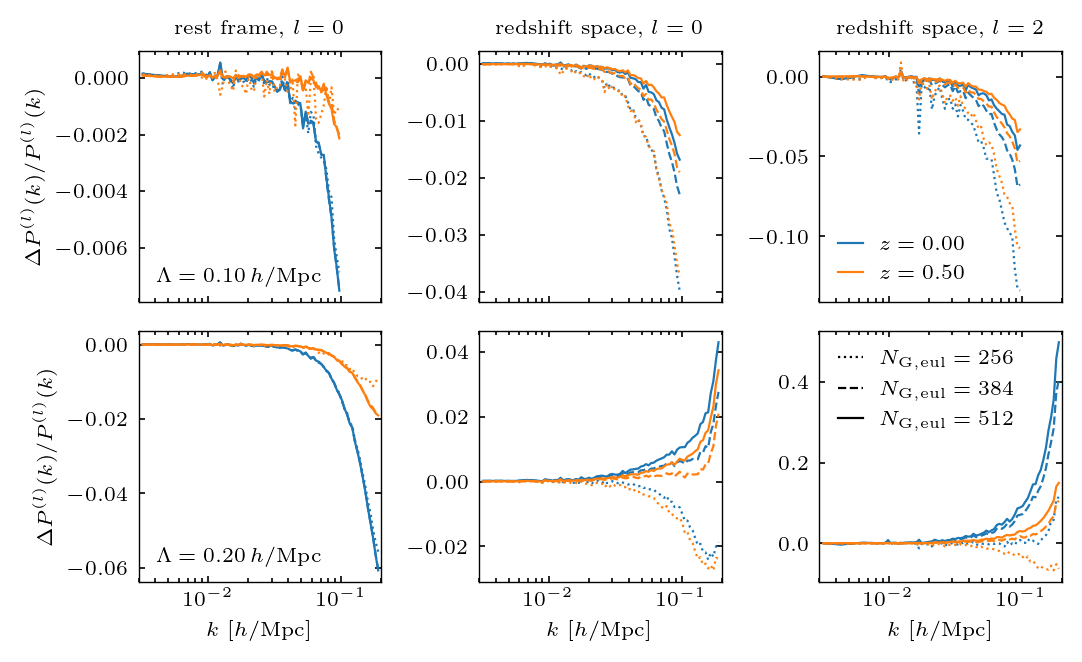}
	\caption{Relative difference between the 3LPT gravity model and N-body simulations for the power spectrum monopole $P^{(0)}$ in the matter rest frame and in redshift space and for the power spectrum quadrupole $P^{(2)}$ in redshift space. The N-body simulations shown here have the same cutoff $\Lambda$ in the initial conditions as used for the perturbative evolution, and thus the difference directly tests the quality of the 3LPT model. We consider two different cutoff values, $\Lambda = 0.10\,,~0.20\,h/\mpc$, shown in the top and bottom row, respectively.
	}
	\label{fig: matter-comparison-pk}
\end{figure}

The two previous tests considered mock data generated from the same forward model as used in the analysis, and as such they where immune to any shortcomings of the 3LPT gravity model. To check explicitly for such potential inconsistencies, we compare the 3LPT predictions for matter in redshift space to N-body simulations which have an identical cutoff in the initial condition. Their relative difference is shown in figure \ref{fig: matter-comparison-pk} for the power spectrum monopole in the matter rest frame and in redshift space, and for the redshift-space quadrupole. We consider two different redshifts ($z=0,\, 0.5$) and cutoff scales ($\Lambda=0.10,\,0.20$). The comparison focuses on the impact of $N_\eul$, i.e. the resolution of the assignment step. See \cite{Schmidt_2021} for a similar comparison as function of the LPT order in the matter rest frame.

Clearly, the redshift-space relative error depends more crucially on $N_\eul$ than the one in the matter rest frame. This is not entirely unexpected. The data density cube is always constructed in a single step of density-assignment, which either considers a particle's rest-frame position $\vec{x}$ or its redshift-space coordinate $\vecs{x}$, to obtain the respective density fields. Similarly, there is a single density assignment when the 3LPT Eulerian-frame density is computed: the model predicts the shift vector on a regular Lagrangian grid, displaces particles accordingly and deposits them on the grid. In this case, the effect of the kernel cancels out between data and model prediction \cite{Schmidt_2021}. In contrast, the 3LPT redshift-space computation involves a second displacement to get from Eulerian to redshift space (see figure \ref{fig: model-flowchart}), and the cancellation is not perfect anymore. 

In the case of $\Lambda=0.10\,h/\mpc$, we see that the relative difference between N-body and 3LPT prediction in redshift space successively decreases as the resolution increases. As expected, the error of the LPT model in Eulerian frame increases for higher cutoffs. For $\Lambda=0.20\,h/\mpc$, there still is a significant dependence on the resolution in redshift space, however the trend with $N_\eul$ now is less clear. For small values, the 3LPT model can yield an excess of power, larger ones result in a deficit. In the transition between the two regimes, $N_\eul=384$ is found to actually give a smaller relative error than $N_\eul=512$. 

In almost all cases, the error is smaller for higher redshifts, where higher-order perturbations are more strongly suppressed and the 3LPT approximation is better. Only for $N_\eul=256$, errors from the density assignment resolution seems to dominate the redshift-space prediction completely, and both redshifts yield a similar relative error. 

As a general trend, figure \ref{fig: matter-comparison-pk} shows that the power spectrum in the rest frame is more accurately predicted by the 3LPT model than the redshift-space monopole. In turn, the accuracy in redshift space is better for the monopole than for the quadrupole. This can be expected to some extent, as flaws in the 3LPT velocity additionally enter the redshift-space density contrast. The displacement to redshift space enhances the monopole and gives rise to the quadrupole, such that the latter indeed is more sensitive to the velocity field. At this point, it is not clear what dominates the error in the 3LPT velocity -- the perturbative order of the 3LPT model, curl terms which we have neglected, or the size of the resolution of the assignment step. 

In the following analyses, we use $N_\eul=384$, which offers a good trade-off between computational cost and accuracy. We verify our choice by a set of convergence tests, which is presented in appendix \ref{sec: appendix-ngeul}. These show very clearly that $N_\eul=256$ is not sufficient for an unbiased inference. In most trials, the results do not change significantly anymore when the particle number is increased from $N_\eul=384$ to $N_\eul=512$. The only exception are results for matter at low redshifts, where Fingers-of-God (FoG) are strongest. We discuss this further in the next subsection. Despite the apparent convergence in $N_\eul$ observed for halos and for matter at higher redshifts, we cannot entirely rule out some residual resolution effects. We plan to improve the 3LPT velocity prediction in a future publication, and to investigate which of the aforementioned factors have the most impact on the redshift-space analysis. In this context, the use of a more sophisticated assignment schemes, which deconvolves the effect of the kernel, is also very interesting.

\section{Results from N-body particles}
\label{sec:matter-inference}

The next test case we consider is the inference of the growth rate from the density field in N-body simulations. By sub-sampling the matter particles, we generate a trivially biased, noisy tracer field. Still, these tracers are subject to the full nonlinear gravitational evolution and thus provide an intermediate scenario between the previously discussed mocks and the halos in section \ref{sec:results}. We have two sets of simulations available that only differ in the initial condition realization. They are the same as used in \cite{Schmidt:2020tao} and described in appendix \ref{sec: appendix-n-body-simulations}. Further, we consider snapshots at three different redshifts, $z=0\,,~0.5\,,~1$. The mean density $\bar{n}$ at which the snapshots are sub-sampled is chosen to give a roughly constant signal-to-noise ratio (SNR) over all redshifts, by keeping the combination $\bar{n}\,D^2(z)$ constant. Our main criterion to judge the performance of an analysis is if the true values of $f$ (see eq.~{\ref{eq:appendix-sim-fvals}}) and $b_\delta = 1$ are recovered within the statistical errors.

\begin{figure}
	\centering
	\includegraphics[]{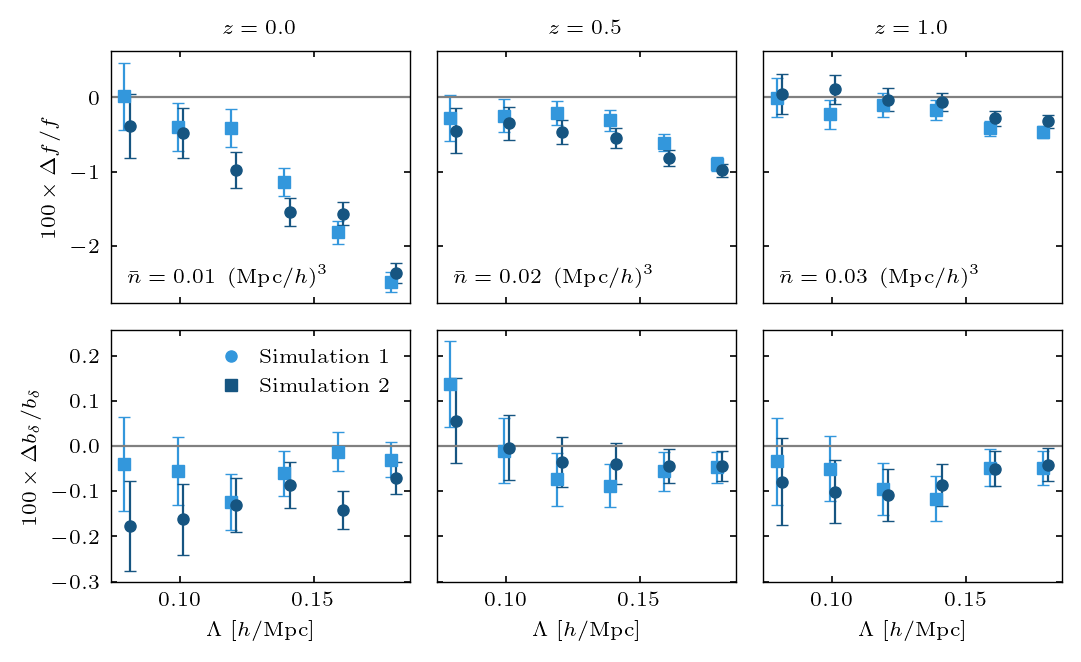}
	\caption{The growth rate and linear bias inferred from N-body particles, sub-sampled to an approximately constant SNR over all redshifts. We consider two realizations of the simulation, which are indicated by different symbols. The results have been obtained for the ``baseline'' analysis scenario which simultaneously infers the growth rate $f$, linear bias parameter $b_\delta$, the higher-derivative bias coefficient $b_{\nabla^2\delta}$, the velocity bias coefficients $\beta_{\partial_\pp\delta}$, $\beta_{\partial_\pp (\sigma^2)}$, $\beta_{\partial_\pp\mathrm{tr}[M^{(1)} M^{(1)}]}$ and the three noise amplitudes from eq.~(\ref{eq:model-noise-parametrization}).}
	\label{fig: matter-overview}
\end{figure}

\begin{figure}
\centering
\includegraphics[]{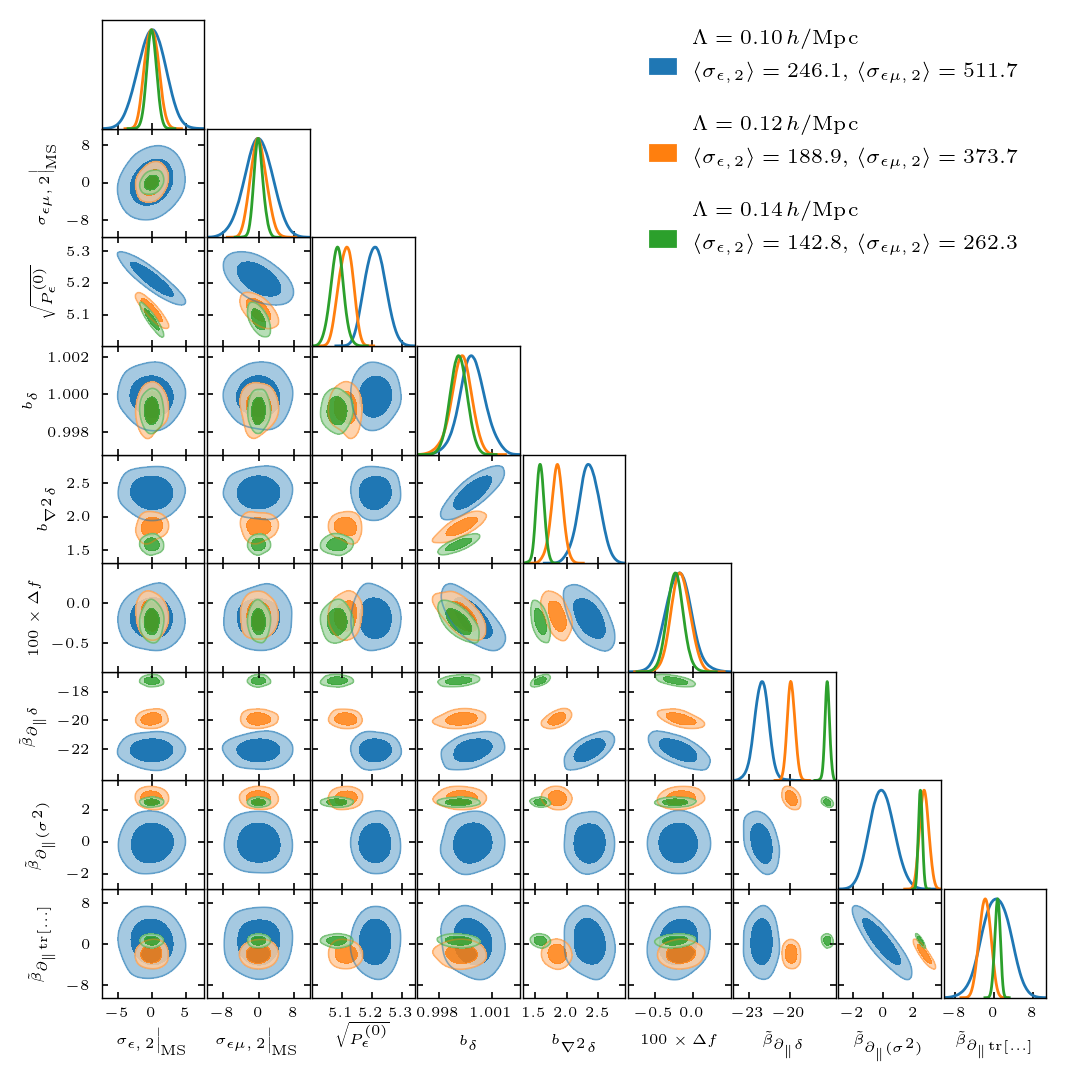}
\caption{Marginalized posterior contours for the analysis of matter in simulation 1 at $z=0.5$ with the ``baseline'' scenario. In case of the velocity bias coefficients, we plot $\tilde{\beta}_\uop = \beta_\uop/f$, and we have abbreviated $\tilde{\beta}_{\partial_\pp\mathrm{tr}[M^{(1)} M^{(1)}]}$ as $\tilde{\beta}_{\partial_\pp\mathrm{tr}[...]}$. We have further subtracted the sample mean as indicated in the legend from the scale-dependent noise amplitudes for better visualization. This is indicated by the notation ``$\left. \ldots \right|_\mathrm{MS}$'' in the axis labels.}
\label{fig:matter-degeneracies}
\end{figure}

The results of applying the ``baseline'' analysis (see section \ref{sec:mocktests-vbias}) are summarized in figure \ref{fig: matter-overview}. Overall, they exhibit the expected scaling with the cutoff scale $\Lambda$ and with redshift. For small $\Lambda$, the inferred parameters asymptote to the ground truth, where the growth rate is recovered with percent-level accuracy and the linear bias at the sub-percent level. The error bars of the individual estimates shrink as $\Lambda$ increases and more modes are included in the analysis. Figure \ref{fig:matter-degeneracies} further illustrates this trend in terms of the two-dimensional marginalized posterior contours of all free parameters, focusing on simulation 1 at $z=0.5$. Apparently, the information from smaller scales in particular tightens the constraints on the higher-order velocity bias coefficients. At some scale, the perturbative forward model breaks down and we note an increasing discrepancy between the inferred parameters and their true value. This effect is particularly apparent in the growth rate. The departure occurs earlier for smaller redshifts, and for the $z=0$ case, we note a significant offset already for cutoffs between $0.12\,h/\mpc$ and $0.14\,h/\mpc$. In contrast to that, we find the model to be robust for much larger values of $\Lambda$ when applied to mock data or halos, see figures \ref{fig: mock-operator-scaling} and \ref{fig: halos-summary}, respectively. 

The appearance of discrepancies at comparably large scales in the matter analysis is most likely caused by Fingers-of-God, i.e. by the random motions of dark matter particles in collapsed viralized structures \cite{Jackson:1971sky}. Indeed, from all tracers we consider in this work, we expect the strongest FoG for matter particles, where the mass fraction of dark matter contained in resolved halos is $\sim 30\%$ at $z=0$, $\sim 20\%$ at $z=0.5$ and $\sim 10\%$ at $z=1$. In contrast, the perturbative mocks do not contain viralized structures, and the contribution from subhalos to the halo samples considered in section \ref{sec:results} is at the level of a few percent at most. FoGs add noise to the velocities, so their leading-order impact is captured by the anisotropic noise term in the likelihood $\propto P_{\epsilon_\mu}^{(2)}$. At scales where this leading-order description is not sufficient anymore, inconsistencies arise as those noted in figure \ref{fig: matter-overview}.  Also the fact that we infer a higher amplitude for the scale-dependent noise contribution from matter particles than from halos at low redshifts agrees with the presence of stronger FoGs in the former sample.

In appendix \ref{sec: appendix-ngeul} and \ref{sec: appendix-transients}, we further check how the results of the ``baseline'' analysis depend on numerical accuracy. In particular, we consider the resolution of the displacement step in the forward model and the starting redshift of the N-body simulations that provide our data. Increasing the resolution beyond $N_\eul=384$ as used here only has an impact on results which are already affected by FoG. An earlier initial time of the numerical simulations, on the other hand, can lead to some shift in the inferred parameters. However, we expect that the residual effect of transients and discretization effects on the results presented here is secondary. 

\begin{figure}
	\centering
	\includegraphics[]{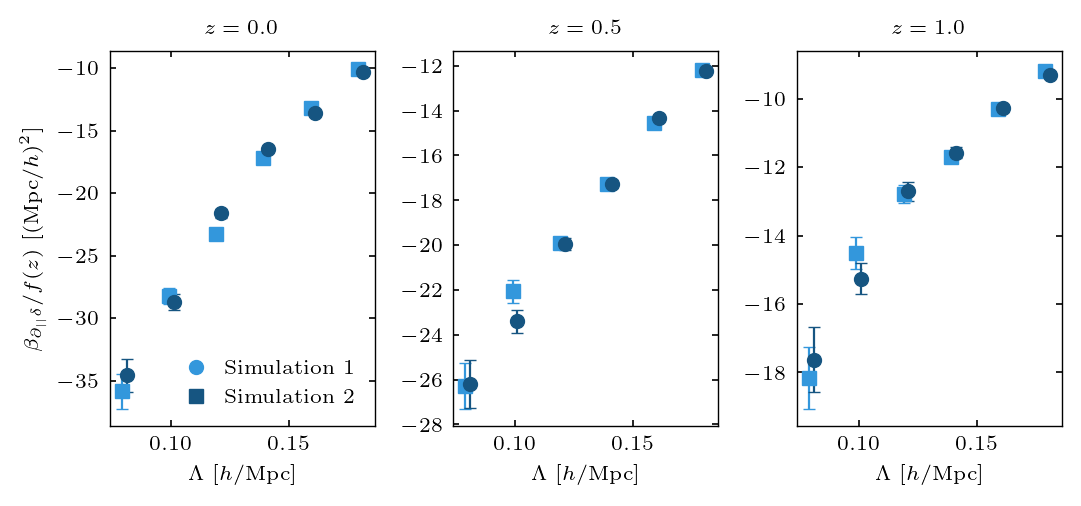}
	\caption{The leading-order velocity bias coefficient inferred from matter using the ``baseline'' analysis scenario. The growth rate and linear bias parameters inferred in the same analyses are depicted in figure~\ref{fig: matter-overview}.}
	\label{fig:matter-vbias}
\end{figure}

The inferred value for the leading-order velocity bias coefficient, $\beta_{\partial_\pp\delta}$, is shown in figure~\ref{fig:matter-vbias}, and it is negative for all cutoffs and redshifts. Since $\partial_\pp\delta \propto - \nabla^2 u_\pp$ at leading order, which corresponds to $k^2 u_\pp(\vec k)$, a negative value for $\beta_{\partial_\pp\delta}$ is consistent with a smoothing of the velocity, i.e. a suppression of the deterministic velocity on small scales. In the density contrast, the velocity bias enters at leading order as
\begin{equation}
\dgds = b_\delta \delta - f\mu^2\delta + \beta_{\partial_\pp\delta}\, (\mu k)^2\, \delta\,,
\end{equation}
and again a negative value of $\beta_{\partial_\pp\delta}$ implies a smoothing. Finally, the leading-order impact of the velocity bias on the power spectrum is a term of the form
\begin{align}
\tilde{P}_{\tracer,\det} (k, \mu)  &\supset  2\,k^2\,P_\mathrm{L}(k) \, \beta_{\partial_\pp\delta} \left(b_\delta\, \mu^2 - f \mu^4\right) \nonumber\\
&\supset 2\,k^2\,P_\mathrm{L}(k)\,\beta_{\partial_\pp\delta} \left[
P_0(\mu)\left(\frac{b_\delta}{3}-\frac{f}{5}\right)
+ P_2(\mu) \left(\frac{2\,b_\delta}{3}-\frac{4f}{7}\right)
- P_4(\mu) \frac{8\,f}{35}
\right]\,.
\label{eq:matter-impact-vbias-leading}
\end{align}
where $P_i\left(\mu\right)$ denotes the $i$-th Legendre polynomial. Since $b_\delta=1$ and $f<1$ (see eq.~\ref{eq:appendix-sim-fvals} for exact values), the negative bias coefficient leads to a reduction of the monopole and quadrupole at high wavenumbers, consistent with the notion of small-scale power suppression, but it enhances the hexadecapole.

\begin{figure}
	\centering
	\includegraphics[]{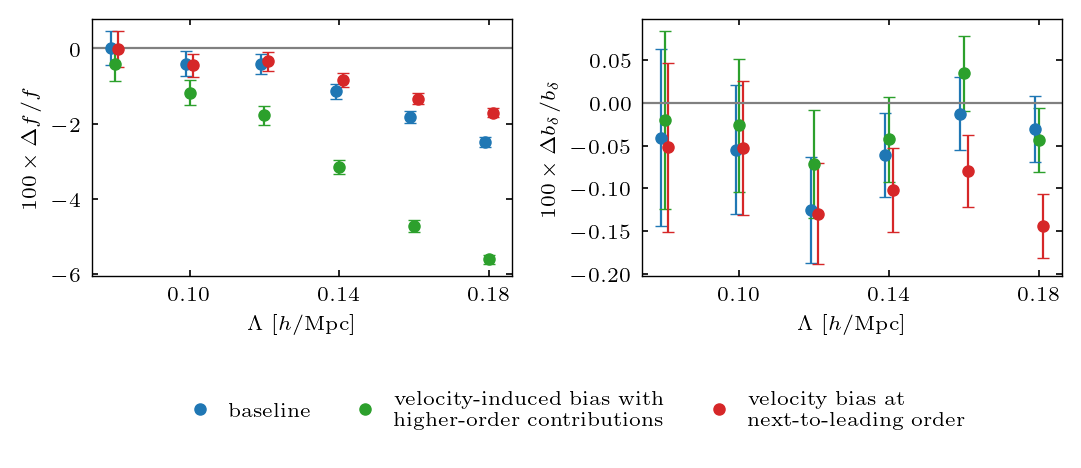}
	\caption{The growth rate and linear bias inferred from matter particles in simulation 1 at $z=0$, sub-sampled to $\bar{n}=0.01\,(\mpc/h)^3$ for different descriptions of velocity bias. The ``baseline'' scenario (blue) is equivalent to the analysis in figure \ref{fig: matter-overview}. Notice that the perturbative incorporation of velocity bias via higher-derivative bias performs poorly on scales above $0.1\,h/\mpc$.}
	\label{fig: matter-z000-operators}
\end{figure}

We now compare with results obtained when replacing the ``baseline'' description of velocity bias with the ``velocity-induced bias with higher-order contributions'', i.e. including the leading anisotropic higher-derivative bias as well as $\partial_\pp^4\delta$. Figure \ref{fig: matter-z000-operators} shows that this approach leads to increasingly biased growth rate inferences for $\Lambda > 0.10\,h/\mpc$, as already anticipated in the mock tests. This has important ramifications for future field-level inferences of the growth rate.

The inclusion of the additional velocity bias term $M\lin_{ji} \partial_i M\lin_{\pp j}$, one the other hand, does lead to a small but significant improvement for high values of $\Lambda$. However, it does not shift the scale where a significant deviation between the inferred value and the ground truth appears first appreciably.

At $z=0.5$, there appears to be a small sub-percent level shift in the inferred growth rate even for small cutoff values (figure \ref{fig: matter-overview}). We have verified in appendix \ref{sec: appendix-ngeul} that the mismatch is not reduced by a higher resolution of the displacement step, nor does the omission of the fourth velocity bias operator $\partial_i M^{(1)}_{\pp j} M^{(1)}_{ji}$ have a significant impact. Given the moderate statistical significance of the effect at the level of $1-2\,\sigma$ and the low number of two realizations which we have tested, a simple statistical fluctuation is a likely explanation. As explained in section \ref{sec: matter-pk}, we plan to improve the 3LPT model of the redshift-space density contrast in a future publication, and this will also allow us to study any possible connection between the precision of the gravity model and the inferred growth rate.

\section{Results from N-body halos}
\label{sec:results}

Finally, we turn to the analysis of fully non-linear, biased tracers, that is halos identified in the N-body simulations. Our halo sample is described in appendix \ref{sec: appendix-n-body-simulations} and consists of four disjoint, logarithmically spaced mass bins covering the range $10^{12.5}-10^{14.5} M_\odot/h$, each identified in snapshots at three redshifts, $z=0,\,0.5,\,1$. Further, we consider two simulation realizations, which only differ in the initial conditions. We focus on those scenarios with a mean number density $\bar{n} \geq 10^{-5} \left(h/\mpc\right)^3$, which effectively excludes the highest mass bin at $z=1$. While nothing sets this data apart fundamentally from all others, the low signal-to-noise ratio leads to substantial parameter degeneracies and makes it difficult to achieve convergence in the posterior sampling.

\subsection{Baseline analysis} 
\label{sec: matter-baseline}

\begin{figure}
	\centering
	\includegraphics[]{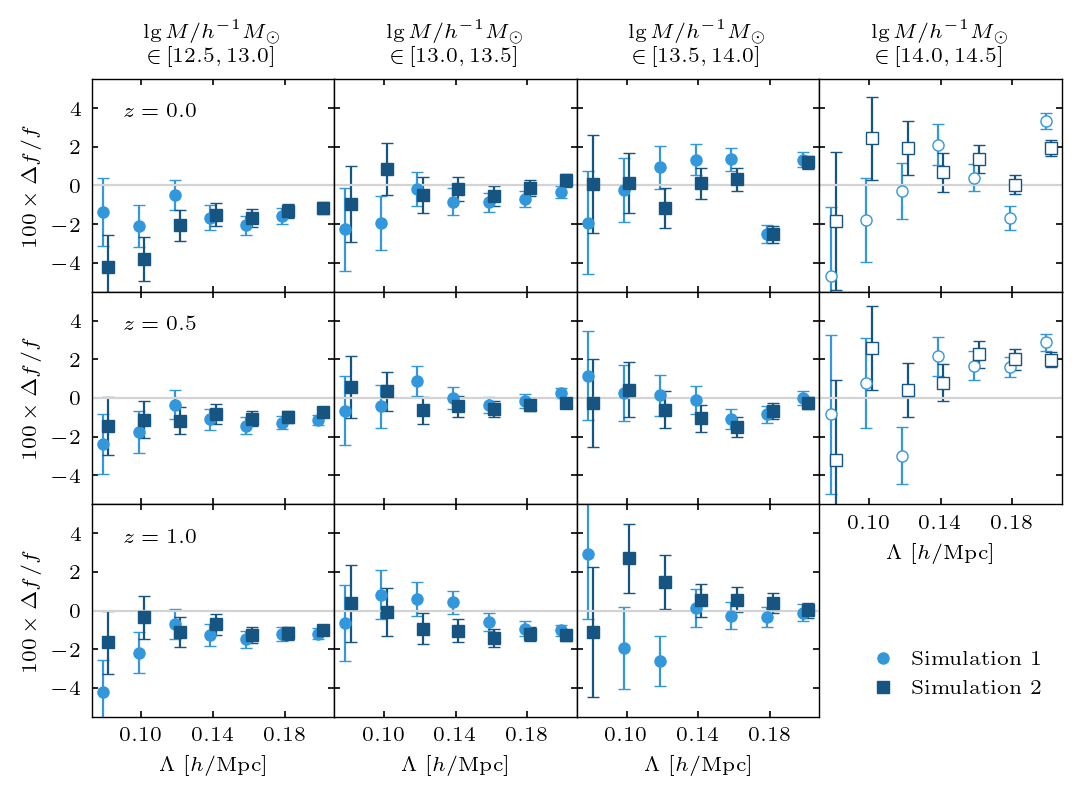}
	\caption{Summary of growth rate inference from halos, obtained with third order bias and the derivative operators from the ``baseline'' scenario in section~\ref{sec:mocktests-vbias}. Open symbols mark those data sets where local posterior maxima might impact the convergence of our sampling chains and the inferred mean value of $f$. In appendix \ref{sec:appendix-convergence} we verify that the global posterior maximum, i.e. the one  with the highest average log-likelihood, also finds $f$ closest to the ground truth.
	}
	\label{fig: halos-summary}
\end{figure}

In our baseline analysis, we use a third order bias expansion, whose operators are listed in appendix \ref{sec:appendix-bias-list}. The higher derivative and velocity operators are the same as in the ``baseline'' matter analysis. Specifically, we have one isotropic higher-derivative contribution $\nabla^2\delta$ and three velocity operators which are $\partial_\pp\delta$, $\partial_\pp(\sigma^2)$ and $\partial_\pp\mathrm{tr}[M^{(1)} M^{(1)}]$. While in the previous analyses all bias coefficients were sampled explicitly, we here make use of analytical marginalization for the coefficients of higher-order density operators, see appendix \ref{sec: appendix-marglh}. Still, the structure of the forward model (figure \ref{fig: model-flowchart}) requires explicit sampling of $b_\delta$ and all velocity coefficients. In addition to the bias parameters, we also infer the noise amplitudes $P_\epsilon^{(0)}$, $\sigma_{\epsilon,\,2}$, $\sigma_{\epsilon\mu,\,2}$, and the growth rate $f$. We then judge our results based on the recovery of the true value of $f$ (see eq.~{\ref{eq:appendix-sim-fvals}}) within its error bars.

Our results are summarized in figure \ref{fig: halos-summary}, and in appendix \ref{sec:appendix-convergence} we present a set of convergence tests. In some cases with a low signal-to-noise ratio, typically $\bar{n}\lesssim 10^{-4}\left(h/\mpc\right)^3$, the posterior exhibits multiple local maxima. Then, there is a chance for the sampling chains to get stuck locally and not explore the full posterior. In most cases, the main difference between individual posterior modes are the preferred values of the velocity bias coefficients, while the growth rate is consistent between them. Then, we expect the mean inferred value for $f$ to be robust with respect to convergence issues, but we caution that its error bar might be underestimated in such cases. However, in the highest mass bin, we could also identify some scenarios in which $f$ shifts significantly between the posterior modes. These cases exhibit a clear hierarchy between the log-likelihood of the posterior modes, allowing to identify a true (global) maximum. Importantly, the global maximum also is the one where $f$ is consistent with the ground truth. Nevertheless, we indicate all scenarios where the inferred mean value of the growth rate might be impacted by posterior multimodality by open symbols in figure \ref{fig: halos-summary}. Note that previous studies \cite{Elsner:2019rql, Schmidt:2020tao, Schmidt:2020viy, Babic:2022dws} employed a profile likelihood analysis and therefore were less affected by local posterior extrema, under the assumption that the minimizer was able to identify the global posterior maximum.

As figure \ref{fig: halos-summary} shows, the ``baseline'' analysis obtains unbiased constraints at a precision of a few percent for a broad range of halo masses, redshifts and cutoff scales. The scatter between the two realizations of the simulations is consistent with the estimated error bars. Moreover, we note the expected scaling behavior with mass and cutoff. That is, for lower values of $\Lambda$ and for lower redshifts and masses, where we expect the halos to be less biased, the estimated value of $f$ converges to the ground truth. However, there remains a small discrepancy in the lowest mass bin at $z=0$, where we consistently infer smaller values than expected. The systematic shift amounts to $\sim4\,\%$ at most and, at a level of $1-2 \sigma$, is only moderately significant, especially given the small number of mock realizations that we studied. The trend is similar to the matter results at low redshift, and might be of the same origin. We plan to investigate its origin and a possible connection with various ``precision aspects'' of the forward model in a future publication (see also the discussion at the end of section~{\ref{sec:matter-inference}}).

In appendix \ref{sec: appendix-ngeul}, we test how these results evolve with the resolution of the displacement step, which is controlled by $N_\eul$. While there is a considerable bias if $N_\eul=256$ is chosen, a further increase beyond $N_\eul=384$ does not significantly impact the estimated value of $f$ for all the cases tried. Thus, the results appear to be converged with respect to the resolution of the displacement steps. Further, we test for residual impacts of the N-body accuracy in appendix \ref{sec: appendix-transients}.

To investigate how FoG impact the results of figure \ref{fig: halos-summary}, we focus on the lowest mass bin, $\lg M/ h^{-1} M_\odot \in \left[12.5, 13.0\right]$, at $z=0$. This is the halo sample for which we expect the strongest FoG. If we exclude subhalos from the analysis, we essentially obtain identical results. This finding agrees with the fact that subhalos make up only a modest fraction ($\sim 2\%$) of the rather high-mass halos. It thus appears that FoG do not have a dominant impact on the halo samples considered here. It would be interesting to apply the analysis to mock catalogs that include satellite galaxies in the future as well. Note, however, that we have already considered a sample with strong FoG in the form of sub-sampled N-body particles in the previous section.

\begin{figure}
\centering
\includegraphics[]{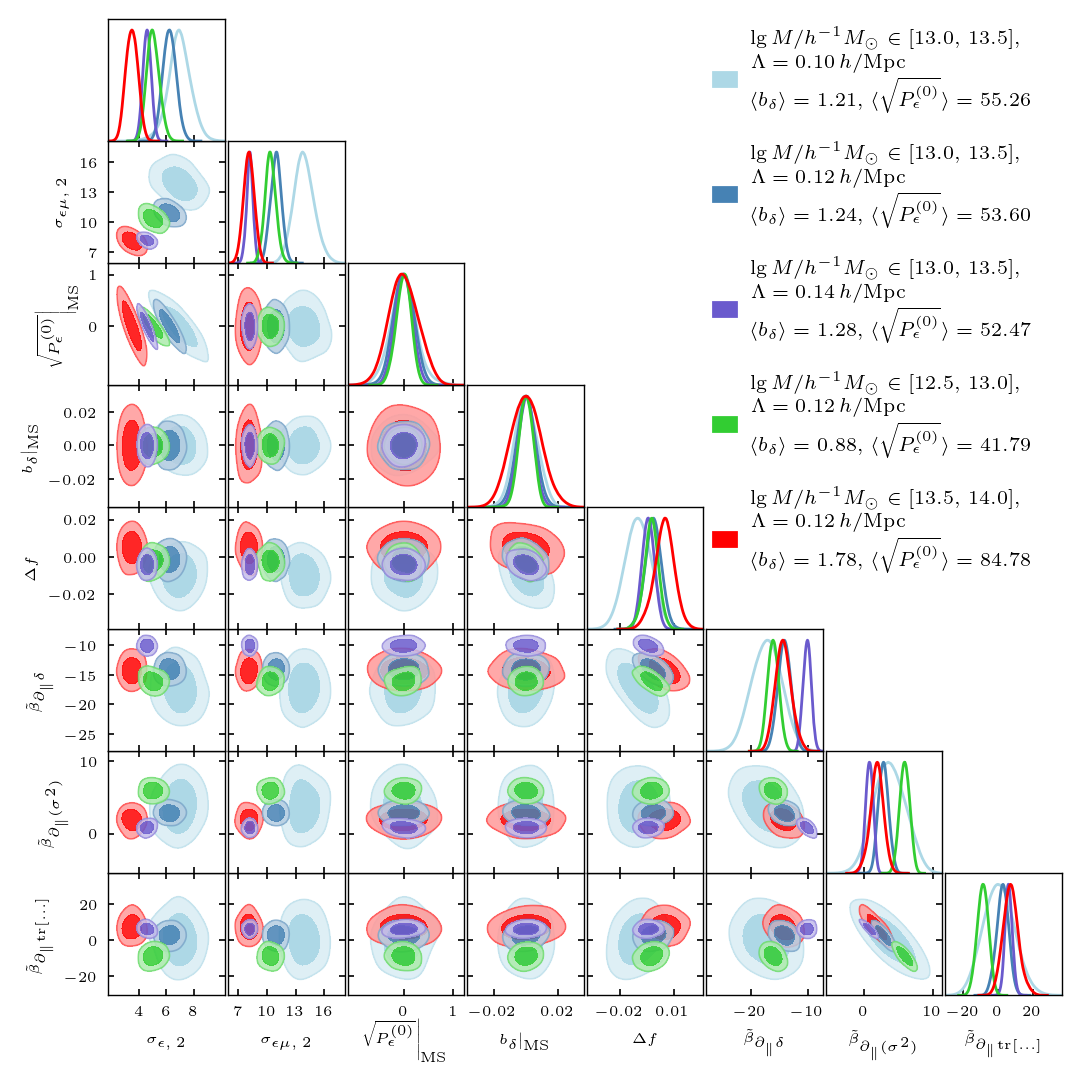}
\caption{Posterior contours from the ``baseline'' analysis for various halo mass bins and cutoff values, all considering simulation 1 at $z=0$. We show all parameters which are sampled explicitly while the higher-order bias coefficients $b_\op$ are marginalized over analytically, see appendix~\ref{sec: appendix-marglh}. In case of the velocity bias coefficients, we plot $\tilde{\beta}_\uop = \beta_\uop/f$ and abbreviate $\tilde{\beta}_{\partial_\pp\mathrm{tr}[M^{(1)} M^{(1)}]}$  as $\tilde{\beta}_{\partial_\pp\mathrm{tr}[...]}$. In case of the noise amplitude and of the linear bias parameter, we subtract the sample mean as given in the legend for better visualization. This is indicated by the notation ``$\left. \ldots\right|_\mathrm{MS}$'' in the axis labels.}
\label{fig: halosM2-degeneracies}
\end{figure}

In figure \ref{fig: halosM2-degeneracies}, we show in blue hues the two-dimensional posterior slices from the ``baseline'' analysis of the second mass bin at $z=0$ for three cutoff values, $\Lambda=0.10\,,~0.12,,~0.14\,h/\mathrm{Mpc}$. In particular the behavior at low cutoffs is representative for most cases. There, the velocity bias coefficient $\beta_{\partial_\pp (\sigma^2)}$ exhibits a degeneracy with both, $\beta_{\partial_\pp\mathrm{tr}[M^{(1)} M^{(1)}]}$ and $\beta_{\partial_\pp\delta}$, and further there is a moderate degeneracy between the leading-order velocity bias coefficient, $\beta_{\partial_\pp\delta}$ and the growth rate. This might also be the reason why some of the inferences at low values of $\Lambda$ tend to under-predict the growth rate. The inclusion of information on smaller scales breaks these degeneracies, in particular the one between $\beta_{\partial_\pp\delta}$ and $f$, and it tightens the constraint on the velocity bias coefficients.

From the legend of figure \ref{fig: halosM2-degeneracies}, it is evident that the inferred linear bias parameter systematically shifts with the cutoff $\Lambda$. We actually observe this trend for all halo samples, but it is absent in the analysis of unbiased mocks and matter particles, figures \ref{fig: mock-operator-scaling} and \ref{fig: matter-overview}, respectively. However, we can reproduce the running in mocks that were generated with a larger cutoff than used in the analysis if these mocks are non-trivially biased. The effect also persists if we carry out the analysis in the halo rest frame (with $f$ fixed to zero). A running of $b_\delta$ with $\Lambda$ is expected, since we do not perform any ``renormalization'' or ``orthogonalization'' on the bias operators here, unlike what was done for the results shown in previous work \cite{Schmidt:2018bkr, Schmittfull:2018yuk, Elsner:2019rql} (see also \cite{abidi/baldauf:2018}). In principle, it is possible to calculate the expected running of bias coefficients with $\Lambda$, and to thus compute the value corresponding to $\Lambda\to 0$ which could be compared with, e.g. separate universe measurements. 
Since we here essentially treat all bias coefficients as nuisance parameters, however, the running has no consequences and we defer such a calculation to future work {\cite{Rubira:2023vzw}}.

We also show in figure \ref{fig: halosM2-degeneracies} in green and red the posterior contours inferred from the lowest- and from the second-highest mass samples at $z=0$ for a cutoff value $\Lambda=0.12\,h/\mpc$. Comparing the three different masses at identical cutoff, one notes that the inferred noise power spectrum decreases for smaller masses as expected. Correspondingly, the constraints on all parameters tighten. While the inferred higher-order velocity bias coefficients are consistent with zero for the two more massive halo samples at $\Lambda=0.12\,h/\mpc$, there is a significant detection in the lowest mass bin. As one proceeds to higher cutoffs, the next-to-leading order velocity bias becomes important also for more massive halos.

\subsection{Velocity bias}
\label{sec: halos-vbias}

The velocity bias coefficients which are inferred for halos actually comprise two contributions: counter terms, i.e. the velocity bias of the effective matter fluid, and halo velocity bias. Explicitly, we can decompose the leading-order term as
\begin{equation}
\beta_{\partial_\pp\delta} = \beta_{\partial_\pp\delta}^\mathrm{c.t.} + \beta_{\partial_\pp\delta}^\mathrm{bias}\,,
\end{equation}
where the first term is expected to be identical for all mass bins and to coincide with the values found from matter in section \ref{sec:matter-inference}. Before we explore how different velocity bias expansions impact the inferred growth rate, we can gain some intuition for $\beta_{\partial_\pp\delta}^\mathrm{bias}$ from the biasing of Lagrangian density peaks.

The reasoning behind Lagrangian density peaks starts from the assumption that dark matter halos form from peaks in the Lagrangian density field, and then follows the conserved evolution of this distribution under gravity \cite{Desjaques_2018, 1970Ap......6..320D, 1984ApJ...284L...9K, 1985MNRAS.217..805P, 1986ApJ...304...15B}. Velocity bias arises from averaging the underling matter field over the finite size of a halo, and because halos occupy special, biased regions of the density and velocity distribution. The latter effect leads to a predicted lowest-order velocity bias coefficient of \cite{Desjacques:2008jj,Desjacques:2009kt, Desjaques_2018}
\begin{equation}
\beta_{\partial_\pp\delta}^\mathrm{LDP} = f(z)\,\frac{\sigma_0^2}{\sigma_1^2}\,.
\label{eq:results-bias-expectation-lpt}
\end{equation}
This expression approximately scales as the Lagrangian radius squared (see eq.~(\ref{eq:model-lagrangian-radius}) for its definition and eq.~(\ref{eq:appendix-nbody-lagrangianradius}) for numerical values),
\begin{equation}
\frac{\sigma_0^2}{\sigma_1^2} = \frac{\int d^3 k~ W_\mathrm{R_\mathrm{L}}(k)\, P_\mathrm{L}(k)}{\int d^3 k~ k^2\, W_\mathrm{R_\mathrm{L}}(k)\, P_\mathrm{L}(k)} \sim R^2_\mathrm{L}(M)\,.
\label{eq:results-ldp-vbias-scaling}
\end{equation}
where $W_\mathrm{R_\mathrm{L}}(k)$ is a spherically symmetric filtering kernel.
The contribution to the velocity bias from smoothing the matter velocity is likewise controlled by $W_\mathrm{R_\mathrm{L}}(k)$. Hence, the overall expected scaling is $\beta_{\partial_\pp\delta}^\mathrm{LDP} \propto R_{\rm L}^2$.

\begin{figure}
\centering
\includegraphics[]{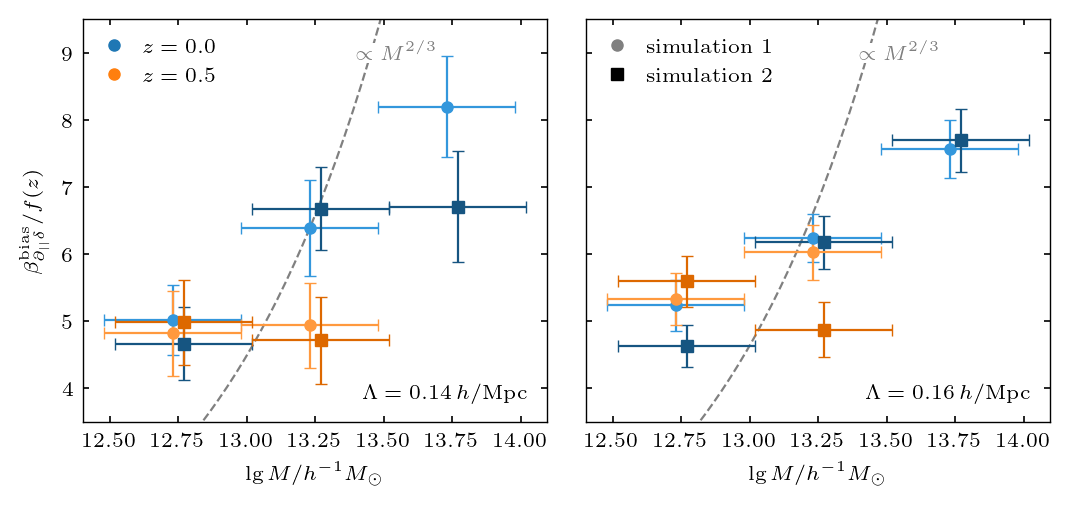}
\caption{The leading-order velocity bias coefficient inferred from halos at $\Lambda=0.14\,h/\mpc$ and $\Lambda=0.16\,h/\mpc$ (left and right panel respectively) for $z=0,\,0.5$. We use the ``baseline'' scenario for the inference and subtract $\beta_{\partial_\pp\delta}^\mathrm{c.t.}$ as obtained in the matter analysis to isolate $\beta_{\partial_\pp\delta}^\mathrm{bias}$. The dashed line illustrates the scaling of $R_\mathrm{L}^2(M) \propto M^{2/3}$ predicted by the biasing of Lagrangian density peaks.}
\label{fig:halos-velocity-bias}
\end{figure}

In section \ref{sec:matter-inference}, we found that the lowest-order velocity counter term comes with a negative coefficient, corresponding to additional smoothing, and that its magnitude decreases for larger values of $\Lambda$. In contrast, the Lagrangian density peaks approach now predicts $\beta_{\partial_\pp\delta}^\mathrm{LDP}$ to be independent of the cutoff, and that the only source of time-dependence is from the growth rate in eq.~(\ref{eq:results-bias-expectation-lpt}). The sign of $\beta_{\partial_\pp\delta}$ depends on whether the contribution in eq.~(\ref{eq:results-ldp-vbias-scaling}) or the smoothing contribution dominates.

For a more detailed view, we show in figure \ref{fig:halos-velocity-bias} the inferred leading-order velocity bias coefficient after subtracting $\beta_{\partial_\pp\delta}^\mathrm{c.t.}$ as found in section \ref{sec:matter-inference}. To focus on cases where the velocity bias is well constrained, we exclude snapshots at $z=1$ and cases with multiple posterior maxima in the velocity bias coefficients (see appendix \ref{sec:appendix-convergence}). Further, we select moderately large cutoff values of $\Lambda=0.14\,,~0.16\,h/\mpc$. Indeed, the qualitative evolution follows the peak-prediction in that we neither note a strong scaling with cutoff nor redshift but a clear tendency of the higher mass bins to larger values. Still, the velocity bias increases less steeply with mass than the simple proportionality in eq.~(\ref{eq:results-ldp-vbias-scaling}) would suggest. Since the model of Lagrangian density peaks makes many simplifying assumptions, such a disagreement on the quantitative level is not entirely unexpected. The values inferred for $\beta_{\partial_\pp\delta}^\mathrm{bias}$ are positive, indicating a partial cancellation between counter terms and bias. As consequence of the mass scaling in $\beta_{\partial_\pp\delta}^\mathrm{bias}$, the inferred coefficients $\beta_{\partial_\pp\delta}$ tend closer to zero for more massive halos.

\begin{figure}
\centering
\includegraphics[]{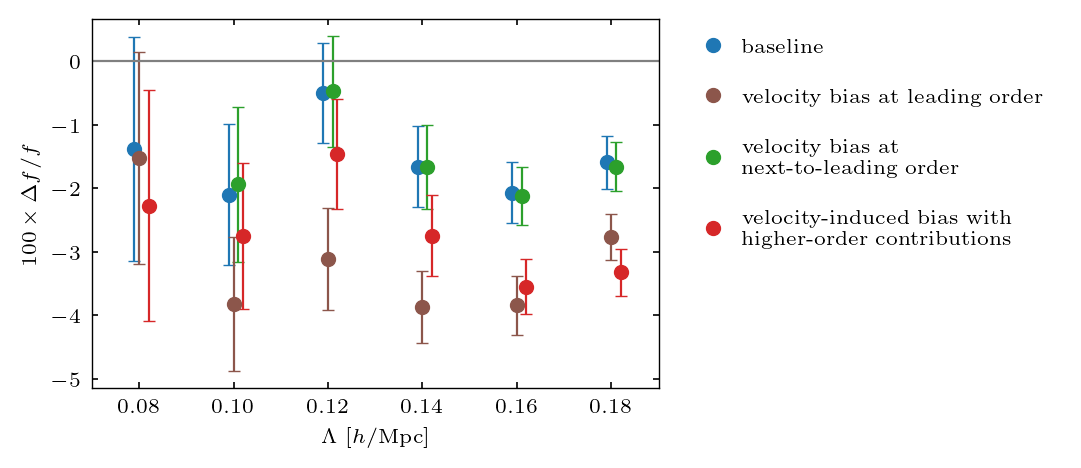}
\caption{The effect of different velocity-bias descriptions on the growth rate inferred from halos in the lowest mass bin, $\lg M/h^{-1}M_\odot \in \left[12.5,13.0\right]$, at $z=0$ for simulation 1. All models assume a third order bias expansion as summarized in appendix \ref{sec:appendix-bias-list} but include different approaches to incorporate velocity bias, see section \ref{sec:mocktests-vbias}. The baseline analysis is identical to figure \ref{fig: halos-summary}.}
\label{fig: halos-z000-operators}
\end{figure}

Finally, we pick the lowest-mass sample at $z=0$ to investigate what effect alternative descriptions of the velocity bias have on the inference. In particular, we compare our ``baseline'' analysis to the ``velocity bias at leading order'', the ``velocity bias at second order'' and the ``velocity-induced bias with higher-order contributions'' scenarios that were introduced in section \ref{sec:mocktests-vbias}. The results are summarized in figure \ref{fig: halos-z000-operators}. There is no significant difference between the ``baseline'' and the ``velocity bias at second order'' analysis, which implies that the operator $\partial_i M^{(1)}_{\pp j} M^{(1)}_{ji}$ has no impact and justifies neglecting it in the ``baseline'' analysis. In contrast, the ``velocity bias at leading order'' analysis agrees with the ``baseline'' results only for very small cutoffs. For higher values of $\Lambda$, the inferred growth rate significantly deviates from the ground truth. Similarly, the results obtained with the ``velocity-induced bias with higher-order contributions'' analysis agree with the ``baseline'' ones for low $\Lambda$ but start to degrade significantly for higher $\Lambda$. Already the matter results (figure \ref{fig: matter-z000-operators}) showed that the velocity-induced density operators cannot capture the effect of velocity bias well, and we now essentially observe the same behavior for halos in the low-mass sample.

\section{Conclusions}
\label{sec:conclusions}

In this study, we extend previous works on field-level cosmology analysis \cite{Schmidt:2020viy, Schmidt:2020tao, Babic:2022dws, Kostic:2022vok} with the EFT likelihood \cite{Schmidt:2018bkr, Elsner:2019rql, Cabass:2019lqx, Cabass:2020jqo} to biased tracers in redshift space. Our most important additions to the forward model are: the computation of the velocity field, the addition of a second displacement step which transforms the biased density field to redshift space, and the introduction of a systematic bias expansion for the velocities. We apply our analysis to mock data and simulations and test how well we can infer the growth rate $f$ (in addition to bias coefficients and noise amplitudes). By fixing the initial conditions in the inference to their known ground truth, we can eliminate cosmic variance to the largest extent possible and perform a stringent test of our model.

To describe the gravitational evolution of the density field, we here use a third order Lagrangian Perturbation Theory (3LPT) model \cite{Schmidt_2021}, that has been implemented in the \texttt{LEFTfield} code previously. In general, however, the implementation is generic and can handle any desired order in perturbations. The bias expansion is constructed in the rest frame from scalar invariants of the Lagrangian distortion tensor \cite{Mirbabayi:2014zca, Schmidt_2021}.
The full forward model is summarized in figure \ref{fig: model-flowchart}. Importantly, we compute the displacements from the Lagrangian to the Eulerian frame and eventually to redshift space fully nonlinearly (in terms of the relation between displacement vector and displaced field). This means that we keep those contributions that are protected by symmetries up to arbitrary order. These symmetries are the equivalence principle in case of the Lagrangian to Eulerian displacement, and number conservation in case of the displacement to redshift space.

For the second displacement, to redshift space, we use the biased velocities as shift vector. This treatment generates additional, higher-order terms in the density contrast, which have not been considered in previous studies. Further new terms arise from the systematic expansion of the velocity bias up to second-order in perturbations in eq.~(\ref{eq:model-vbias-list}). We argue that these new operators indeed are more relevant than the $\partial_\pp^4\delta$ bias in the density contrast, which is often introduced by redshift-space models (e.g. \cite{Schmittfull:2020trd, Chudaykin:2020aoj, Chudaykin:2020hbf}).

Our analysis correctly recovers the growth rate for a diverse set of test cases. Specifically, we have considered mock data that were generated from the perturbative forward model with a higher cutoff than used in the analysis, matter particles in N-body simulations that we sub-sampled to generate a trivially biased, noisy but fully non-linear tracer density, and halos in N-body simulations. In the case of simulations, we consider redshift snapshots at $z=0\,,~0.5\,,~1$ and for the halos a broad range of masses $10^{12.5} \leq \lg M/h^{-1}M_\odot \leq 10^{14.5}$, divided into four mass bins. The resulting constraints on the growth rate are at the level of a few percent, accentuating both the stringency of the tests and the accuracy of the redshift-space forward model. Furthermore, we find the expected scaling with halo mass, redshift and cutoff scale: when lowering the cutoff, the inferred growth rate converges to the ground truth, and this happens at higher cutoffs for less biased samples. This even is the case if we analyze tracers with strong Fingers-of-God, in the form of sub-sampled matter particles at low redshifts. To put these results into context, the anticipated precision on the RSD measurement by DESI is $\sigma\left(f\sigma_8\right)/\left(f\sigma_8\right) \sim 2.6\,\%$ in the most constraining redshift range around $z\sim1.2$ {\cite{DESI:2023dwi}}. Before we can make a meaningful comparison of the constraining power between field-level and standard analyses, we have to address the sampling of initial conditions. Further, note that the number density of tracers, the effective volume and scale cuts differ between the forecast and this work. Our results achieve an accuracy which is comparable to DESI's statistical power in a power spectrum analysis (but within a smaller volume); hence, the systematic accuracy of the forward model presented here already appears to be sufficient for such an analysis, but not necessarily for a field-level inference on DESI data. More detailed studies including sampling the initial conditions will be required for such an assessment.

We compare the explicit treatment of velocity bias to the standard approach, where its impact on the density contrast is expanded to leading order (captured by the operator $\partial^2_\pp\delta$), and to extended models which consider one higher-order contribution of the form $\partial^4_\pp\delta$. For all cases considered -- mock data, dark matter particles and low mass halos --  we find that the explicit treatment performs much better. This has important implications for future field-level analyses in redshift space, but likely also for higher $n$-point functions. Given the considerable improvement from the explicit treatment of velocity bias, we further conclude that our analysis yields a significant and robust detection of velocity bias. That is, the field-level analysis is able to detect velocity bias purely from the tracer distribution in redshift space without explicit access to the tracers' velocities.

The success of the field-level analysis in redshift-space depends on various aspects of the forward model. Here, we found the most crucial to be an adequate description for the bias and in particular velocity bias, and the resolution at which the displacements from Lagrangian to Eulerian and to redshift space are computed. In a future publication, we plan to study the impact of further ``precision aspects'' of the forward model -- such as the choice of the assignment kernel, the resolution of the displacement, perturbative order and curl contributions to the displacement vector. The use of more involved assignment schemes, that can deconvolve the effect of the kernel, is in particular interesting in this context to keep the computational costs manageable.

This work is an important first step towards the full field-level inference in redshift space. Eventually, the parameter inference presented here has to be combined with the sampling of initial phases in order for the technique to be applicable to actual observational data. In the meantime, the stringency of the field-level analysis at fixed initial conditions also provides important tests for the forward model, and for the capability of the bias expansion in particular, as well as the opportunity to study tracer velocity bias at high precision.

\acknowledgments
We want to thank Raúl Angulo, Ivana Babi\'c, José Luis Bernal, Giovanni Cabass,  Andrija Kosti\'c,  Minh Nguyen and Beatriz Tucci for useful discussion. This work was supported by the Deutsch Forschungsgemeinschaft (DFG, German Research Foundation) under Germany's Excellence Strategy - EXC-2094 - 390783311 and performed in part at Aspen Center for Physics, which is supported by National Science Foundation grant PHY-1607611. FS acknowledges support from the Starting Grant (ERC-2015-STG 678652) ``GrInflaGal'' of the European Research Council.

\appendix

\section{The marginalized EFT-likelihood}
\label{sec: appendix-marglh}

In section \ref{sec:model-likelihood}, we have introduced the EFT-likelihood, which is the likelihood of the observed data conditional on initial conditions, cosmological parameters and bias coefficients. Further, we focus in this work on the Fourier-space version and apply it to the redshift-space density contrast. That is, we evaluate
\begin{equation}
\ln \mathcal{P}\left(\left.\dg\right|\dgd\right) = -\frac{1}{2} \int_{|\vec{k}|<k_{\max}}\, \frac{d^3\vec{k}}{(2\pi)^3} \left\{ \frac{\left|\tilde{\delta}_\tracer(\vec{k})-\dgds(\vec{k})\right|^2}{P_\epsilon(k,\mu)} 
+ \ln\left[2\pi P_\epsilon(k,\mu)\right]\right\}\,,
\end{equation}
which is eq.~(\ref{eq:model-likelihood-fourier}) with the density fields replaced by their redshift-space equivalents. The noise power spectrum is defined in eq.~(\ref{eq:model-noise-power-pectrum}).

In many contexts, the bias coefficients are nuisance parameters, which are marginalized over to derive constraints on the cosmological parameters. The likelihood has a very simple dependence on the higher-order density bias coefficients. According to eq.~(\ref{eq:model-deltaDet-zspace}), they enter the model prediction $\dgds$ only as multiplicative factors of the higher-order operators. It is therefore possible to perform the marginalization over these parameters analytically \cite{Elsner:2019rql}. Note that the same is not true for the linear bias coefficient $b_\delta$ and for the velocity operators $\beta_\uop$, which all appear in the nonlinear transformation from the rest frame to redshift space.

To perform the analytic marginalization, we write $\dgds$ as
\begin{align}
\dgds(\vec{k}) &= \tilde{\mu}(\vec{k}) + \sum_{{\op\in\{\op\}_\mathrm{marg.}}} b_\op\, \tilde{\op}(\vec{k})\,, 
\quad \mathrm{where}\quad
\tilde\op = \left[\frac{\op}{1+\eta}\right]\left(\vec{k}\right)\\
\mathrm{and}\quad
\tilde{\mu}\left(\vec{k}\right) &= \left[\frac{1+b_\delta \delta}{1+\eta} -1\right](\vec{k}) + \sum_{\op\in\{\op\}/\{\op\}_\mathrm{marg.}}\,b_\op\, \tilde{\op}(\vec{k}) .
\label{eq:appendix-marglh-lh}
\end{align}
That is, we split the higher-order contributions in eq.~(\ref{eq:model-deltaDet-zspace}) into a marginalized and an unmarginalized part. See appendix \ref{sec:appendix-bias-list} for the full list of $\{\op\}_\mathrm{marg.}$ adopted for this work. With this decomposition, the analytically marginalized likelihood is given by \cite{Elsner:2019rql}
\begin{align}
&\ln \mathcal{P}\left(\left.\tilde{\delta}\right| \dgds\,, \{b_\op\}_{\op\in\{\op\}/\{\op\}_\mathrm{marg.}}\right) = \ln \left[ \left(\prod_{\op\in\{\op\}_\mathrm{marg}} \int d b_\op \right) \mathcal{P}\left(\left.\tilde{\delta}\right| \dgds\,, \{b_\op\}\right) \right]
\nonumber \\
&\quad =-\frac{1}{2} \left\{ \ln \left|A_{\op\op'}\right| + 
\sum_{\vec{k}\neq 0}^{k_\mathrm{max}} \ln P_\epsilon(k,\mu) + 
 C  
- \sum_{{\op,\op'\in\{\op\}_\mathrm{marg}}} B_\op\, \left[A^{-1}\right]_{\op \op'}\, B_{\op'}
\right\} + \mathrm{const.}\,,
\end{align}
where we have omitted the contribution from the normalization constant that is independent of all free parameters. We also have expressed the integral in eq.~(\ref{eq:appendix-marglh-lh}) as sum over a finite number of modes, appropriate for the numerical evaluation on a discrete grid. The quantities $A$, $B$, $C$ are a square matrix, vector and scalar, respectively. The dimensionality of the former two is given by the number of marginalized bias coefficients, while the latter only depends on the data and the unmarginalized part of the bias expansion. Explicitly, they are given by
\begin{align}
C &= \sum_{\vec{k}\neq 0}^{k_\mathrm{max}} \frac{\left|\tilde{\delta}_\tracer(\vec{k}) - \tilde{\mu}(\vec{k}) \right|^2}{P_\epsilon(k,\mu)}\,,
\\
B_\op &= \sum_{\vec{k}\neq 0}^{k_\mathrm{max}}  \frac{\Re\left\{\left[\tilde{\delta}_\tracer(\vec{k}) - \tilde{\mu}(\vec{k})\right]\tilde{\op}^{*}\right\}}{P_\epsilon(k,\mu)}\,,
\label{eq:appendix-marglh-B}
\\
A_{\op \op'} &= \sum_{\vec{k}\neq 0}^{k_\mathrm{max}} \frac{\tilde{\op}(\vec{k})\, \tilde{\op}'^*(\vec{k})}{P_\epsilon(k,\mu)}\,,
\label{eq:appendix-marglh-A}
\end{align}
with $\op,\op'\in\{\op\}_\mathrm{marg}$. Note that $A$ and $B$ in eqs.~(\ref{eq:appendix-marglh-B}) and (\ref{eq:appendix-marglh-A}) assume uninformative priors, as done throughout this work. Gaussian priors, however, can easily be incorporated \cite{Schmidt:2020tao}.

\section{N-body simulations and halo catalogs}
\label{sec: appendix-n-body-simulations}

\begin{table}
	\centering
	\begin{tabular}{l|c|l}
		\hline
		matter density in units of the critical density & $\Omega_\mathrm{m}$ & $0.30$\\
		dark energy density in units of the critical density & $\Omega_\Lambda$ & $0.70$\\
		Hubble constant in $100\,\mathrm{km}/\mpc/\mathrm{s}$ & $h$ & $0.70$ \\
		perturbation amplitude & $\sigma_8$ & $0.84$\\
		spectral index of scalar primordial perturbations & $n_\mathrm{s}$ & $0.967$\\[.5\baselineskip]
		\hline
		simulation box size in units of $h/\mpc$ & $L_\mathrm{box}$ & $2,000$\\
		number of dark matter particles & $N_\mathrm{part}$ & $1536^3$\\
		mass of dark matter particles in units of $M_\odot/h$ & $M_\mathrm{part}$ & $1.8\times10^{11}$\\
		starting redshift & $z_\mathrm{in}$ & $24$\\
		LPT order to set initial conditions && 2\\
		\hline
	\end{tabular}
	\caption{Parameters for the Gadget-2 N-body simulations used in this work.}
	\label{tap: appendix-nbody-parameters}
\end{table}

We use the same N-body simulations as previous studies \cite{Schmidt:2020tao, Babic:2022dws}. They were obtained with Gadget-2 \cite{Springel:2005mi} assuming a flat \LCDM~cosmology whose parameters are summarized in table~\ref{tap: appendix-nbody-parameters}. Two realizations of these simulations are available, which differ only in their random initial conditions. Further we consider snapshots at redshift $z=0\,,~0.5\,,~1$. At these redshifts, the growth rate of the fiducial cosmology is
\begin{equation}
f(z=0) = 0.5128\,,\quad
f(z=0.5) = 0.7491\,,\quad
f(z=1) = 0.8690\,.
\label{eq:appendix-sim-fvals}
\end{equation}
Halos were subsequently identified in these snapshots as spherical overdensities \cite{Press:1973iz, 1992ApJ...399..405W, Lacey:1994su} with the Amiga Halo Finder (AHF) algorithm \cite{Gill:2004km, 2009ApJS..182..608K}, using an overdensity threshold of 200 times the background. We displace the halos to redshift space using their peculiar velocity as reported by AHF, which corresponds to the bulk velocity of the halo. For the analysis, the halos are divided into four logarithmically spaced, non-overlapping mass bins between $10^{12.5}\, M_\odot/h$ and $10^{14.5}\, M_\odot/h$. The properties of the resulting halo samples are summarized in table \ref{tab: appendix-nbody-halosamples}. Note that we do not consider the highest mass bin at $z=1$ in our analyses due to its low signal-to-noise ratio.

Table \ref{tab: appendix-nbody-halosamples} also lists the root-mean-square displacement between Eulerian and redshift space. Apparently, there is no strong evolution with redshift, but the displacement seems to be largest at $z=0.5$ and smallest at $z=0$. This is indeed what one would expect from the linear velocity relation in eq.~(\ref{eq:model-linear-velocity}). While $\delta\lin$ grows proportionally to the growth factor $D(z)$, the growth rate $f(z)$ decreases with time. For the cosmology given in table~\ref{tap: appendix-nbody-parameters} the product $f(z) \times D(z)$ is $0.513$ at $z=0$, $0.579$ at $z=0.5$ and $0.532$ at $z=1$, in agreement with the trends apparent in table~\ref{tab: appendix-nbody-halosamples}. The RMS in table \ref{tab: appendix-nbody-halosamples} would also be impacted by the random motion of subhalos within their parent. The fact that there is no increase for low masses and low redshifts matches the fact that subhalos do not contribute significantly to the total halo number; even in the lowest-mass sample at $z=0$ they make up $2\%$ only.

\begin{table}
	\centering
	\begin{tabular}{c|c|c|c}
		\makecell{\bf redshift \\ ~} & 
		\makecell{\bf mass \\ $[\lg M/h^{-1} M_\odot]$}  & 
		\makecell{\bf number density \\ $[10^{-4}\,\left(h/\mpc\right)^3]$} & 
		\makecell{\bf RMS displacement \\ $[\mpc/h]$}  \\
		\hline\hline 
		\multirow{4}{*}{$z=0$} & $12.5-13.0$ & 7.64 & 3.17  \\
		& $13.0-13.5$ & 3.65 & 3.31 \\
		& $13.5-14.0$ & 1.17 & 3.28  \\
		& $14.0-14.5$ & 0.30 & 3.18 \\
		\hline 
		\multirow{4}{*}{$z=0.5$} 
		& $12.5-13.0$ & 6.55 & 3.61 \\
		& $13.0-13.5$ & 2.88 & 3.69\\
		& $13.5-14.0$ & 0.75 & 3.67 \\
		& $14.0-14.5$ & 0.13 & 3.61 \\
		\hline
		\multirow{4}{*}{$z=1$} 
		& $12.5-13.0$ & 5.19 & 3.33 \\
		& $13.0-13.5$ & 1.95 & 3.39 \\
		& $13.5-14.0$ & 0.38 & 3.39 \\
		& $14.0-14.5$ & 0.04 & 3.34  \\
	\end{tabular}
	\label{tab: appendix-nbody-halosamples}
	\caption{Summary of halo samples used in this work. We quote the average over two simulations that only differ in the realization of the initial conditions. In addition to the number density, we also measure the RMS displacement between halo rest-frame and redshift space (right column).
	}
\end{table}

For convenience, we also give the Lagrangian radius, eq.~(\ref{eq:model-lagrangian-radius}), at the edges of our mass bins
\begin{align}
&R_\mathrm{L}\left(10^{12.5}\,M_\odot\right) = 2.1\, \mpc/h\,,\quad
 R_\mathrm{L}\left(10^{13.0}\,M_\odot\right) = 3.0\,\mpc/h\,,\quad \nonumber\\
&R_\mathrm{L}\left(10^{13.5}\,M_\odot\right) = 4.4\,\mpc/h\,,\quad
 R_\mathrm{L}\left(10^{14.0}\,M_\odot\right) = 6.5\,\mpc/h\,,\quad \nonumber\\
&R_\mathrm{L}\left(10^{14.5}\,M_\odot\right) = 9.5\,\mpc/h \,.
\label{eq:appendix-nbody-lagrangianradius}
\end{align}

In the last stage of this project, we became aware of a small defect in the catalogs, where $\simeq 0.14\%$ of the volume does not contain any halos. These ``holes'' are roughly rectangular slabs that are very narrow in the second Cartesian coordinate. This affects both simulation realizations and all redshifts and mass bins. In contrast, there are no such holes in the underlying matter snapshots. We where unable to identify the source of error in AHF. To test which impact it might have on our analysis, we use the \texttt{Rockstar} \cite{Behroozi2013} catalogs of the same simulations that where introduced in \cite{Schmidt:2020tao}, and do not exhibit such defects. We then analyze the \texttt{Rockstar} catalogs and compare the results with two further runs, where we have introduced holes (a) of similar size and shape as in the AHF sample and (b) of twice that size. Apart from a small shift in the estimated noise amplitudes, all three catalogs yield equal results for the growth rate and the inferred bias parameters. Since the impact of the defects is negligible and the analyses consume considerable computational time, we present our original results based on the AHF catalogs in section \ref{sec:results}.

\section{Baseline bias model}
\label{sec:appendix-bias-list}
The baseline analysis assumes a third-order bias expansion, which contains the following operators: 
\begin{gather}
\delta\,,\quad
\sigma^2\,,\quad
\sigma^3\,,\quad
\sigma\, \tr[M^{(1)} M^{(1)}]\,,\quad\nonumber\\
\tr[M^{(1)} M^{(1)} M^{(1)}]\,,\quad
\tr[M^{(1)} M^{(1)}]\,,\quad
\tr[M^{(1)} M^{(2)}]\,.
\end{gather}
In addition, there is one higher-derivative operator,
\begin{equation}
\nabla^2\delta\,.
\end{equation}
Unless noted otherwise, we only use a subset of the four velocity bias operators in eq.~(\ref{eq:model-vbias-list}), which is given by
\begin{equation}
\partial_\pp\delta\,,\quad
\partial_\pp\left(\sigma^2\right)\,,\quad
\partial_\pp \mathrm{tr}\left[M^{(1)} M^{(1)}\right]\,.
\end{equation}
We stress that there is no a-priori reason to exclude the operator $M^{(1)}_{ji}\partial_i M^{(1)}_{\pp j}$ from the velocity bias expansion. We test the impact of this operator for a selected subset of scenarios.

The noise parametrization is given in eq.~(\ref{eq:model-noise-parametrization}) and contains the lowest-order white component, an isotropic $k^2$-dependent component, and a LOS-dependent $k^2$-component. We express the power in the zeroth-order contribution as
\begin{equation}
P_\epsilon^{(0)} = \left(\frac{L_\mathrm{box}}{N_\mathrm{G,like}}\right)^3 \sigma_{\epsilon,\,0}^2\,,
\end{equation}
where $L_\mathrm{box}$ is the physical size of our simulation box and $N_\mathrm{G,like}^3$ the size of the grid on which the likelihood is evaluated. Doing so, we ensure that the noise power spectrum assumes only positive values. The free parameters to describe the noise in our inference then are
\begin{equation}
\sigma_{\epsilon,\,0}\,,\quad
\sigma_{\epsilon,\,2}\,,\quad
\sigma_{\epsilon\mu,\,2}\,.
\end{equation}
In total this amounts to 14 free parameters describing the halo bias and stochasticity to be inferred in the baseline configuration, in addition to the cosmological parameters (in our case, only $f$).

\section{Posterior sampling methodology and convergence}
\label{sec:appendix-convergence}

\begin{figure}
	\includegraphics[]{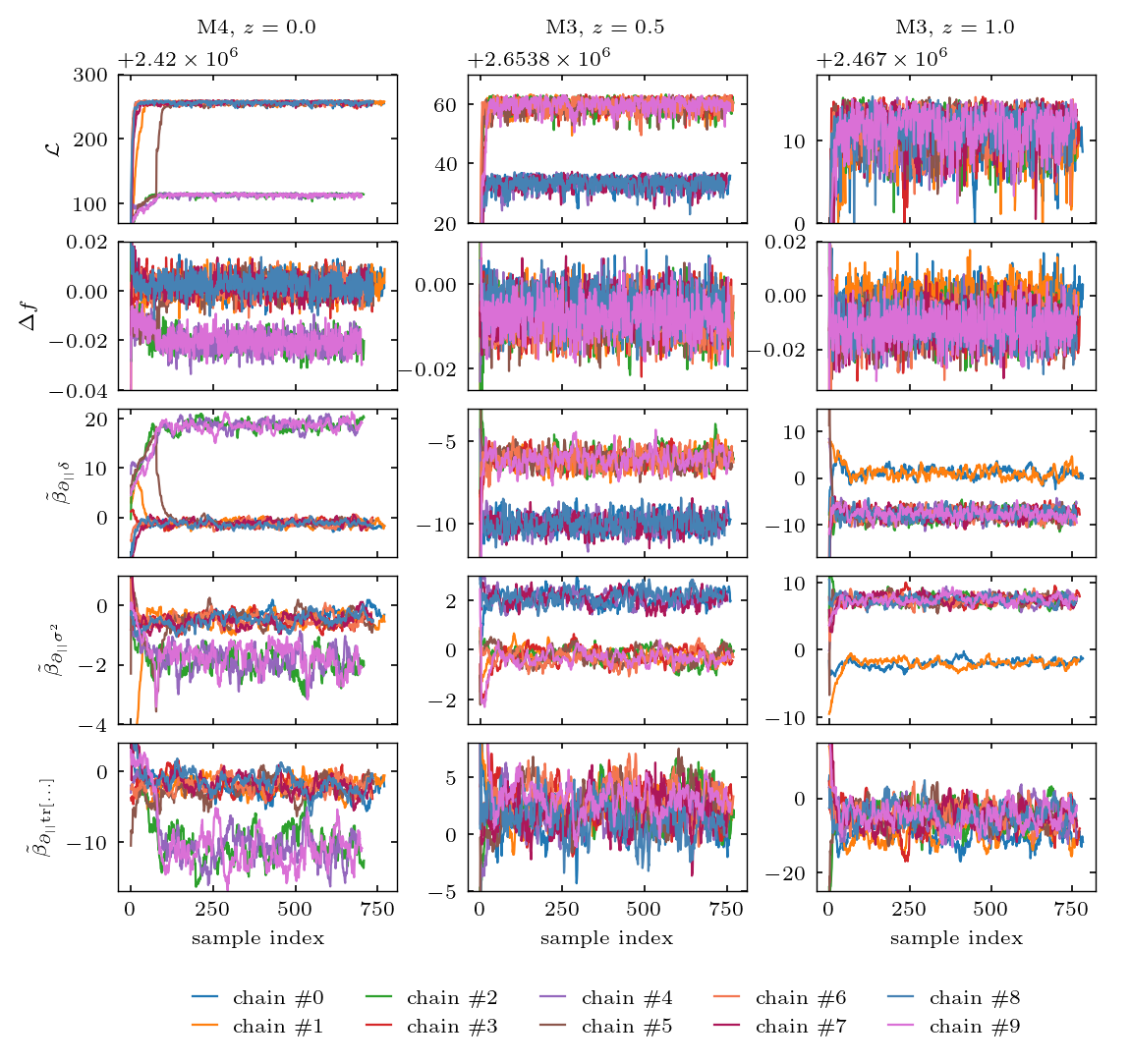}
	\caption{Trace plots over the log-likelihood (denoted as $\mathcal{L}$), growth rate and velocity bias coefficients for ten independent sampling chains per data set. All chains where initialized with different random seeds from randomly selected starting points. The analyses use the ``baseline'' configuration of the velocity bias expansion and consider a cutoff of $\Lambda=0.16\,h/\mpc$. In the labels, we have abbreviated $\lg M/h^{-1}M_\odot\in\left[13.5,14.0\right]$ as ``M3'' and $\lg M/h^{-1}M_\odot\in\left[13.5,14.0\right]$ as ``M4''. In case of the velocity bias coefficients, we plot $\tilde{\beta}_\uop = \beta_\uop/f$ and write $\tilde{\beta}_{\partial_\pp\mathrm{tr}[M^{(1)} M^{(1)}]}$ as $\tilde{\beta}_{\partial_\pp\mathrm{tr}[...]}$ for brevity. See figure \ref{fig:appendix-convergence-meanofcahins} for the mean and standard deviation of parameters inferred from the individual chains.}
	\label{fig:appendix-convergence-traces}
\end{figure}

We use slice sampling \cite{2000physics...9028N} to explore the joint posterior distribution of the cosmological parameters, bias coefficients and noise amplitudes. After inspecting the trace plots of each chain, we remove at least 100 samples for burn in and more if we recognize a significant drift beyond this threshold. This, however, only occurs in the case of mock analyses with the two ``velocity-induced bias'' scenarios (see section \ref{sec:mocktests-vbias}). For each chain, we obtain the effective sample size using \texttt{NumPyro} \cite{phan2019composable, bingham2019pyro}. We demand at least 100 effective samples in the growth rate and all further parameters of interest. The latter correspond to $b_\delta$ for all analyses of mock data sets and sub-sampled matter particles and $\beta_{\partial_\pp\delta}$ for those scenarios depicted in figures \ref{fig:matter-vbias} and \ref{fig:halos-velocity-bias}. Note that, since the bias coefficients typically exhibit longer correlation length, the effective sample size in $f$ significantly exceeds the threshold of 100 in most cases. Typical computation times for one analysis are on the order of a few days on a single computing node.

For the resolution tests in appendix \ref{sec:appendix-convergence}, and in particular for the high-resolution runs with $N_\eul=512$, we modify the criteria on burn in and sample size. Since these runs are computationally expensive, we want to discard as little samples as possible, and hence we reduce the burn-in cut to 50. Further, we demand an effective sample size of at least $50$ samples in the parameters of interest.

One-dimensional parameter constraints, as the ones depicted in figures \ref{fig: mock-operators}, \ref{fig: matter-overview} or \ref{fig: halos-summary}, are obtained as the mean and standard deviation over all samples after discarding the burn-in. Further, we use \texttt{GetDist} \cite{Lewis:2019xzd} to generate two-dimensional marginalized posterior slices as shown in figures \ref{fig:matter-degeneracies} and \ref{fig: halosM2-degeneracies}.

\newcolumntype{L}[1]{>{\raggedright\let\newline\\\arraybackslash\hspace{0pt}}m{#1}}
\begin{table}
	\centering
	\begin{tabular}{L{4.cm}|p{1.4cm}||p{1cm}|p{1cm}|p{1cm}|p{1cm}|p{1cm}}
		\multirow[b]{2}{*}{data set and analysis} &
		\multirow[b]{2}{*}{$N$} &
		\multicolumn{4}{c}{Gelman-Rubin criterion} \\
		& & $f$ & $b_\delta$ & $\tilde{\beta}_{\partial_\pp\delta}$ & $\tilde{\beta}_{\partial_\pp (\sigma^2)}$ & $\tilde{\beta}_{\partial_\pp\mathrm{tr}[\ldots]}$ \\
		\hline \hline
		mock 4 (crosses), $\Lambda{=}0.10\,h/\mpc$ & 
		$4.7\times10^3$ & $1.007$ & $1.009$ & $1.01$ & $1.07$ & $1.06$\\
		\hline
		mock 1 (circles), $\Lambda{=}0.16\,h/\mpc$ & 
		$4.6\times10^3$ & $1.007$ & $1.006$ & $1.04$ & $1.01$ & $1.02$\\
		\hline \hline
		matter, sim. 1, $z{=}0$, $\Lambda{=}0.12\,h/\mpc$ & 
		$4.6\times10^3$ & $1.002$ & $1.003$ & $1.01$ & $1.02$ & $1.02$\\
		\hline
		matter, sim. 1, $z{=}1$, $\Lambda{=}0.12\,h/\mpc$ & 
		$4.5\times10^3$ & $1.01$ & $1.01$ & $1.01$ & $1.007$ & $1.004$\\
		\hline \hline
		halos, sim. 1, M2, $z{=}0$, $\Lambda{=}0.12\,h/\mpc$ & 
		$3.2\times 10^3$ & $1.004$ & $1.001$ & $1.01$ & $1.02$ & $1.03$\\
		\hline
		halos, sim. 1, M4, $z{=}0$, $\Lambda{=}0.12\,h/\mpc$ & 
		$1.2\times10^3$ & $1.01$ & $1.07$ & $2.4$ & $2.9$ & $1.3$\\
		\hline
		halos, sim. 1, M4, \hfill$\bm{\ast}$\newline $z{=}0$, $\Lambda{=}0.16\,h/\mpc$ &
		$6.5\times10^3$ & $3.0$ & $1.4$ & $12.8$ & $2.7$ & $2.9$\\
		\hline
		halos, sim. 1, M3, $z{=}0.5$, $\Lambda{=}0.12\,h/\mpc$ &
		$3.4\times10^3$ & $1.002$ & $1.000$ & $1.007$ & $1.02$ & $1.01$\\
		\hline
		halos, sim. 1, M3,\hfill$\bm{\ast}$\newline $z{=}0.5$, $\Lambda{=}0.16\,h/\mpc$ &
		$7.6\times10^3$ & $1.04$ & $1.04$ & $4.2$ & $4.5$ & $1.9$ \\
		\hline
		halos, sim. 1, M4, $z{=}0.5$, $\Lambda{=}0.12\,h/\mpc$ &
		$2.7\times10^3$ & $1.2$ & $1.08$ & $1.2$ & $1.4$ & $1.8$\\
		\hline
		halos, sim. 1, M4, $z{=}0.5$, $\Lambda{=}0.16\,h/\mpc$&
		$7.1\times 10^3$ & $1.3$ & $1.07$ & $1.06$ & $2.8$ & $3.7$ \\
		\hline
		halos, sim. 1, M1, $z{=}1$, $\Lambda{=}0.14\,h/\mpc$ &
		$2.5\times 10^3$ & $1.001$ & $1.000$ & $1.009$ & $1.02$ & $1.02$ \\
		\hline
		halos, sim. 1, M3, $z{=}1$, $\Lambda{=}0.10\,h/\mpc$ &
		$2.4\times10^3$ & $1.03$ & $1.03$ & $1.2$ & $1.2$ & $1.02$ \\
		\hline
		halos, sim. 1, M3, \hfill$\bm{\ast}$\newline $z=1$, $\Lambda{=}0.16\,h/\mpc$&
		$7.8\times10^3$ & $1.2$ & $1.2$ & $4.2$ & $7.4$ & $1.3$\\
	\end{tabular}
	\caption{Gelman-Rubin diagnostics for a subset of scenarios. For each data set, we run 10 chains of the ``baseline'' analysis, and $N$ denotes the total number of samples included in the Gelman-Rubin test. In the leftmost column, we abbreviate the four halo mass bins listed in table \ref{tab: appendix-nbody-halosamples} as ``M1'' to ``M4'', where M1 is the sample with the lowest mass. In case of the velocity bias coefficients, we consider $\tilde{\beta}_\uop = \beta_\uop/f$ and write $\tilde{\beta}_{\partial_\pp\mathrm{tr}[M^{(1)} M^{(1)}]}$ as $\tilde{\beta}_{\partial_\pp\mathrm{tr}[...]}$ for brevity. The cases marked by an asterisk are further analyzed in figures \ref{fig:appendix-convergence-traces} and \ref{fig:appendix-convergence-meanofcahins}.}
	\label{tab:appendix-gelman-rubin}
\end{table}

\begin{figure}
	\includegraphics[]{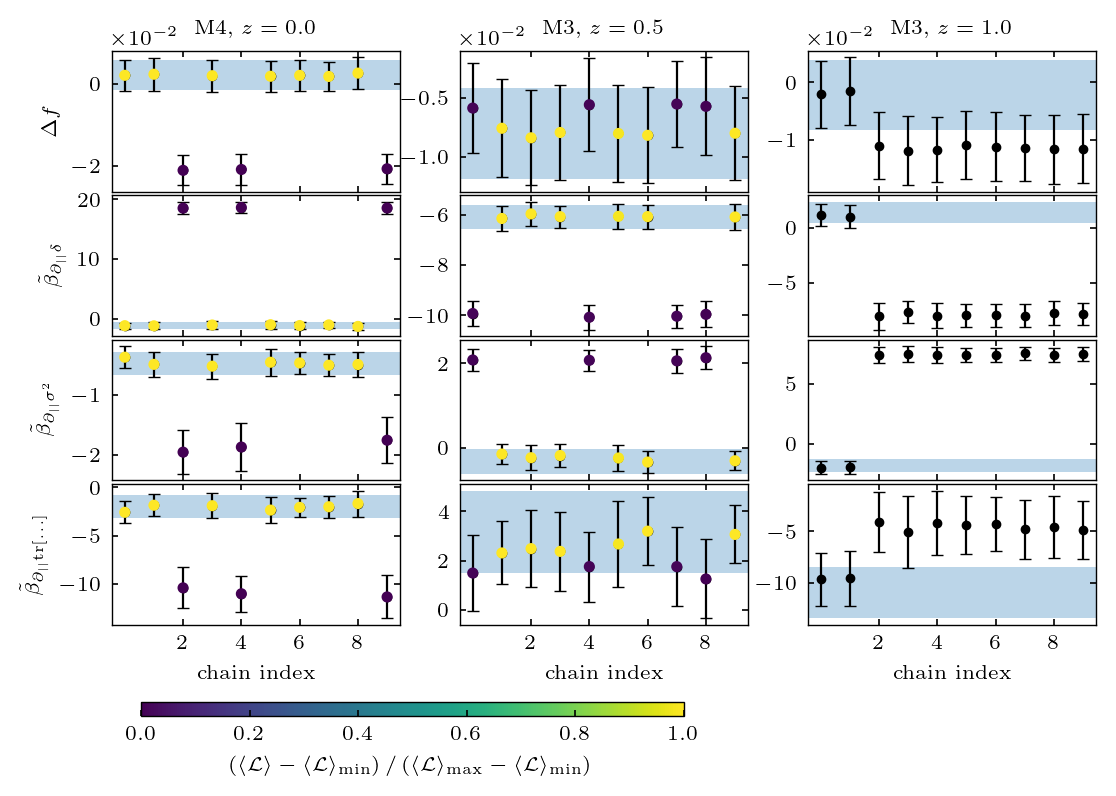}
	\caption{Parameters inferred from individual sampling chains, whose traces are shown in figure \ref{fig:appendix-convergence-traces}. The blue shaded region marks the respective estimate from one longer chain, which we also give as our main result in figure \ref{fig: halos-summary}. The color coding in the first two rows refers to the mean log-likelihood of the chain (denoted as $\langle\mathcal{L}\rangle$), in relation to the minimum and maximum of means encountered over the ten runs. In the third case, depicted in the right row, we forgo such color-coding since the variance within a single chain is comparable to the difference between chains.}
	\label{fig:appendix-convergence-meanofcahins}
\end{figure}

To verify that the typical set is adequately explored by our chains, we pick a sub-set of analyses for a Gelman-Rubin test \cite{Gelman:1992zz}. These are listed in the leftmost column of table \ref{tab:appendix-gelman-rubin}. The selection was motivated on the one hand to pick a range of representative analysis choices and one the other hand to check conspicuous features, such as the outlier data set in figure \ref{fig: mock-operators}, chains with particularly long correlation lengths or transient-like features in the trace plots of individual parameters. For each scenario, we run ten additional, independent chains from randomly selected starting values. We discard the first 50 steps for burn in, and more if we note significant drifts beyond this threshold. Then, we prune all chains to the same length and use \texttt{NumPyro} to evaluate the Gelman-Rubin criterion, which we list in table \ref{tab:appendix-gelman-rubin}. 

Most scenarios that we analyze indeed show satisfactory convergence. The notable exception are massive halo samples at higher redshifts. These are also the data sets where we expect the highest noise and the most complex bias relation. We have picked three representative examples from the scenarios with a high Gelman-Rubin criterion, and for them we show in figure \ref{fig:appendix-convergence-traces} trace plots over all ten chains from the Gelman-Rubin test. Further, we compare in figure \ref{fig:appendix-convergence-meanofcahins} the parameter estimates from individual chains amongst each other. Apparently, the poor convergence statistics originates from different chains exploring different local maxima of the full posterior distribution.

In case of the third mass bin, $\lg M/h^{-1}M_\odot \in \left[13.5, 14.0\right]$, the local maxima mainly differ in the velocity bias coefficients, while the growth rate estimate is largely consistent between the individual chains. Most cases where we found signs of multiple maxima in the third mass bin are qualitatively similar to the middle panel of figures \ref{fig:appendix-convergence-traces} and \ref{fig:appendix-convergence-meanofcahins} in that there is a clear hierarchy between the likelihood in the different posterior modes, allowing to identify a true (global) maximum. In contrast, there is no such hierarchy apparent in the right column of figure \ref{fig:appendix-convergence-traces}. The main criterion in section \ref{sec:results}, where we present the results of the halo analyses, is the recovery of the true growth rate. The presence of multiple maxima in the third mass bin seems not to shift the mean value of $f$ appreciably, but as figure \ref{fig:appendix-convergence-meanofcahins} indicates, it might likely lead to a slight under-estimation of the error bars. With respect to the estimate of velocity bias coefficients, we focus the discussion on only those scenarios where the sampling appears well converged.

In contrast to that, we encounter some cases with a considerable shift of the growth rate between local posterior maxima in the highest mass bin, $\lg M/h^{-1}M_\odot \in \left[14.0, 14.5\right]$. Importantly, here we always find a clear hierarchy between the posterior modes, and the global maximum coincides with $f$ being much better aligned with the expected ground truth. Thus, the EFT-likelihood in combination with our perturbative forward model indeed is able to identify the correct cosmological parameters even in those high-noise cases. Nevertheless, we indicate all scenarios where the estimated mean value of $f$ might be affected by multiple maxima in the posterior distribution, and related to that by the convergence of sampling chains, by open symbols in figure \ref{fig: halos-summary}.

\section{Resolution of the displacement steps in the forward model}
\label{sec: appendix-ngeul}

\begin{figure}
\centering
\includegraphics[]{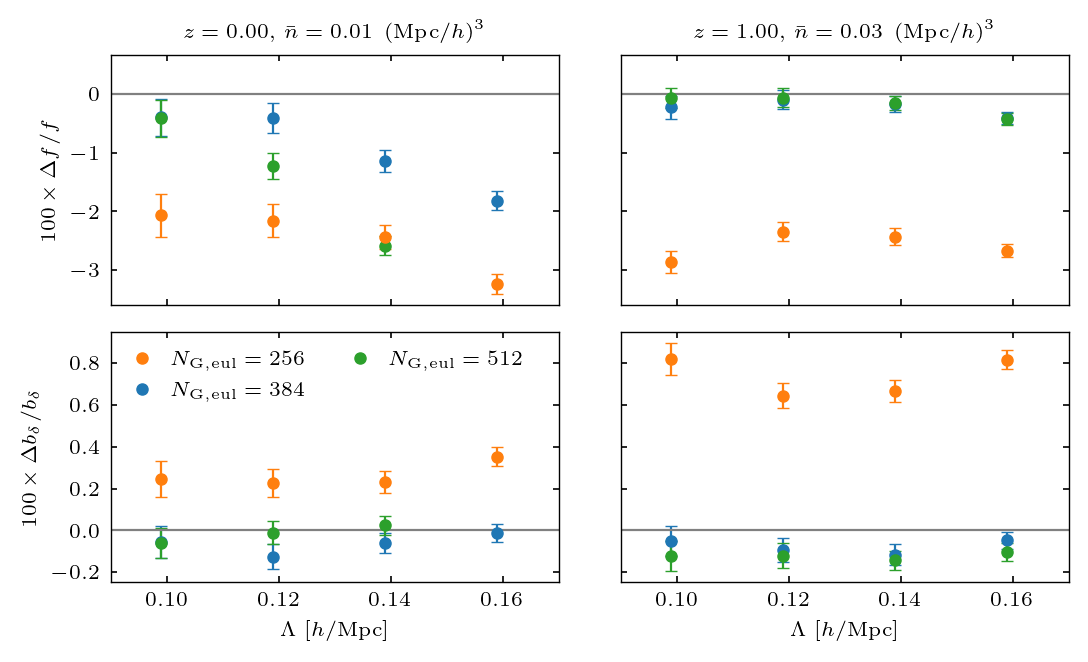}
\caption{Impact of the resolution at which the displacement step is computed on the ``baseline'' analysis of matter particles in N-body simulations. The matter particles from the simulation are sub-sampled to obtain an unbiased, noisy tracer and their density is chosen to give roughly constant signal-to-noise ratio at all redshifts; here we specifically consider simulation 1 at $z=0\,,~1$. The indicated value of $N_\eul$ was used in the forward model of the analysis and for generating the data cube. Analyses with $N_\eul=384$ are identical to those of section \ref{sec:matter-inference}. For $N_\eul=512$ we only consider a sub-set of cutoff values due to the increased numerical cost of these runs.}
\label{fig: appendix-ngeul-matter}
\end{figure}

The data considered in the EFT-likelihood is a three dimensional cube, containing the tracer overdensity. It is constructed from the catalog of tracer positions in a preparatory step where the density is assigned to the grid with a CIC kernel. The same kernel is also used in the forward model, when we compute the displacement of pseudo-particles from Lagrangian to Eulerian and from Eulerian to redshift space. However, as discussed in section~\ref{sec: matter-pk}, the cancellation of the kernel between model and data is not exact. We therefore test here explicitly how the resolution of the displacement affects the inference of the growth rate from sub-sampled matter particles and from halos in N-body simulations. To this end, we lower $N_\eul$ to $256$ and increase it to $512$. In all these cases, we assume the same value for $N_\eul$ in generating the data cube and in the analysis. Note that the mock data sets of section \ref{sec:mocktests} are generated from the same forward model as used in the analysis, and correspondingly we could not detect any resolution effects there.

The results for matter particles are summarized in figure \ref{fig: appendix-ngeul-matter} for simulation 1 at two redshifts. Evidently, a too low resolution of the displacement step leads to a significant under-estimation of the growth rate and a significant over-estimation of $b_\delta$. The results for the two larger values of $N_\eul$ at $z=0$ agree well initially but depart for higher cutoffs. In this regime, we believe that FoG affect the analysis significantly, and apparently their effect is further enhanced by the refined resolution. In agreement with this interpretation, there is no significant difference between $N_\eul=384$ and $N_\eul=512$ at higher redshift, where FoG are more suppressed.

\begin{figure}
	\centering
	\includegraphics[]{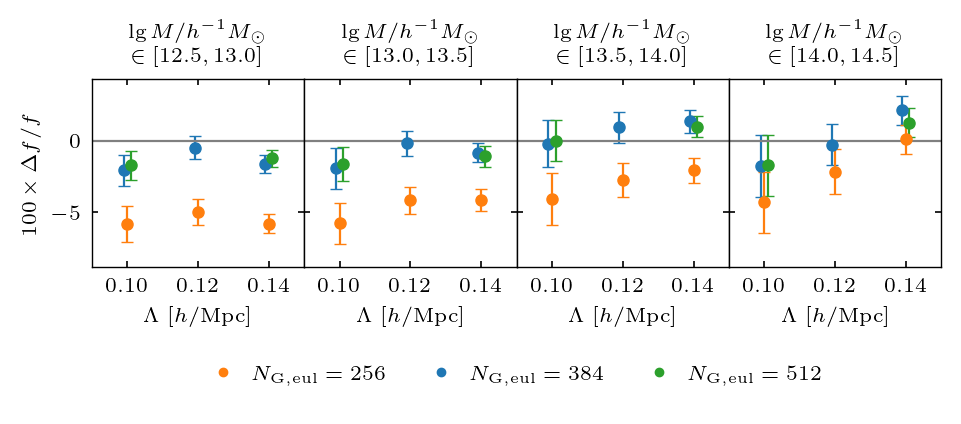}
	\caption{Impact of the resolution at which the displacement step is computed on the ``baseline'' analysis of halos from simulation 1 at $z=0$. The indicated value of $N_\eul$ was used in the forward model of the analysis and for generating the data cube. Analyses with $N_\eul=384$ are identical to those presented in section \ref{sec:results}. Due to the increased numerical cost of these analyses, we only consider a subset of cutoffs for $N_\eul=512$, namely $\Lambda=0.10\,,~0.14\,h/\mpc$.}
	\label{fig: appendix-ngeul-halos}
\end{figure}

In the analysis of resolution effects with halos, we focus on the $z=0$ snapshot, and show in figure \ref{fig: appendix-ngeul-halos} results for all four halo mass bins. As it was the case for matter, too low resolution leads to a significant bias of $f$ towards lower values. The results at $N_\eul=384$ and $N_\eul=512$, on the other hand, are very consistent. 

To conclude, we do not find strong resolution effects at $N_\eul=384$, apart for the case of matter at low redshifts which is strongly affected by FoG.

\section{N-body accuracy}
\label{sec: appendix-transients}

\begin{figure}
	\centering
	\includegraphics[]{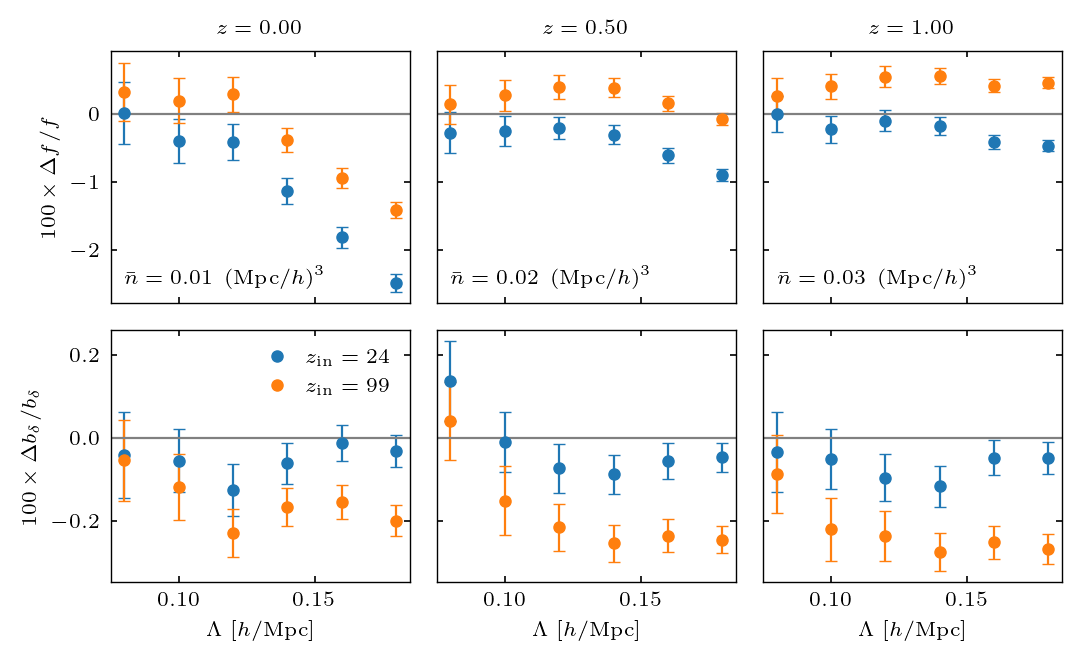}
	\caption{The effect of the starting redshift in the N-body simulations, $z_\mathrm{in}$, on the growth rate inferred of sub-sampled matter particles. We focus on the results from simulation 1 and compare two different values for $z_\mathrm{in}$. Results with $z_\mathrm{in}=24$ are identical to those presented in section \ref{sec:matter-inference}.}
	\label{fig: appendix-transients-matter}
\end{figure}

\begin{figure}
	\centering
	\includegraphics[]{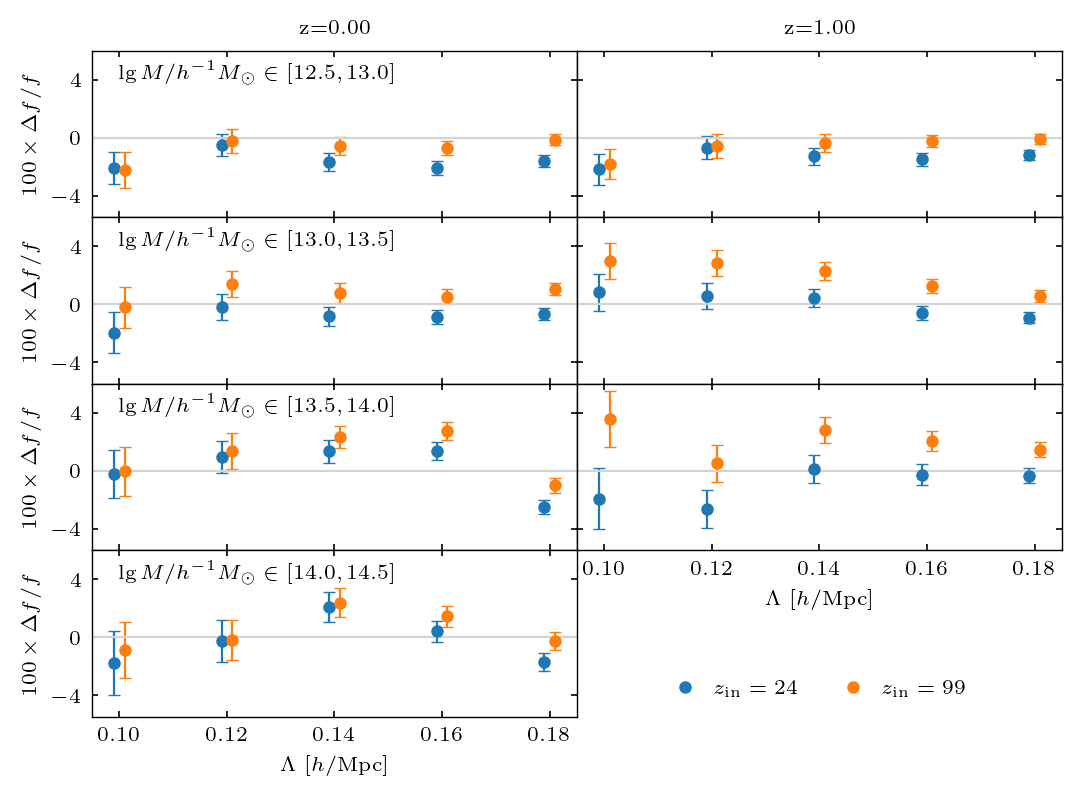}
	\caption{The effect of the starting redshift in the N-body simulations, $z_\mathrm{in}$, on the growth rate inferred from halos. We focus on the results from simulation 1 at $z=0,\,1$ and compare two different choices $z_\mathrm{in}$. Results with $z_\mathrm{in}=24$ are identical to those presented in section \ref{sec:results}.}
	\label{fig: appendix-transients-halos}
\end{figure}

The growth rate inference from matter particles and from halos in sections \ref{sec:matter-inference} and \ref{sec:results} can also be affected by the accuracy of the underlying N-body simulations. Given the generality of the bias expansion used in the forward model, we expect that any local effect or small scale noise can be captured and should not bias the inferred growth rate. In the context of previous studies inferring $\sigma_8$ in the halo rest frame, the time-stepping in the simulations and the halo finder were found to have negligible impact. But the choice of the starting time of the N-body simulations, $z_\mathrm{in}$, led to percent-level shifts of the inferred value in some cases \cite{Schmidt:2020tao}. Therefore, we here repeat the test of different values for $z_\mathrm{in}$ for the growth rate inference in redshift space.

The N-body simulations used in this work are initialized at the starting time $z_\mathrm{in}=24$ from second-order LPT particle positions and velocities. However, the choice of $z_\mathrm{in}$ is a subtle issue. Late starting times lead to transients from initial conditions, caused by the truncation of the early-time evolution at some finite LPT order. Early starting times enhance discreteness effects, i.e. deviations from the fluid characteristics in the continuum limit \cite{Michaux:2020yis}.

To assess what impact the choice of $z_\mathrm{in}$ has on the growth rate inference from matter, we compare our baseline results for simulation 1, to an otherwise identical simulation with earlier starting time, $z_\mathrm{in}=99$ in figure \ref{fig: appendix-transients-matter}. As expected, the starting time has the strongest impact on results inferred at high redshifts. In all cases, the earlier starting time shifts $f$ to higher and $b_\delta$ to lower values, away from the ground truth. The effect is at the sub-percent level, but given the high precision of the inferred growth rate, it can still lead to a significant mismatch.

The analogous results for halos at $z=0$ and $z=1$ are shown in figure \ref{fig: appendix-transients-halos} and show the same trend: the earlier starting time leads to an upwards shift in the inferred value of $f$ that is stronger for higher redshifts, but differs between the individual mass bins. Overall, we find that the parameters inferred from simulations with $z_\mathrm{in}=24$ are closer to the ground truth, in agreement with recommendations in \cite{Michaux:2020yis}. A more systematic investigation of the impact of transients on the inferred growth rate would be very interesting but is deferred to future work.

\bibliographystyle{JHEP} 
\bibliography{rsd-field-based-inference}

\providecommand{\href}[2]{#2}\begingroup\raggedright\begin{thebibliography}{10}

\bibitem{BOSS:2016wmc}
{\scshape BOSS} collaboration, \emph{{The clustering of galaxies in the
  completed SDSS-III Baryon Oscillation Spectroscopic Survey: cosmological
  analysis of the DR12 galaxy sample}},
  \href{https://doi.org/10.1093/mnras/stx721}{\emph{Mon. Not. Roy. Astron.
  Soc.} {\bfseries 470} (2017) 2617}
  [\href{https://arxiv.org/abs/1607.03155}{{\ttfamily 1607.03155}}].

\bibitem{eBOSS:2020yzd}
{\scshape eBOSS} collaboration, \emph{{Completed SDSS-IV extended Baryon
  Oscillation Spectroscopic Survey: Cosmological implications from two decades
  of spectroscopic surveys at the Apache Point Observatory}},
  \href{https://doi.org/10.1103/PhysRevD.103.083533}{\emph{Phys. Rev. D}
  {\bfseries 103} (2021) 083533}
  [\href{https://arxiv.org/abs/2007.08991}{{\ttfamily 2007.08991}}].

\bibitem{DESI:2016fyo}
{\scshape DESI} collaboration, \emph{{The DESI Experiment Part I:
  Science,Targeting, and Survey Design}},
  \href{https://arxiv.org/abs/1611.00036}{{\ttfamily 1611.00036}}.

\bibitem{Amendola:2016saw}
L.~Amendola et~al., \emph{{Cosmology and fundamental physics with the Euclid
  satellite}}, \href{https://doi.org/10.1007/s41114-017-0010-3}{\emph{Living
  Rev. Rel.} {\bfseries 21} (2018) 2}
  [\href{https://arxiv.org/abs/1606.00180}{{\ttfamily 1606.00180}}].

\bibitem{2014PASJ...66R...1T}
M.~{Takada}, R.S.~{Ellis}, M.~{Chiba}, J.E.~{Greene}, H.~{Aihara}, N.~{Arimoto}
  et~al., \emph{{Extragalactic science, cosmology, and Galactic archaeology
  with the Subaru Prime Focus Spectrograph}},
  \href{https://doi.org/10.1093/pasj/pst019}{\emph{Publications of the
  Astronomical Society of Japan} {\bfseries 66} (2014) R1}
  [\href{https://arxiv.org/abs/1206.0737}{{\ttfamily 1206.0737}}].

\bibitem{Dore:2014cca}
O.~Dor\'e et~al., \emph{{Cosmology with the SPHEREX All-Sky Spectral Survey}},
  \href{https://arxiv.org/abs/1412.4872}{{\ttfamily 1412.4872}}.

\bibitem{Desjaques_2018}
V.~Desjacques, D.~Jeong and F.~Schmidt, \emph{Large-scale galaxy bias},
  \href{https://doi.org/https://doi.org/10.1016/j.physrep.2017.12.002}{\emph{Physics
  Reports} {\bfseries 733} (2018) 1}.

\bibitem{Baumann:2010tm}
D.~Baumann, A.~Nicolis, L.~Senatore and M.~Zaldarriaga, \emph{{Cosmological
  Non-Linearities as an Effective Fluid}},
  \href{https://doi.org/10.1088/1475-7516/2012/07/051}{\emph{JCAP} {\bfseries
  07} (2012) 051} [\href{https://arxiv.org/abs/1004.2488}{{\ttfamily
  1004.2488}}].

\bibitem{Carrasco:2012cv}
J.J.M.~Carrasco, M.P.~Hertzberg and L.~Senatore, \emph{{The Effective Field
  Theory of Cosmological Large Scale Structures}},
  \href{https://doi.org/10.1007/JHEP09(2012)082}{\emph{JHEP} {\bfseries 09}
  (2012) 082} [\href{https://arxiv.org/abs/1206.2926}{{\ttfamily 1206.2926}}].

\bibitem{Planck:2018nkj}
{\scshape Planck} collaboration, \emph{{Planck 2018 results. I. Overview and
  the cosmological legacy of Planck}},
  \href{https://doi.org/10.1051/0004-6361/201833880}{\emph{Astron. Astrophys.}
  {\bfseries 641} (2020) A1}
  [\href{https://arxiv.org/abs/1807.06205}{{\ttfamily 1807.06205}}].

\bibitem{Schmidt:2018bkr}
F.~Schmidt, F.~Elsner, J.~Jasche, N.M.~Nguyen and G.~Lavaux, \emph{{A rigorous
  EFT-based forward model for large-scale structure}},
  \href{https://doi.org/10.1088/1475-7516/2019/01/042}{\emph{JCAP} {\bfseries
  01} (2019) 042} [\href{https://arxiv.org/abs/1808.02002}{{\ttfamily
  1808.02002}}].

\bibitem{Elsner:2019rql}
F.~Elsner, F.~Schmidt, J.~Jasche, G.~Lavaux and N.-M.~Nguyen, \emph{{Cosmology
  inference from a biased density field using the EFT-based likelihood}},
  \href{https://doi.org/10.1088/1475-7516/2020/01/029}{\emph{JCAP} {\bfseries
  01} (2020) 029} [\href{https://arxiv.org/abs/1906.07143}{{\ttfamily
  1906.07143}}].

\bibitem{Cabass:2019lqx}
G.~Cabass and F.~Schmidt, \emph{{The EFT Likelihood for Large-Scale
  Structure}}, \href{https://doi.org/10.1088/1475-7516/2020/04/042}{\emph{JCAP}
  {\bfseries 04} (2020) 042}
  [\href{https://arxiv.org/abs/1909.04022}{{\ttfamily 1909.04022}}].

\bibitem{2011hmcm.book..113N}
R.~{Neal}, \emph{{MCMC Using Hamiltonian Dynamics}},  in \emph{Handbook of
  Markov Chain Monte Carlo}, pp.~113--162 (2011),
  \href{https://doi.org/10.1201/b10905}{DOI}.

\bibitem{2013MNRAS.432..894J}
J.~{Jasche} and B.D.~{Wandelt}, \emph{{Bayesian physical reconstruction of
  initial conditions from large-scale structure surveys}},
  \href{https://doi.org/10.1093/mnras/stt449}{\emph{Mon. Not. Roy. Astron.
  Soc.} {\bfseries 432} (2013) 894}
  [\href{https://arxiv.org/abs/1203.3639}{{\ttfamily 1203.3639}}].

\bibitem{2013MNRAS.429L..84K}
F.S.~{Kitaura}, \emph{{The initial conditions of the universe from constrained
  simulations.}}, \href{https://doi.org/10.1093/mnrasl/sls029}{\emph{Mon. Not.
  Roy. Astron. Soc.} {\bfseries 429} (2013) L84}
  [\href{https://arxiv.org/abs/1203.4184}{{\ttfamily 1203.4184}}].

\bibitem{2013ApJ...772...63W}
H.~{Wang}, H.J.~{Mo}, X.~{Yang} and F.C.~{van den Bosch}, \emph{{Reconstructing
  the Initial Density Field of the Local Universe: Methods and Tests with Mock
  Catalogs}},
  \href{https://doi.org/10.1088/0004-637X/772/1/63}{\emph{Astrophys. J.}
  {\bfseries 772} (2013) 63} [\href{https://arxiv.org/abs/1301.1348}{{\ttfamily
  1301.1348}}].

\bibitem{Wang:2014hia}
H.~Wang, H.J.~Mo, X.~Yang, Y.P.~Jing and W.P.~Lin, \emph{{ELUCID - Exploring
  the Local Universe with reConstructed Initial Density field I: Hamiltonian
  Markov Chain Monte Carlo Method with Particle Mesh Dynamics}},
  \href{https://doi.org/10.1088/0004-637X/794/1/94}{\emph{Astrophys. J.}
  {\bfseries 794} (2014) 94} [\href{https://arxiv.org/abs/1407.3451}{{\ttfamily
  1407.3451}}].

\bibitem{Modi:2018cfi}
C.~Modi, Y.~Feng and U.~Seljak, \emph{{Cosmological Reconstruction From Galaxy
  Light: Neural Network Based Light-Matter Connection}},
  \href{https://doi.org/10.1088/1475-7516/2018/10/028}{\emph{JCAP} {\bfseries
  10} (2018) 028} [\href{https://arxiv.org/abs/1805.02247}{{\ttfamily
  1805.02247}}].

\bibitem{Shallue:2022mhf}
C.J.~Shallue and D.J.~Eisenstein, \emph{{Reconstructing cosmological initial
  conditions from late-time structure with convolutional neural networks}},
  \href{https://doi.org/10.1093/mnras/stad528}{\emph{Mon. Not. Roy. Astron.
  Soc.} {\bfseries 520} (2023) 6256}
  [\href{https://arxiv.org/abs/2207.12511}{{\ttfamily 2207.12511}}].

\bibitem{Modi:2022pzm}
C.~Modi, Y.~Li and D.~Blei, \emph{{Reconstructing the Universe with Variational
  self-Boosted Sampling}},  in \emph{{39th International Conference on Machine
  Learning Conference}}, 6, 2022
  [\href{https://arxiv.org/abs/2206.15433}{{\ttfamily 2206.15433}}].

\bibitem{Dai:2022dso}
B.~Dai and U.~Seljak, \emph{{Translation and rotation equivariant normalizing
  flow (TRENF) for optimal cosmological analysis}},
  \href{https://doi.org/10.1093/mnras/stac2010}{\emph{Mon. Not. Roy. Astron.
  Soc.} {\bfseries 516} (2022) 2363}
  [\href{https://arxiv.org/abs/2202.05282}{{\ttfamily 2202.05282}}].

\bibitem{Qin:2023dew}
F.~Qin, D.~Parkinson, S.E.~Hong and C.G.~Sabiu, \emph{{Reconstructing the
  cosmological density and velocity fields from redshifted galaxy distributions
  using V-net}},  \href{https://arxiv.org/abs/2302.02087}{{\ttfamily
  2302.02087}}.

\bibitem{Jindal:2023qew}
V.~Jindal, D.~Jamieson, A.~Liang, A.~Singh and S.~Ho, \emph{{Predicting the
  Initial Conditions of the Universe using Deep Learning}},
  \href{https://arxiv.org/abs/2303.13056}{{\ttfamily 2303.13056}}.

\bibitem{Charnock:2019rbk}
T.~Charnock, G.~Lavaux, B.D.~Wandelt, S.~Sarma~Boruah, J.~Jasche and
  M.J.~Hudson, \emph{{Neural physical engines for inferring the halo mass
  distribution function}},
  \href{https://doi.org/10.1093/mnras/staa682}{\emph{Mon. Not. Roy. Astron.
  Soc.} {\bfseries 494} (2020) 50}
  [\href{https://arxiv.org/abs/1909.06379}{{\ttfamily 1909.06379}}].

\bibitem{Jasche:2018oym}
J.~Jasche and G.~Lavaux, \emph{{Physical Bayesian modelling of the non-linear
  matter distribution: new insights into the Nearby Universe}},
  \href{https://doi.org/10.1051/0004-6361/201833710}{\emph{Astron. Astrophys.}
  {\bfseries 625} (2019) A64}
  [\href{https://arxiv.org/abs/1806.11117}{{\ttfamily 1806.11117}}].

\bibitem{Lavaux:2019fjr}
G.~Lavaux, J.~Jasche and F.~Leclercq, \emph{{Systematic-free inference of the
  cosmic matter density field from SDSS3-BOSS data}},
  \href{https://arxiv.org/abs/1909.06396}{{\ttfamily 1909.06396}}.

\bibitem{Neyrinck:2013ezr}
M.C.~Neyrinck, M.A.~Aragon-Calvo, D.~Jeong and X.~Wang, \emph{{A halo bias
  function measured deeply into voids without stochasticity}},
  \href{https://doi.org/10.1093/mnras/stu589}{\emph{Mon. Not. Roy. Astron.
  Soc.} {\bfseries 441} (2014) 646}
  [\href{https://arxiv.org/abs/1309.6641}{{\ttfamily 1309.6641}}].

\bibitem{Schmittfull:2018yuk}
M.~Schmittfull, M.~Simonovi\'c, V.~Assassi and M.~Zaldarriaga, \emph{{Modeling
  Biased Tracers at the Field Level}},
  \href{https://doi.org/10.1103/PhysRevD.100.043514}{\emph{Phys. Rev. D}
  {\bfseries 100} (2019) 043514}
  [\href{https://arxiv.org/abs/1811.10640}{{\ttfamily 1811.10640}}].

\bibitem{Schmittfull:2020trd}
M.~Schmittfull, M.~Simonovi\'c, M.M.~Ivanov, O.H.E.~Philcox and M.~Zaldarriaga,
  \emph{{Modeling Galaxies in Redshift Space at the Field Level}},
  \href{https://doi.org/10.1088/1475-7516/2021/05/059}{\emph{JCAP} {\bfseries
  05} (2021) 059} [\href{https://arxiv.org/abs/2012.03334}{{\ttfamily
  2012.03334}}].

\bibitem{Kostic:2022vok}
A.~Kosti\'c, N.-M.~Nguyen, F.~Schmidt and M.~Reinecke, \emph{{Consistency tests
  of field level inference with the EFT likelihood}},
  \href{https://arxiv.org/abs/2212.07875}{{\ttfamily 2212.07875}}.

\bibitem{Schmidt:2020viy}
F.~Schmidt, G.~Cabass, J.~Jasche and G.~Lavaux, \emph{{Unbiased Cosmology
  Inference from Biased Tracers using the EFT Likelihood}},
  \href{https://doi.org/10.1088/1475-7516/2020/11/008}{\emph{JCAP} {\bfseries
  11} (2020) 008} [\href{https://arxiv.org/abs/2004.06707}{{\ttfamily
  2004.06707}}].

\bibitem{Schmidt:2020tao}
F.~Schmidt, \emph{{Sigma-Eight at the Percent Level: The EFT Likelihood in Real
  Space}}, \href{https://doi.org/10.1088/1475-7516/2021/04/032}{\emph{JCAP}
  {\bfseries 04} (2021) 032}
  [\href{https://arxiv.org/abs/2009.14176}{{\ttfamily 2009.14176}}].

\bibitem{Schmidt_2021}
F.~Schmidt, \emph{An n-th order lagrangian forward model for large-scale
  structure},
  \href{https://doi.org/10.1088/1475-7516/2021/04/033}{\emph{Journal of
  Cosmology and Astroparticle Physics} {\bfseries 2021} (2021) 033}.

\bibitem{Babic:2022dws}
I.~Babi\'c, F.~Schmidt and B.~Tucci, \emph{{BAO scale inference from biased
  tracers using the EFT likelihood}},
  \href{https://doi.org/10.1088/1475-7516/2022/08/007}{\emph{JCAP} {\bfseries
  08} (2022) 007} [\href{https://arxiv.org/abs/2203.06177}{{\ttfamily
  2203.06177}}].

\bibitem{1987MNRAS.227....1K}
N.~{Kaiser}, \emph{{Clustering in real space and in redshift space}},
  \href{https://doi.org/10.1093/mnras/227.1.1}{\emph{Mon. Not. Roy. Astron.
  Soc.} {\bfseries 227} (1987) 1}.

\bibitem{1992ApJ...385L...5H}
A.J.S.~{Hamilton}, \emph{{Measuring Omega and the Real Correlation Function
  from the Redshift Correlation Function}},
  \href{https://doi.org/10.1086/186264}{\emph{Astrophys. J. Lett.} {\bfseries
  385} (1992) L5}.

\bibitem{Cabass:2020jqo}
G.~Cabass, \emph{{The EFT Likelihood for Large-Scale Structure in Redshift
  Space}}, \href{https://doi.org/10.1088/1475-7516/2021/01/067}{\emph{JCAP}
  {\bfseries 01} (2021) 067}
  [\href{https://arxiv.org/abs/2007.14988}{{\ttfamily 2007.14988}}].

\bibitem{Baldauf:2015tla}
T.~Baldauf, E.~Schaan and M.~Zaldarriaga, \emph{{On the reach of perturbative
  descriptions for dark matter displacement fields}},
  \href{https://doi.org/10.1088/1475-7516/2016/03/017}{\emph{JCAP} {\bfseries
  03} (2016) 017} [\href{https://arxiv.org/abs/1505.07098}{{\ttfamily
  1505.07098}}].

\bibitem{Taruya:2018jtk}
A.~Taruya, T.~Nishimichi and D.~Jeong, \emph{{Grid-based calculation for
  perturbation theory of large-scale structure}},
  \href{https://doi.org/10.1103/PhysRevD.98.103532}{\emph{Phys. Rev. D}
  {\bfseries 98} (2018) 103532}
  [\href{https://arxiv.org/abs/1807.04215}{{\ttfamily 1807.04215}}].

\bibitem{2012JCAP...04..013T}
S.~{Tassev} and M.~{Zaldarriaga}, \emph{{The mildly non-linear regime of
  structure formation}},
  \href{https://doi.org/10.1088/1475-7516/2012/04/013}{\emph{JCAP} {\bfseries
  2012} (2012) 013} [\href{https://arxiv.org/abs/1109.4939}{{\ttfamily
  1109.4939}}].

\bibitem{Tassev:2013rta}
S.~Tassev, \emph{{Lagrangian or Eulerian; Real or Fourier? Not All Approaches
  to Large-Scale Structure Are Created Equal}},
  \href{https://doi.org/10.1088/1475-7516/2014/06/008}{\emph{JCAP} {\bfseries
  06} (2014) 008} [\href{https://arxiv.org/abs/1311.4884}{{\ttfamily
  1311.4884}}].

\bibitem{Buchert:1992ya}
T.~Buchert, \emph{{Lagrangian theory of gravitational instability of
  Friedman-Lemaitre cosmologies and the 'Zel'dovich approximation'}},
  {\emph{Mon. Not. Roy. Astron. Soc.} {\bfseries 254} (1992) 729}.

\bibitem{Rampf:2012up}
C.~Rampf, \emph{{The recursion relation in Lagrangian perturbation theory}},
  \href{https://doi.org/10.1088/1475-7516/2012/12/004}{\emph{JCAP} {\bfseries
  12} (2012) 004} [\href{https://arxiv.org/abs/1205.5274}{{\ttfamily
  1205.5274}}].

\bibitem{Zheligovsky:2013eca}
V.~Zheligovsky and U.~Frisch, \emph{{Time-analyticity of Lagrangian particle
  trajectories in ideal fluid flow}},
  \href{https://doi.org/10.1017/jfm.2014.221}{\emph{J. Fluid Mech.} {\bfseries
  749} (2014) 404} [\href{https://arxiv.org/abs/1312.6320}{{\ttfamily
  1312.6320}}].

\bibitem{Matsubara:2015ipa}
T.~Matsubara, \emph{{Recursive Solutions of Lagrangian Perturbation Theory}},
  \href{https://doi.org/10.1103/PhysRevD.92.023534}{\emph{Phys. Rev. D}
  {\bfseries 92} (2015) 023534}
  [\href{https://arxiv.org/abs/1505.01481}{{\ttfamily 1505.01481}}].

\bibitem{Ehlers:1996wg}
J.~Ehlers and T.~Buchert, \emph{{Newtonian cosmology in Lagrangian formulation:
  Foundations and perturbation theory}},
  \href{https://doi.org/10.1023/A:1018885922682}{\emph{Gen. Rel. Grav.}
  {\bfseries 29} (1997) 733}
  [\href{https://arxiv.org/abs/astro-ph/9609036}{{\ttfamily
  astro-ph/9609036}}].

\bibitem{1970A&A.....5...84Z}
Y.B.~{Zel'dovich}, \emph{{Gravitational instability: An approximate theory for
  large density perturbations.}}, {\emph{Astron. Astrophys.} {\bfseries 5}
  (1970) 84}.

\bibitem{White:2014gfa}
M.~White, \emph{{The Zel'dovich approximation}},
  \href{https://doi.org/10.1093/mnras/stu209}{\emph{Mon. Not. Roy. Astron.
  Soc.} {\bfseries 439} (2014) 3630}
  [\href{https://arxiv.org/abs/1401.5466}{{\ttfamily 1401.5466}}].

\bibitem{Hockney_1988}
R.W.~{Hockney} and J.W.~{Eastwood}, \emph{{Computer simulation using
  particles}}, Bristol: Hilger, 1988 (1988).

\bibitem{Mirbabayi:2014zca}
M.~Mirbabayi, F.~Schmidt and M.~Zaldarriaga, \emph{{Biased Tracers and Time
  Evolution}}, \href{https://doi.org/10.1088/1475-7516/2015/07/030}{\emph{JCAP}
  {\bfseries 07} (2015) 030} [\href{https://arxiv.org/abs/1412.5169}{{\ttfamily
  1412.5169}}].

\bibitem{1986ApJ...304...15B}
J.M.~{Bardeen}, J.R.~{Bond}, N.~{Kaiser} and A.S.~{Szalay}, \emph{{The
  Statistics of Peaks of Gaussian Random Fields}},
  \href{https://doi.org/10.1086/164143}{\emph{Astrophys. J.} {\bfseries 304}
  (1986) 15}.

\bibitem{peacock/lumsden/heavens:1987}
J.A.~{Peacock}, S.L.~{Lumsden} and A.F.~{Heavens}, \emph{{Cosmological
  streaming velocities and large-scale density maxima}}, {\emph{\mnras}
  {\bfseries 229} (1987) 469}.

\bibitem{percival/schaefer:2008}
W.J.~{Percival} and B.M.~{Sch{\"a}fer}, \emph{{Galaxy peculiar velocities and
  evolution-bias}},
  \href{https://doi.org/10.1111/j.1745-3933.2008.00437.x}{\emph{\mnras}
  {\bfseries 385} (2008) L78}
  [\href{https://arxiv.org/abs/0712.2729}{{\ttfamily 0712.2729}}].

\bibitem{Desjacques:2008jj}
V.~Desjacques, \emph{{Baryon acoustic signature in the clustering of density
  maxima}}, \href{https://doi.org/10.1103/PhysRevD.78.103503}{\emph{Phys. Rev.
  D} {\bfseries 78} (2008) 103503}
  [\href{https://arxiv.org/abs/0806.0007}{{\ttfamily 0806.0007}}].

\bibitem{Desjacques:2018pfv}
V.~Desjacques, D.~Jeong and F.~Schmidt, \emph{{The Galaxy Power Spectrum and
  Bispectrum in Redshift Space}},
  \href{https://doi.org/10.1088/1475-7516/2018/12/035}{\emph{JCAP} {\bfseries
  12} (2018) 035} [\href{https://arxiv.org/abs/1806.04015}{{\ttfamily
  1806.04015}}].

\bibitem{Perko:2016puo}
A.~Perko, L.~Senatore, E.~Jennings and R.H.~Wechsler, \emph{{Biased Tracers in
  Redshift Space in the EFT of Large-Scale Structure}},
  \href{https://arxiv.org/abs/1610.09321}{{\ttfamily 1610.09321}}.

\bibitem{Desjacques:2009kt}
V.~Desjacques and R.K.~Sheth, \emph{{Redshift space correlations and
  scale-dependent stochastic biasing of density peaks}},
  \href{https://doi.org/10.1103/PhysRevD.81.023526}{\emph{Phys. Rev. D}
  {\bfseries 81} (2010) 023526}
  [\href{https://arxiv.org/abs/0909.4544}{{\ttfamily 0909.4544}}].

\bibitem{2000physics...9028N}
R.M.~Neal, \emph{{Slice sampling}},
  \href{https://doi.org/10.1214/aos/1056562461}{\emph{The Annals of Statistics}
  {\bfseries 31} (2003) 705 }.

\bibitem{Matsubara:2007wj}
T.~Matsubara, \emph{{Resumming Cosmological Perturbations via the Lagrangian
  Picture: One-loop Results in Real Space and in Redshift Space}},
  \href{https://doi.org/10.1103/PhysRevD.77.063530}{\emph{Phys. Rev. D}
  {\bfseries 77} (2008) 063530}
  [\href{https://arxiv.org/abs/0711.2521}{{\ttfamily 0711.2521}}].

\bibitem{Carlson:2012bu}
J.~Carlson, B.~Reid and M.~White, \emph{{Convolution Lagrangian perturbation
  theory for biased tracers}},
  \href{https://doi.org/10.1093/mnras/sts457}{\emph{Mon. Not. Roy. Astron.
  Soc.} {\bfseries 429} (2013) 1674}
  [\href{https://arxiv.org/abs/1209.0780}{{\ttfamily 1209.0780}}].

\bibitem{Wang:2013hwa}
L.~Wang, B.~Reid and M.~White, \emph{{An analytic model for redshift-space
  distortions}}, \href{https://doi.org/10.1093/mnras/stt1916}{\emph{Mon. Not.
  Roy. Astron. Soc.} {\bfseries 437} (2014) 588}
  [\href{https://arxiv.org/abs/1306.1804}{{\ttfamily 1306.1804}}].

\bibitem{Vlah:2016bcl}
Z.~Vlah, E.~Castorina and M.~White, \emph{{The Gaussian streaming model and
  convolution Lagrangian effective field theory}},
  \href{https://doi.org/10.1088/1475-7516/2016/12/007}{\emph{JCAP} {\bfseries
  12} (2016) 007} [\href{https://arxiv.org/abs/1609.02908}{{\ttfamily
  1609.02908}}].

\bibitem{Vlah:2018ygt}
Z.~Vlah and M.~White, \emph{{Exploring redshift-space distortions in
  large-scale structure}},
  \href{https://doi.org/10.1088/1475-7516/2019/03/007}{\emph{JCAP} {\bfseries
  03} (2019) 007} [\href{https://arxiv.org/abs/1812.02775}{{\ttfamily
  1812.02775}}].

\bibitem{Chen:2019lpf}
S.-F.~Chen, Z.~Vlah and M.~White, \emph{{The reconstructed power spectrum in
  the Zeldovich approximation}},
  \href{https://doi.org/10.1088/1475-7516/2019/09/017}{\emph{JCAP} {\bfseries
  09} (2019) 017} [\href{https://arxiv.org/abs/1907.00043}{{\ttfamily
  1907.00043}}].

\bibitem{Chen:2020zjt}
S.-F.~Chen, Z.~Vlah, E.~Castorina and M.~White, \emph{{Redshift-Space
  Distortions in Lagrangian Perturbation Theory}},
  \href{https://doi.org/10.1088/1475-7516/2021/03/100}{\emph{JCAP} {\bfseries
  03} (2021) 100} [\href{https://arxiv.org/abs/2012.04636}{{\ttfamily
  2012.04636}}].

\bibitem{Pellejero-Ibanez:2022efv}
M.~Pellejero-Ibanez, R.E.~Angulo, M.~Zennaro, J.~Stuecker, S.~Contreras,
  G.~Arico et~al., \emph{{The Bacco Simulation Project: Bacco Hybrid Lagrangian
  Bias Expansion Model in Redshift Space}},
  \href{https://arxiv.org/abs/2207.06437}{{\ttfamily 2207.06437}}.

\bibitem{Chudaykin:2020hbf}
A.~Chudaykin, M.M.~Ivanov and M.~Simonovi\'c, \emph{{Optimizing large-scale
  structure data analysis with the theoretical error likelihood}},
  \href{https://doi.org/10.1103/PhysRevD.103.043525}{\emph{Phys. Rev. D}
  {\bfseries 103} (2021) 043525}
  [\href{https://arxiv.org/abs/2009.10724}{{\ttfamily 2009.10724}}].

\bibitem{Chudaykin:2020aoj}
A.~Chudaykin, M.M.~Ivanov, O.H.E.~Philcox and M.~Simonovi\'c, \emph{{Nonlinear
  perturbation theory extension of the Boltzmann code CLASS}},
  \href{https://doi.org/10.1103/PhysRevD.102.063533}{\emph{Phys. Rev. D}
  {\bfseries 102} (2020) 063533}
  [\href{https://arxiv.org/abs/2004.10607}{{\ttfamily 2004.10607}}].

\bibitem{Jackson:1971sky}
J.C.~Jackson, \emph{{Fingers of God: A critique of Rees' theory of primoridal
  gravitational radiation}},
  \href{https://doi.org/10.1093/mnras/156.1.1P}{\emph{Mon. Not. Roy. Astron.
  Soc.} {\bfseries 156} (1972) 1P}
  [\href{https://arxiv.org/abs/0810.3908}{{\ttfamily 0810.3908}}].

\bibitem{abidi/baldauf:2018}
M.M.~{Abidi} and T.~{Baldauf}, \emph{{Cubic Halo Bias in Eulerian and
  Lagrangian Space}}, {\emph{ArXiv e-prints} (2018) }
  [\href{https://arxiv.org/abs/1802.07622}{{\ttfamily 1802.07622}}].

\bibitem{Rubira:2023vzw}
H.~Rubira and F.~Schmidt, \emph{{Galaxy bias renormalization group}},
  \href{https://arxiv.org/abs/2307.15031}{{\ttfamily 2307.15031}}.

\bibitem{1970Ap......6..320D}
A.G.~{Doroshkevich}, \emph{{Spatial structure of perturbations and origin of
  galactic rotation in fluctuation theory}},
  \href{https://doi.org/10.1007/BF01001625}{\emph{Astrophysics} {\bfseries 6}
  (1970) 320}.

\bibitem{1984ApJ...284L...9K}
N.~{Kaiser}, \emph{{On the spatial correlations of Abell clusters.}},
  \href{https://doi.org/10.1086/184341}{\emph{Astrophys. J. Lett.} {\bfseries
  284} (1984) L9}.

\bibitem{1985MNRAS.217..805P}
J.A.~{Peacock} and A.F.~{Heavens}, \emph{{The statistics of maxima in
  primordial density perturbations}},
  \href{https://doi.org/10.1093/mnras/217.4.805}{\emph{Mon. Not. Roy. Astron.
  Soc.} {\bfseries 217} (1985) 805}.

\bibitem{DESI:2023dwi}
{\scshape DESI} collaboration, \emph{{Validation of the Scientific Program for
  the Dark Energy Spectroscopic Instrument}},
  \href{https://arxiv.org/abs/2306.06307}{{\ttfamily 2306.06307}}.

\bibitem{Springel:2005mi}
V.~Springel, \emph{{The Cosmological simulation code GADGET-2}},
  \href{https://doi.org/10.1111/j.1365-2966.2005.09655.x}{\emph{Mon. Not. Roy.
  Astron. Soc.} {\bfseries 364} (2005) 1105}
  [\href{https://arxiv.org/abs/astro-ph/0505010}{{\ttfamily
  astro-ph/0505010}}].

\bibitem{Press:1973iz}
W.H.~Press and P.~Schechter, \emph{{Formation of galaxies and clusters of
  galaxies by selfsimilar gravitational condensation}},
  \href{https://doi.org/10.1086/152650}{\emph{Astrophys. J.} {\bfseries 187}
  (1974) 425}.

\bibitem{1992ApJ...399..405W}
M.S.~{Warren}, P.J.~{Quinn}, J.K.~{Salmon} and W.H.~{Zurek}, \emph{{Dark Halos
  Formed via Dissipationless Collapse. I. Shapes and Alignment of Angular
  Momentum}}, \href{https://doi.org/10.1086/171937}{\emph{Astrophys. J.}
  {\bfseries 399} (1992) 405}.

\bibitem{Lacey:1994su}
C.G.~Lacey and S.~Cole, \emph{{Merger rates in hierarchical models of galaxy
  formation. 2. Comparison with N body simulations}},
  \href{https://doi.org/10.1093/mnras/271.3.676}{\emph{Mon. Not. Roy. Astron.
  Soc.} {\bfseries 271} (1994) 676}
  [\href{https://arxiv.org/abs/astro-ph/9402069}{{\ttfamily
  astro-ph/9402069}}].

\bibitem{Gill:2004km}
S.P.D.~Gill, A.~Knebe and B.K.~Gibson, \emph{{The Evolution substructure 1: A
  New identification method}},
  \href{https://doi.org/10.1111/j.1365-2966.2004.07786.x}{\emph{Mon. Not. Roy.
  Astron. Soc.} {\bfseries 351} (2004) 399}
  [\href{https://arxiv.org/abs/astro-ph/0404258}{{\ttfamily
  astro-ph/0404258}}].

\bibitem{2009ApJS..182..608K}
S.R.~{Knollmann} and A.~{Knebe}, \emph{{AHF: Amiga's Halo Finder}},
  \href{https://doi.org/10.1088/0067-0049/182/2/608}{\emph{The Astrophysical
  Journal Supplement Series} {\bfseries 182} (2009) 608}
  [\href{https://arxiv.org/abs/0904.3662}{{\ttfamily 0904.3662}}].

\bibitem{Behroozi2013}
P.S.~{Behroozi}, R.H.~{Wechsler} and H.-Y.~{Wu}, \emph{{The ROCKSTAR
  Phase-space Temporal Halo Finder and the Velocity Offsets of Cluster Cores}},
  \href{https://doi.org/10.1088/0004-637X/762/2/109}{\emph{Astrophys. J.}
  {\bfseries 762} (2013) 109}
  [\href{https://arxiv.org/abs/1110.4372}{{\ttfamily 1110.4372}}].

\bibitem{phan2019composable}
D.~Phan, N.~Pradhan and M.~Jankowiak, \emph{Composable effects for flexible and
  accelerated probabilistic programming in numpyro}, {\emph{arXiv preprint
  arXiv:1912.11554} (2019) }.

\bibitem{bingham2019pyro}
E.~Bingham, J.P.~Chen, M.~Jankowiak, F.~Obermeyer, N.~Pradhan, T.~Karaletsos
  et~al., \emph{Pyro: Deep universal probabilistic programming}, {\emph{J.
  Mach. Learn. Res.} {\bfseries 20} (2019) 28:1}.

\bibitem{Lewis:2019xzd}
A.~Lewis, \emph{{GetDist: a Python package for analysing Monte Carlo samples}},
   \href{https://arxiv.org/abs/1910.13970}{{\ttfamily 1910.13970}}.

\bibitem{Gelman:1992zz}
A.~Gelman and D.B.~Rubin, \emph{{Inference from Iterative Simulation Using
  Multiple Sequences}},
  \href{https://doi.org/10.1214/ss/1177011136}{\emph{Statist. Sci.} {\bfseries
  7} (1992) 457}.

\bibitem{Michaux:2020yis}
M.~Michaux, O.~Hahn, C.~Rampf and R.E.~Angulo, \emph{{Accurate initial
  conditions for cosmological N-body simulations: Minimizing truncation and
  discreteness errors}},
  \href{https://doi.org/10.1093/mnras/staa3149}{\emph{Mon. Not. Roy. Astron.
  Soc.} {\bfseries 500} (2020) 663}
  [\href{https://arxiv.org/abs/2008.09588}{{\ttfamily 2008.09588}}].

\end{thebibliography}\endgroup

\end{document}